\documentclass[11pt,a4paper]{article}

\usepackage{amsmath,amsfonts,amssymb,amsthm,cite,a4}
\usepackage{feynmf}
\input amssym.def
\input amssym.tex
\usepackage[dvips]{graphics}
\usepackage[hypertex]{hyperref}

\allowdisplaybreaks[4]

\newtheorem{thm}{Theorem}[section]
\newtheorem{lem}[thm]{Lemma}

\newtheorem{prop}[thm]{Proposition}
\theoremstyle{definition}
\newtheorem{defn}[thm]{Definition}

\theoremstyle{remark}

\numberwithin{equation}{section}        

\DeclareMathOperator{\Tr}{Tr}           

\newcommand{\A}{\mathcal{A}}            
\renewcommand{\b}{\beta}
\newcommand{\B}{\mathcal{B}}            
\newcommand{\C}{\mathbb{C}}             
\newcommand{\D}{\mathcal{D}}
\newcommand{\eps}{\varepsilon}          
\newcommand{\E}{\mathcal{E}}
\newcommand{\Ga}{\Gamma}                
\newcommand{\ga}{\gamma}
\renewcommand{\H}{\mathcal{H}}          

\renewcommand{\L}{\mathcal{L}}          
\newcommand{\N}{\mathbb{N}}             
\newcommand{\ox}{\otimes}
\newcommand{\pa}{\partial}
\newcommand{\R}{\mathbb{R}}             
\newcommand{\sepword}[1]{\quad\mbox{#1}\quad} 
\renewcommand{\SS}{\mathcal{S}}         
\newcommand{\T}{\mathbb{T}}             
\newcommand{\Th}{\Theta}                
\renewcommand{\th}{\theta}
\newcommand{\thalf}{\tfrac{1}{2}}       
\newcommand{\tri}{\triangle}

\newcommand{\x}{\times}                 
\newcommand{\Z}{\mathbb{Z}}             

\newcommand{\vf}{\varphi}
\renewcommand{\:}{\colon}               

\begin{document}

\thispagestyle{empty}
CPT-P67-2006

\begin{center}

$\left.\right.$

\vspace{2cm}

{\huge
\textbf{Quantum field theory on projective modules}} \\

\vspace{1.5cm}

{\large\bfseries V. Gayral$^1$, J.-H.\ Jureit$^{2,3}$, 
T. Krajewski${}^{2,4}$ and R. Wulkenhaar${}^5$}  
 
\end{center}

\vspace{3cm}

{\large\textbf{Abstract:}} We propose a general formulation of
perturbative quantum field theory on (finitely generated) projective
modules over noncommutative algebras. This is the analogue of scalar
field theories with non-trivial topology in the noncommutative realm.
We treat in detail the case of Heisenberg modules over noncommutative
tori and show how these models can be understood as large rectangular
$p\times q$ matrix models, in the limit $p/q\rightarrow\theta$, where
$\theta$ is a possibly irrational number. We find out that the model
is highly sensitive to the number-theoretical aspect of $\theta$ and
suffers from an UV/IR-mixing. We give a way to cure the entanglement
and prove one-loop renormalizability.

\qquad

 
\begin{quote}

\end{quote}

\vspace{5cm}

\noindent 
$^1$ Matematisk Afdeling, K\o benhavns Universitet, 
gayral@math.ku.dk\\
$^2$ Centre de Physique Th\'eorique\\
-- Unit\'e Mixte de Recherche du CNRS (UMR 6207) et des
Universit\'es Aix-Marseille I, Aix-Marseille II et de l'Universit\'e
du Sud Toulon-Var\\
-- Laboratoire affili\'e \`a la FRUMAM -- FR 2291\\
$^3$ Also at Christian-Albrechts-Universit\"at Kiel,  
jureit@cpt.univ-mrs.fr\\
$^4$ Also at Universit\'e de Provence,
krajew@cpt.univ-mrs.fr \\
$^5$ Mathematisches Institut, Westf\"alische Wilhelms-Universit\"at
M\"unster, \\ raimar@math.uni-muenster.de

\newpage

\tableofcontents

\section{Introduction}

In its most general acceptance, quantum field theory (QFT) can be defined as
the theory of infinite-dimensional dynamical systems which are based
on geometrical concepts like, for instance,  locality and invariance
principles. Within the path integral approach, one first defines an
action functional $S[\phi]$, the argument of which is a field  $\phi$ that
belongs to a fixed configuration space ${\cal C}$. Then, one defines
and studies the (euclidian) path integral
\begin{equation}
\langle{\cal F}\rangle=
\frac{\displaystyle \int_{{\cal C}}[\D\phi]{{\cal F}[\phi]}e^{-S[\phi]}}
{\displaystyle \int_{{\cal C}}[\D\phi]e^{-S[\phi]}},
\end{equation}
where ${\cal F}$ belongs to a suitable class of functionals of the
field. Such a framework is at the root of our current understanding of
the dynamics of elementary particles, where the configuration space
contains matter fields (section of various bundles over space-time)
and gauge fields (connections over these bundles). For a trivial
topology of these bundles, all these fields can be understood as
functions over the space-time manifold ${\cal M}$. Then the action
functional involves the integral over ${\cal M}$ of a Lagrangian
density ultimately constructed out of pointwise products of the
fields.

Over the last decade, a radical modification of this construction has
proven to be of interest in mathematical physics: instead of being
grounded in ordinary differential geometry, QFT can also fruitfully
use the concepts introduced in {\it Noncommutative Geometry}. The
latter is a branch of mathematics pioneered in the eighties by A.
Connes (see \cite{connes} and \cite{garden} for an overview of recent
developments), that extends geometrical concepts to a wide class of
spaces whose coordinate algebras are noncommutative, instead of being
merely functions with the pointwise product. For instance,
noncommutative analogues of vector bundles, which are the natural
receptacles for matter fields, are defined as projective modules over
the noncommutative algebra of coordinates. Noncommutative field theory
(NCFT) has grown up from the need of new methods in understanding a
wide range of problems in theoretical physics, ranging from the
construction of open string field theory \cite{sft} to the
understanding of the dynamics of string theory \cite{seiberg} and
M-theory \cite{CDS} in magnetic backgrounds. Though not directly
related to NCFT, other applications of noncommutative geometry include
the geometry of aperiodic solids \cite{aperiodic} as well as the
standard model of particle physics \cite{sm}.

The NCFT's involved in most of these applications are based on
configuration spaces ${\cal C}$ which consist of matrices with
coeffcients in the algebra of quantum coordinates ${\cal A}$. The latter are
deformations of the commutative algebra of coordinates over
space-time, with deformation parameters depending on some background.
In the language of noncommutative geometry, these configuration spaces
correspond to free modules, which are the analogues of trivial vector
bundles. There is first success with renormalisation to all orders of
such NCFT's \cite{Grosse:2004yu}. Since noncommutative geometry is
versatile enough to include non-trivial bundles as projective modules,
the restriction to free modules appears to be rather unnatural. While
projective modules already appear in the string theory literature (see
for instance \cite{Ktheory} and \cite{seiberg}), the NCFT's they
naturally define has not yet been investigated. The present work aims
at filling this gap, in the case of the simplest non-trivial 
projective modules over noncommutative tori.

This paper is organized as follows.

In section 2, we first give a general construction of the configuration
space and the action functional in the framework of noncommutative
geometry, making use of a spectral triple $\left({\cal A},{\cal
    H},{\cal D}\right)$. Then, we illustrate the general theory for
some Heisenberg modules over $d$-dimensional noncommutative
tori. These modules are constructed using representations of the
Heisenberg group that can be formulated either using functions over 
${\Bbb R}^{\frac{d}{2}}$ or holomorphic functions on ${\Bbb
  C}^{\frac{d}{2}}$. In this last case, we illustrate how such a
module appears naturally in the study of an electron on the plane in
an external magnetic field, confined to its ground state in the
presence of a periodic potential.

Section 3 is devoted to the construction of the NCFT based on
Heisenberg modules. We first give the perturbative expansion of the
path integral using a Hubbard-Stratonovitch transformation. The
resulting theory makes use of planar diagrams reminiscent of
rectangular matrix models and exhibits a duality symmetry. Then, we
use the position space formulation to strengthen this analogy with
rectangular matrices. Finally, we give the general rules for the
computation of Feynman diagrams in the holomorphic representation.
 
In section 4 we use the holomorphic representation to compute one-loop
diagrams. While the divergent planar diagrams turn out to be
renormalizable by standard methods, new phenomena occur in the
non-planar case. If the deformation parameter fulfills a Diophantine
condition, we give explicit bounds on the amplitude showing that they
are renormalized by an extra counterterm.

Section 5 deals with the effect of the new counterterm. It is shown
that it is harmless and does not spoil the general properties of the
NCFT.

Finally, in section 6 we compute the $\beta$ functions for the two 
interacting terms.

\section{Classical field theory}

\subsection{Projective modules in noncommutative geometry}

To begin with, let us recall some basic facts about projective modules in
noncommutative geometry and the way they enter in the construction of
a noncommutative field theory. A much more thorough presentation of the
subject can be found in \cite{connes} (see also \cite{garden} for an
overview of recent developments in noncommutative geometry). Here, we 
restrict ourselves to the
amount of material that is necessary in
order to construct our noncommutative field theory.

First, we recall that the basic idea of noncommutative geometry is to
replace the commutative data of a space ${\cal X}$ by a possibly
noncommutative algebra ${\cal A}$ that plays the role of the
complex-valued functions on ${\cal X}$. In general, it is necessary to
assume that ${\cal A}$ is a C$^{*}$-algebra, which means that it is
equipped with an involution $*$ and a norm which are compatible. This
is motivated by the Gelfand-Naimark theorem which asserts that
commutative C$^{*}$-algebras with unit are equivalent to compact
Hausdorff spaces.

In the same spirit, the notion of a finite-dimensional complex vector
bundle is extended to the noncommutative realm by first looking at the
structure of its space of sections. By the Serre-Swan theorem, spaces
of sections of vector bundles over $\cal X$ are equivalent to finitely
generated projective modules ${\cal E}$ (projective modules for short)
over the algebra of continuous functions over $\cal X$. Up to an
isomorphism, such a module can always be realized as the right ${\cal
  A}$-module $e{\cal A}^{N}$, where $e$ is a projection in the algebra
of $N\times N$ matrices with coefficients in ${\cal A}$. The relation
between the idempotent and the standard construction of a vector
bundle using transition functions goes as follows. Cover the compact
space ${\cal X}$ by $N$ open sets $\left\{ U_{i}\right\}$ defining a
good cover and let $\sum_i|f_{i}|^{2}$ be a partition of unity
associated to this cover. Using the transition functions $g_{ij}$, we
define the $N\times N$ matrix $e_{ij}=f_{i}^{*}g_{ij}f_{j}$, whose
entries are complex-valued functions over ${\cal X}$. Then, it is
straightforward to check that $e$ is a projection using the cocycle
condition $g_{ij}=g_{ik}g_{kj}$. Accordingly, in the general setting
one defines by duality a noncommutative vector bundle to be a
finitely generated projective module over a possibly noncommutative
$C^{*}$-algebra. From a classical field theoretical perspective, the
module ${\cal E}$ is the natural receptacle for the matter fields with
non-trivial topology. If the topology is trivial, these fields simply
live in a free module ${\cal A}^{N}$, the simplest version being the
algebra itself.

Rather than at the $C^*$-level, we will work here at the smooth one,
i.e.\ we will assume that $\A$ is a Fr\'echet pr\'e-$C^*$-algebra.
This means that $\A$ is a dense sub-algebra of a $C^*$-algebra $A$,
which is stable under holomorphic functional calculus and which is
endowed with a Fr\'echet topology.  We will assume that the topology
comes from a set of semi-norms, say $\{p_i\}$.  Without any further
structure, $\E$ is simply a projective right module over $\A$; it will
be promoted to a topological vector space later on.  There are many
reasons why it is preferable to work at the smooth level.  First, our
construction of perturbative field theory relies on the theory of
distributions, useless at the level of $C^*$-algebras and
$C^*$-modules (our projective modules will be soon endowed with an
$\A$-valued hermitian paring promoting it to a pre-$C^*$-module). But
also for many interesting examples (e.g.\ Heisenberg modules and
$C^*$-dynamical systems \cite{connesc*}), one does not lose any
geometrical information passing from $C^*$- to smooth structures.

To construct a noncommutative field theory out of the previous data,
one further needs a suitable space of differential forms as well as
some scalar products on these forms. Differential forms are defined
through a graded differential algebra: 
\begin{equation}
\Omega({\cal A}):=\bigoplus_{n\in\N}\,\Omega_{n}({\cal A}),
\end{equation}
which is a graded bi-module over $\A$,
together with a nilpotent differential operator
\begin{equation}
\mathrm{d}\:\Omega_{n}({\cal A})\rightarrow \Omega_{n+1}({\cal A}),
\end{equation}
fulfilling the graded Leibniz rule. Note that in the general noncommutative
setting, it is not assumed that $\Omega({\cal A})$ is graded
commutative. Usually, one assumes that $\Omega_{0}({\cal A})={\cal
A}$ and that $\Omega({\cal A})$ is equipped with an involution
$*$ compatible with those of $\A$. 
The scalar product $\langle \cdot,\cdot\rangle_{n}$ on
$\Omega_{n}({\cal A})$ is assumed to be compatible with the left and
right actions of ${\cal A}$,
\begin{equation}
\left\{
\begin{array}{r@{\;}c@{\;}l}
\langle \omega,\eta a\rangle_{n}&=&\langle \omega a^{*},\eta
\rangle_{n},
\\[1ex]
\langle \omega,a\eta \rangle_{n}&=&\langle a^{*}\omega ,\eta \rangle_{n},
\end{array}
\right. 
\end{equation}
for all $\omega,\eta\in\Omega_{n}({\cal A})$ and $a\in{\cal A}$. In
full generality, the scalar product  in degree $0$ is always obtained from a
faithful state $\Psi$, i.e.\ a normalized positive-definite 
linear functional on $\A$:
\begin{equation}
\langle a,b\rangle_{0}:=\Psi\left( a^{*}b\right), \quad a,b\in\A.
\end{equation}
In view of the application we have in mind, it is more appropriate to
require that the state is actually a faithful trace on $\A$, so that
\begin{equation}
\langle a,b\rangle_{0}=\mathrm{Tr}_{{\cal A}}\left( a^{*}b\right).
\end{equation}

For instance, in the simplest version of a (commutative or not) 
Yang-Mills theory, one
starts with a topologically trivial connection given by an anti-hermitian
element $A$ of $\Omega_{1}({\cal A})$, out of which we define the
curvature as $F=dA+A^{2}\in\Omega_2(\A)$. 
The Yang-Mills action is then constructed
as $\langle F,F\rangle_{2}$, where $\langle \cdot,\cdot\rangle_{2}$ is an
invariant scalar product on $\Omega_{2}({}\cal A)$. This invariance
condition simply states that the left and right actions of ${\cal A}$
are compatible with the scalar product. This ensures gauge invariance,
where gauge transformations are given by unitary elements
$u\in{\cal A}$, acting as
\begin{equation}
\left\{
\begin{array}{rcl}
A&\rightarrow&u^{-1}Au+u^{-1}du,\cr
F&\rightarrow&u^{-1}Fu.
\end{array}
\right.
\end{equation}
Such a scalar product encodes an information that goes beyond the
topological level, contained in the algebra ${\cal A}$ alone. For
instance, in a four-dimensional Yang-Mills theory, it amounts to the
choice of a conformal structure. Before we come to grips with such an
issue, let us note that one can also define in full generality a
noncommutative Chern-Simons theory out of a cyclic 3-cocycle
\cite{chern-simons} which is a convenient setting to develop open string field
theory \cite{sft}. As a topological theory, the construction of
Chern-Simons theory does not rely on such a scalar product.

Differential forms and their scalar products are conveniently
constructed out of a spectral triple $\left( {\cal A}, {\cal D}, {\cal
    H} \right)$. This involves a Hilbert space ${\cal H}$ carrying a 
representation $\pi$ of the algebra ${\cal A}$ by bounded operators 
and a self-adjoint
unbounded operator ${\cal D}$ with compact resolvent. This data is
constrained by compatibility conditions, allowing to reconstruct a
smooth Riemannian manifold when $\A$ is commutative.  For instance,
the commutators of $\D$ with the elements of ${\cal A}$ must extend to
bounded operators.  For a spectral triple, there is a notion of
dimension, given by the growth of the eigenvalues of $|\D|$. More
precisely, such a triple has spectral dimension $d$ if the resolvent
of $\D$ belongs to the $d$-th weak Schatten ideal $\L^{d,\infty}(\H)$.
The latter is the ideal of compact operators whose sequence singular
values (in decreasing order and counted with multiplicity) are
$\mathcal{O}(n^{-1/d})$. With a spectral triple, one defines
differential forms as a representation of the universal differential
algebra
\begin{equation}
a_{0}\,da_{1}\cdots da_{n}\quad\mapsto\quad 
\pi(a_{0})
\left[{\cal D},\pi(a_{1}) \right]\cdots\left[{\cal D},\pi(a_{n})\right].
\end{equation}
Such representation can be quite pathological since it may happen
that the image of an element of the universal differential algebra 
is zero, whereas the image
of the differential of this element is not! To overcome this problem,
one has to divide the resulting algebra by a graded differential ideal
(the so-called junk ideal). Fortunately we can ignore this point here.
This is because the first non-trivial component of this ideal occurs at the
level of two-forms only. However, this is highly relevant for a proper
formulation of noncommutative Yang-Mills theory in this framework.\\
For a triple of spectral dimension $d$, there is a canonical way to
define the scalar product between two differential forms in $\Omega_{n}(\cal
A)$:
\begin{equation}
\langle\omega,\eta\rangle_{n}=
\mathrm{Tr}_{\,\mathrm{Dix}}\big( \omega^{*}\eta\,
(1+|{\cal D}|)^{-d}\big).
\end{equation}
Here, $\mathrm{Tr}_{\,\mathrm{Dix}}$ is any of the Dixmier traces.
Such an object is a singular trace defined on the ideal
$\L^{1,\infty}(\H)$ and
is heuristically given by the coefficient of the 
logarithmic divergence of the ordinary operator trace (see
\cite{connes} for more details on Dixmier traces). There is no point to
enter here in the mathematical subtleties of the theory of
Dixmier traces. This is because the operator $(1+|\D|)^{-d}$ we will
consider in our example belongs to the class of `measurable
operators'. The latter consists of elements of $\L^{1,\infty}(\H)$
for which any Dixmier trace gives the same result.\\
It is important to know that the Dirac operator ${\cal D}$
encodes the metric aspect of noncommutative geometry and allows also to
define the fermionic action for spinors $\psi\in{\cal H}$ as
$\langle\psi,{\cal D}\psi\rangle_\H$.\\
It is worthwhile to mention that the use of a spectral triple to
construct noncommutative field theory is highly convenient (mainly
because it allows to define differential forms with scalar products
in a canonical way) but not necessary. Indeed, 
our construction is much more general and  
works for any differential calculus with scalar products on each
component.

Turning back to the projective module ${\cal E}$, one has to define a
connection $\nabla$. This is an operator that 
extends to elements of ${\cal E}$ the
differential of a given differential calculus
$\big(\Omega(\A),d\big)$. In general, a connection is a linear map 
\begin{equation}
\nabla\:{\cal E}\rightarrow
{\cal E}\otimes_{{\cal A}}\Omega_{1}\left(\cal A\right),
\end{equation}
fulfilling the Leibniz rule
\begin{equation}
\nabla\left( \phi a\right)=\nabla\left(\phi\right)a+\phi\otimes da,
\end{equation} 
for any $\phi\in{\cal E}$ and $a\in{\cal A}$. The projective module
${\cal E}$ is further equipped with a hermitian structure
$(\cdot,\cdot)_{{\cal A}}:\mathcal{E} \times \mathcal{E} \to
\mathcal{A}$, which is an ${\cal A}$-valued sesquilinear form on
${\cal E}$, satisfying the following compatibility and positivity
conditions:
\begin{equation}
\label{eq:paring}
\left\{
\begin{array}{r@{\;}c@{\;}l}
\left(\phi,\chi a\right)_{{\cal A}}&=&\left(\phi,\chi\right)_{{\cal
    A}}a, \\[1ex]
\left(\phi,\chi \right)_{{\cal A}}&=&\left(\chi,\phi\right)_{{\cal
    A}}^{*}, \\[1ex]
\left(\phi,\phi\right)_{{\cal A}}&\geq&0,
\end{array}
\right.
\end{equation}
for all $\phi,\chi\in{\cal E}$ and $a\in{\cal A}$. We recall that $a$
is a positive element of ${\cal A}$ if it can be written as $a=b^{*}b$
with $b\in{\cal A}$. Besides, the connection has to be compatible with
the hermitian structure in the sense that
\begin{equation}
\mathrm{d}\left(\phi,\chi\right)_{{\cal A}}=
\left(\nabla\phi,\chi\right)_{{\cal A}}+
\left(\phi,\nabla\chi\right)_{{\cal A}},
\end{equation}
for all $\phi,\chi\in{\cal E}$. We would like to stress that the field
theory we are constructing is Euclidean, precisely because the paring 
\eqref{eq:paring} is positive definite.\\
As a side remark, let us note that
one can define the curvature of  $\nabla$ as $F=\nabla^{2}$, where
$\nabla$ has been extended by the Leibniz rule to 
${\cal E}\otimes_{{\cal A}}\Omega\left({\cal A}\right)$. This is the
starting point for the development of a general gauge theory in the
framework of noncommutative geometry.

Within the supplementary structure of an hermitian $\A$-valued paring, $\E$
can be promoted to a topological vector space in different ways. 
When $\A$ comes from 
a $C^*$-algebra $A$, $\E$ is called a pr\'e-Hilbert module and it can
be completed with respect to the norm
$|||\phi|||:=\sqrt{\|\langle\phi,\phi\rangle_{_\A}\|}$, where $\|.\|$ 
denotes the
$C^*$-norm of $A$. The resulting Banach space is obviously called a
right Hilbert- or $C^*$-module. One can also defined a
Fr\'echet topology on $\E$, finer than the $C^*$ one, 
via the semi-norms $\{p_i\}$ of $\A$, setting
$q_i(\phi):=\sqrt{p_i(\langle\phi,\phi\rangle_{_\A})}$, for
  all $\phi\in\E$. When $\E$
is complete with respect to this topology, that we will assume from
now on, we will call it a Fr\'echet projective right $\A$-module.
Finally, when $\A$ possesses a faithful trace $\Tr_\A$, $\E$ can be
completed to an Hilbert space $\H_\E$ using the scalar product
$\langle.,.\rangle_\E$, obtained
by composition of the trace with the hermitian structure:
\begin{equation}
\langle\phi,\chi\rangle_\E:=\Tr_\A\left[(\phi,\chi)_\A\right].
\label{scalarE}
\end{equation}
We use the notation $\H_\E$ for this Hilbert space, in order to
emphasize its canonical nature, once a trace and an hermitian structure
are given.

\quad

If we realize the projective module as $e{\cal A}^{N}$ for some
hermitian idempotent $e\in M_{N}({\cal A})$, then a connection is easily
defined as $\nabla(e\xi):=e\mathrm{d}\left(e\xi\right)+eAe\xi$,
$\xi\in\A^N$, where $A$ is an anti-hermitian matrix in $M_{N}(\cal A)$. In this
case, the hermitian structure is given by 
$\left(e\xi,e\zeta\right)_{{\cal A}}:=\xi^{*}e\,\zeta$,
and the compatibility of the connection follows from $(eAe)^{*}=-eAe$.

Finally, let us construct a functional action for a classical field
$\phi$ in a projective
module ${\cal E}$. To this aim, we have to extend the scalar
product $\langle.,.\rangle_{1}$ on $\Omega_{1}({\cal A})$ to a scalar
product on $\E\ox_\A\Omega_1(\A)$. It will allow 
to construct the kinetic term out of $\nabla\phi$.
Let us construct this scalar product on $\E\ox_\A\Omega_n(\A)$, for
any $n\in\N$. 
For $\Phi,\Psi\in\E\ox_\A\Omega_n(\A)$, we write 
\begin{equation}
\Phi=\sum_{i}\phi_{i}\otimes \omega_{i}, \quad
\Psi=\sum_{i}\psi_{i}\otimes \eta_{i},
\end{equation}
with $\phi_{i},\psi_i\in{\cal E}$ and 
$\omega_{i},\eta_i\in\Omega_{n}(\cal A)$.
Then, one can define the scalar product $\langle.,.\rangle_{\A,n}$  
using both $\langle.,.\rangle_n$ and $(.,.)_\A$:
\begin{equation}
\label{scalarfe}
\langle\Phi,\Psi\rangle_{\A,n}:=
\sum_{i,j}\langle\omega_{i},
\left(\phi_{i},\psi_{j}\right)_{{\cal A}}\eta_{j}\rangle_{n}.
\end{equation}
In particular, it allows to define the kinetic term as
\begin{equation}
\langle\nabla\phi,\nabla\phi\rangle_{\A,1}.
\end{equation}
To construct the mass term, we simply use the scalar product
\eqref{scalarE}:
\begin{equation}
\mu_0^2\,\langle\phi,\phi\rangle_\E=
\mu_0^{2}\,\mathrm{Tr}_{\cal A}[\left( \phi,\phi\right)_{{\cal A}}].
\end{equation}
For the interaction term, which should be a polynomial in the field,
we can extend the former construction. For instance,
the basic $\lambda\phi^{4}$ interaction reads
\begin{equation}
\frac{\lambda}{2}\,\mathrm{Tr}_{\cal A}
\big[\left( \phi,\phi\right)_{{\cal A}}^2\big].
\end{equation}
More generally, one can construct arbitrary monomials
\begin{equation}
\label{monom}
\frac{\lambda}{n}\,\mathrm{Tr}_{\cal A}
\big[\left( \phi,\phi\right)_{{\cal A}}^n\big].
\end{equation}
We may also consider products of such terms. Despite they look very
unnatural with respect to ordinary QFT, they should be needed to
obtain a stable quantum theory; stable with respect to the
renormalization group flow. We will see in the next sections that for
a $\phi_4^4$ theory on the Heisenberg module, such a `product of
trace' term is precisely the missing term which will allow to cure the
UV/IR mixing problem.

The basic action we shall use in the sequel reads
\begin{equation}
S[\phi,\phi^{\dagger}]:=
\langle \nabla\phi,\nabla\phi\rangle_{\A,1}+
\mu_0^{2}\,\mathrm{Tr}_{\cal A}\big[\left( \phi,\phi\right)_{{\cal A}}
\big]+
\frac{\lambda}{2}\,\mathrm{Tr}_{\cal A}
\left[\big( \phi,\phi\right)_{{\cal A}}^2\big].
\label{actiongeneral}
\end{equation}

Here we have followed the traditional notation of QFT where a complex
field $\phi$ and its conjugate $\phi^{\dagger}$ are treated as
independent variables. In our setting, $\phi^{\dagger}$ has to be
considered as an element of the dual module ${\cal E}^{*}$ which
consists of ${\cal A}$-linear forms on $\E$, defined by
\begin{equation}
\phi^{\dagger}\left( \chi\right):=\left(\phi,\chi\right)_{\cal A},
\end{equation}
for any $\chi\in{\cal A}$. This distinction between $\phi$ and
$\phi^{\dagger}$ will prove to be very convenient when developing the
perturbative path integral approach, for instance in getting the right
symmetry factors. However,
it is important to notice that $\E^*$ is not the topological dual of
$\E$. Moreover, it should be clear that $\E^*\simeq\E$ as a linear
space, the identification being given by
$\phi\leftrightarrow(\phi,.)_\A^*=(.,\phi)_\A$. Also, $\E^*$ carries the
same Fr\'echet topology than $\E$. 

In the case of a field theory constructed out of a spectral triple,
we would like to stress that the choice of 
the order $n$ of the monomials \eqref{monom} involved in the interaction term
depends on the spectral dimension of the triple $d$. 
In the general setting, $n$ should be related 
to the spectral properties of the propagator, which is the bounded
operator on $\H_\E$ given by the inverse of the (densely defined) positive 
operator $H$, corresponding to the quadratic part of the action
\eqref{actiongeneral}. More precisely, $H$ is defined in terms of the
following quadratic form on $\E$
\begin{equation}
\langle\phi,H\chi\rangle_\E:=\langle
\nabla\phi,\nabla\chi\rangle_{\A,1}, \quad \phi,\chi\in\E\subset\H_\E.
\end{equation}
It is precisely because we are going to study a model coming from a
spectral triple of spectral dimension 4 
that we focus on quartic interaction.

It may seems to be quite restrictive to ask for the existence of a
faithful trace to define the classical action. Faithfulness is
required in order that the quantum theory has a power-counting
properly related to the spectral properties of the propagator of the
model. But traciality is not needed at all, it simply makes the
noncommutative models closer to the commutative one. For instance,
there are numerous noncommutative algebras giving rise to non-trivial
spectral triples (e.g.\ $SU_q(2)$ \cite{trieste}), that do not posses
any faithful trace. In such circumstance, one can define the classical
theory by replacing everywhere the trace $\Tr_\A$ by a faithful state
$\Psi$ on $\A$.

It is also worthwhile to notice that this theory is naturally coupled
to gauge fields since it is invariant under
\begin{equation}
\left\{
\begin{array}{rcl}
\phi&\rightarrow&u\,\phi\cr
\nabla&\rightarrow&u\ox1_{\Omega_1(\A)}\nabla u^{-1},
\end{array}
\right.
\end{equation}
where $u$ is a unitary element of the algebra ${\cal
  B}:=\mathrm{End}_{{\cal A}}({\cal E})$ of ${\cal A}$-linear
transformations of ${\cal E}$. Thus the gauge invariant action for the
field $\phi$ coupled to a Yang-Mills connection
$\nabla$ reads
\begin{equation}
S[\phi,\phi^{\dagger},\nabla]:=
\langle F,F\rangle_{2}+
\langle \nabla\phi,\nabla\phi\rangle_{\A,1}+
\mu_0^{2}\,\mathrm{Tr}_{\cal A}\big[\left( \phi,\phi\right)_{{\cal A}}
\big]+
\frac{\lambda}{2}\,\mathrm{Tr}_{\cal A}
\big[\left( \phi,\phi\right)_{{\cal A}}^2\big],
\label{actiongauge}
\end{equation}
which involves the scalar product $\langle\cdot,\cdot\rangle_{2}$ on the
the space of 2-forms $\Omega_{2}({\cal A})$. As already mentioned, 
in the context of spectral triple, the curvature is in principle  
an equivalence class of 2-forms (modulo the junk ideal). However, we
still ignore this point since we are
not going to include a gauge degree of freedom when developing the
quantum theory. For that reason, it is preferable to stick to
the action functional \eqref{actiongeneral}.

\subsection{Noncommutative tori}
\subsubsection{Geometric structures}

Let us now work out the previous construction in the case of a
$d$-dimensional smooth noncommutative torus. The latter is defined through
its algebra of coordinates, ${\cal A}_{\Theta}$, which is the algebra of
all power series
$a=\sum_{\gamma\in{\Bbb Z}^{d}}a_{\gamma}\,U_{\gamma}$,
 with fast decreasing coefficients $\{a_{\gamma}\}\in\SS(\Z^d)$.
Here, $U_{\gamma}$ are unitary elements of ${\cal A}_{\Theta}$ fulfilling
\begin{equation}
U_{\gamma}U_{\gamma'}=e^{-i\pi\Theta(\gamma,\gamma')}\,U_{\gamma+\gamma'},
\end{equation}
where $\Theta$ is a skew-symmetric real $d\times d$ matrix which defines a
2-cocycle on the group ${\Bbb Z}^{d}\subset{\Bbb R}^{d}$. 
The algebra $\A_\Th$ is endowed with its natural Fr\'echet topology, 
coming from the set of semi-norms
$p_n(a)=\sup_{\ga\in\Z^d}\big((1+|\ga|^2)^n|a_\ga|\big)$, 
$n\in\N$, $a\in\A_\Th$.
In analogy with its commutative counterpart, where the unitaries
$U_\ga$'s are nothing but the Fourier modes, the smoothness condition
relies on the rapid decay of the sequence $\{a_\ga\}$. 

The structure of this algebra depends on the arithmetical properties of the
entries of the matrix $\Theta$. Three typical cases have to be distinguished.
\begin{itemize}
\item
If all the entries of $\Theta$ are integers, then  the algebra ${\cal
  A}_{\Theta}$ is commutative and can be identified with the algebra
of smooth functions on an ordinary torus.
\item
If all the entries of are rational numbers, then ${\cal A}_{\Theta}$
can be realized as a bundle of matrix algebras over an ordinary
torus. This is clear for $d=2$, where the unique parameter
$\theta=p/q$ in $\Theta$
determines the size of the matrices to be $q$. The general case
follows from reducing the matrix $\Theta$ to a direct sum of $2\times
2$ matrices plus a zero
matrix (that corresponds to the null space of $\Th$), 
using a transformation in $SL(d,{\Bbb Z})$. The algebra
of functions
over the underlying torus is nothing but the center of ${\cal A}_{\Theta}$.  
\item
If all the entries of $\Theta$ are irrational numbers and $\Th$ is
invertible,  then the center
of the algebra is trivial and ${\cal A}_{\Theta}$ is a noncommutative
space that cannot be reduced to an ordinary space. 
\end{itemize}

In our analysis of noncommutative field theories, we are mostly
interested in the irrational case that
exhibits some new phenomena. However, it is also interesting to keep in
mind the first two cases since they can always be understood in the
context of commutative field theories with matrix-valued fields. In
the discussion of the non-planar diagrams, it will be
necessary to single out irrational numbers that fulfill a Diophantine 
condition. The latter are irrational numbers that are `far away from
the rationals'. The use of the Diophantine condition is
not infrequent in noncommutative geometry. For instance, it appears to
be crucial in the computation of the Hochschild cohomology of ${\cal
  A}_{\Theta}$ \cite{ihes}, as well as for the heat-invariants \cite{GIV}.

The differential algebra can be constructed out of the $d$ commuting
derivations
\begin{equation}
\label{derivation}
\delta_{\mu}\left( U_{\gamma}\right):=2i\pi\gamma_{\mu}
\, U_{\gamma},\quad \gamma=(\gamma_{1},\dots,\gamma_{d})\in\Z^d,
\end{equation}
which is the infinitesimal form of the proper action of $\T^d$ on $\A_\Th$
given by
\begin{equation}
(e^{2i\pi\alpha_1},\cdots,e^{2i\pi\alpha_d})\cdot U(\gamma):=
e^{2i\pi\alpha\cdot\gamma}U(\gamma).\label{tor}
\end{equation} 
Elements of $\Omega_{n}\left({\cal A}_{\Theta}\right)$ are completely
antisymmetric multiplets  $\omega_{\mu_{1},\cdots,\mu_{n}}$ 
of ${\cal A}_{\Theta}$. The multiplication and the
differential in $\Omega\left({\cal A}_{\Theta}\right)$ obey the same
  algebraic rules as the wedge product and the de Rham
  differential. The algebra ${\cal A}_{\Theta}$ has a faithful trace 
$\mathrm{Tr}_{{\cal A}_{\Theta}}$ defined by
\begin{equation}
\mathrm{Tr}_{{\cal A}_{\Theta}}
\Big( \sum_{\gamma\in{\Bbb Z}^{d}}a_{\gamma}\,U_{\gamma}
\Big):=a_{0}.
\end{equation}
If ${\cal A}_{\Theta}$ is commutative, this trace is nothing but the
integral over the underlying torus, with a volume normalized to 1.
In the irrational case,  $\mathrm{Tr}_{{\cal A}_{\Theta}}$ is the
unique faithful trace up to normalization. The scalar product on 
$\Omega_{n}\left({\cal A}_{\Theta}\right)$ is made out of the trace
\begin{equation}
\langle \omega,\eta\rangle_{n}:=\mathrm{Tr}_{{\cal A}_{\Theta}}
\left( \omega^{*}_{\mu_{1}\cdots\mu_{n}}\eta^{\mu_{1},\cdots,\mu_{n}}\right),
\end{equation}
where the euclidian metric and Einstein's summation convention have
been used.  This construction follows readily from the general
  principles, using the spectral triple $\left( {\cal A}_{\Theta},{\cal
    H},{\cal D}\right)$ where ${\cal H}=A_{\Theta}^{2^{[d/2]}}$, with
$A_\Th$ the completion of $\A_\Th$ with respect to the norm induced
by the scalar product $\langle a,b\rangle_{\A_\Th}:=\Tr_{\A_\Th}(a^*b)$ 
and where ${\cal   D}=i\Gamma^{\mu}\delta_{\mu}$ is the standard 
euclidian Dirac operator. Note that a noncommutative torus is a noncommutative 
manifold without
boundary, in the sense that the integral of a derivative always
vanishes: 
$\mathrm{Tr}_{{\cal A}_{\Theta}}\left(\delta_{\mu}U_{\gamma}\right)=0$.
This relation is particularly useful in the study of classical field
theories, because it allows to derive classical field equations and
invariance laws.  


\subsubsection{Heisenberg modules}

A wide class of projective modules over ${\cal A}_{\Theta}$ can be
constructed as projective representations of groups of the type
$G\times\widehat{G}$ with $G={\Bbb R}^{p}\times{\Bbb
  Z}^{q}\times F$, where $F$ is a finite Abelian group and
$\widehat{G}$ the dual of $G$ \cite{rieffel}. Equivalently, these
projective representations can be thought of as representations of the
Heisenberg groups associated to the corresponding central extensions.
 The group $G\times\widehat{G}$ acts by unitary operators on the Hilbert space 
$L^2(G,dg)$ obtained by completion of the space of smooth
fast decreasing functions ${\cal S}(G)$ on $G$. For $(g,\mu)\in G\x
\widehat{G}$ and $\psi\in L^2(G,dg)$, the action is
\begin{equation}
\label{projrep}
T_{g,\mu}\psi(x):=\mu(g)^{1/2}\,\mu(x)\,\psi(x+g).
\end{equation}
Then, given a lattice 
$\Gamma$ isomorphic to ${\Bbb Z}^{d}$ in $G\times\widehat{G}$, one can
represent the algebra $\A_\Th$ (acting on the right) on $L^2(G,dg)$ 
by restricting the action \eqref{projrep} to the sub-group $\Ga$: 
\begin{equation}
\phi U_\ga:=T_{g,\mu}\phi,\quad \phi\in L^2(G,dg), \,\,
\ga=(g,\mu)\in\Ga.
\end{equation}
The
multiplication law of the algebra ${\cal A}_{\Theta}$ is satisfied
with the so-called Heisenberg cocycle:
$e^{-2i\pi\Theta(\gamma,\gamma')}:=\mu(g')\,\mu'(g)^{-1}$.
If we assume that $(G\times\widehat{G})/\Gamma$ is compact, then 
$\E_\SS:=\SS(G)$ is a finitely generated
projective module called the Heisenberg module. In particular, this
forces the dimension of the noncommutative torus $d$ to be even.
This means that for $a\in\A_\Th$ and $\phi\in\E_\SS$ then
$\phi a=\sum_\ga a_\ga\,\phi U_\ga$ is well 
defined as an element of $\E_\SS$. 
This can be proven by elementary Fourier analysis.
This module is
equipped with a ${\cal A}_{\Theta}$-valued scalar
product defined by
\begin{equation}
\left(\phi,\chi\right)_{{\cal A}_{\Theta}}:=
\mathop{\sum}\limits_{\gamma\in\Gamma}\,
\langle \phi,\chi U_{\gamma}\rangle_{L^2(G,dg)}\,
U_{\!-\gamma}.\label{scalarA}
\end{equation}
This paring takes values in the smooth algebra $\A_\Th$ and not in its
$C^*$-completion. Again,
basic Fourier analysis shows that the sequence
$\{\langle\phi,\chi U_\ga\rangle_{L^2(G,dg)}\}_{\ga\in\Z^d}$  
is of Schwartz class whenever $\phi,\chi\in\E_\SS$. 

A connection on $\E_\SS$ is entirely specified by its covariant
derivatives $\nabla_{\mu}$, once we have identified
$\Omega_{1}\left({\cal A}_{\Theta}\right)$ with 
$\left({\cal A}_{\Theta}\right)^{d}$. The canonical Heisenberg
connection $\nabla$ is
obtained from the infinitesimal action of the continuous part of the
group $G\times\widehat{G}$. In general, it both involves
partial derivatives $\frac{\partial}{\partial x_{i}}$ (action of
${\Bbb R}^{p}$) and multiplication by $x_{i}$ (action of $({\Bbb
  R}^{p})^{*}$). Finally, it is also useful to note that the endomorphism
algebra $\mathrm{End}_{{\cal A}_{\Theta}}({\cal E}_\SS)$ is nothing but
another noncommutative torus generated by the dual lattice
$\widehat{\Gamma}\in \widehat{G}\times G$, which pairs trivially with
$\Gamma$ with respect to the Heisenberg cocycle. \\
Besides, the curvature of $\nabla$ is defined as
$F_{\mu\nu}:=\left[\nabla_{\mu},\nabla_{\nu}\right]$,
which is always an anti-hermitian element of $\mathrm{End}_{{\cal
    A}_{\Theta}}({\cal E}_\SS)$. 

To construct the action functional for a classical field $\phi\in\E_\SS$, 
we choose a constant positive definite matrix
$g_{\mu\nu}$ to define a scalar product between 1-forms, so that the
general form of the action given by
(\ref{actiongeneral}) is
\begin{equation}
S[\phi,\overline{\phi}]=
\sqrt{g}g^{\mu\nu}\,\mathrm{Tr}_{{\cal A}_{\Theta}}\big[
\left( \nabla_{\mu}\phi,\nabla_{\nu}\phi\right)_{{\cal A}_{\Theta}}\big]+
\sqrt{g}
\mu_0^{2}\,\mathrm{Tr}_{{\cal A}_{\Theta}}
\big[\left( \phi,\phi\right)_{{\cal A}_{\Theta}}\big]+
\sqrt{g}\frac{\lambda}{2}\,\mathrm{Tr}_{{\cal A}_{\Theta}}
\big[\left( \phi,\phi\right)_{{\cal A}_{\Theta}}^2\big],
\label{actiontorus}
\end{equation}
with $g^{\mu\nu}$ denoting the inverse of $g_{\mu\nu}$ and $g$ its
determinant. 
Since the module we consider is made out of complex-valued functions,
from now on, we denote an element of the dual module $\E^*_\SS$ by
$\overline{\phi}$.
Note that more general terms such as
\begin{equation}
\left[\mathrm{Tr}_{{\cal A}_{\Theta}}
\big[\left( \phi,\phi\right)_{{\cal A}_{\Theta}}\big]^{n}\right]^{k},
\end{equation}  
can be introduced. In the last section, we shall see that for $d=4$, the
renormalization forces the introduction of such a term with $n=1$, $k=2$.

The simplest example of a Heisenberg module is constructed explicitly
for $d=2$ as follows. In the two-dimensional case there is only one
deformation parameter in the matrix $\Th$, namely $\th$. 
We will denote by $\A_\th$
the corresponding noncommutative torus algebra. 
In this case, we start with the projective representation \eqref{projrep}
of ${\Bbb R}\times (\Bbb R)^{*}$ on the Schwartz space of the real
line. Then, one defines the lattice $\Gamma$ and its dual
$\widehat\Ga$ as 
\begin{equation}
\left\{
\begin{array}{r@{\;}c@{\;}l}
\Gamma&:=&\big\{ \gamma=(\theta m,2\pi n)\quad\mathrm{with}\quad
  m,n\in{\Bbb Z} \big\},\\[1ex]
\widehat{\Gamma}&:=&\big\{
\widehat{\gamma}=(m',\frac{2\pi n'}{\theta})
\quad\mathrm{with}\quad m',n'\in{\Bbb Z} \big\},
\end{array}
\right.
\end{equation}
with $\theta>0$. The lattices $\Gamma$ and $\widehat{\Gamma}$ define
two commuting noncommmutative torus algebras ${\cal A}_{\theta}$ and
$\mathrm{End}_{\cal A_\th}({\cal E}_\SS)={\cal
  A}_{1/\theta}$, acting on $\phi\in{\cal E}_\SS$ as
\begin{equation}
\left\{
\begin{array}{r@{\;}c@{\;}l}
\big(\phi U_{\gamma}\big)(x)
&:=&e^{i\pi\theta mn}\,e^{2i\pi nx}\,\phi(x+m\theta),\\[1ex]
\big(U_{\widehat{\gamma}}\phi\big)(x)&:=&
e^{i\frac{\pi mn}{\theta}}\,e^{\frac{2i\pi n'x}{\theta}}\,\phi(x+m').
\end{array}
\right.
\end{equation}
In the previous equation, we have identified 
$({\Bbb R}\times{\Bbb R}^{*})^{*}={\Bbb R}^{*}\times{\Bbb R}$ with
${\Bbb R}\times{\Bbb R}^{*}$ so that the lattice and its dual are
both subsets of ${\Bbb R}\times{\Bbb R}^{*}$. 

The $\A_\th$-valued scalar product follows from the general 
form \eqref{scalarA} and is explicitly given by
\begin{equation}
\left(\phi,\chi\right)_{{\cal A}_{\theta}}=
\mathop{\sum}\limits_{\gamma\in\Gamma}
\left(e^{i\pi\theta mn}\int_{{\Bbb R}}dx \,\overline{\phi}(x)
\,e^{2i\pi nx}\,\chi(x{+}m\theta)\right)\,U_{\!-\gamma}
.\label{scalarx}
\end{equation}

The Heisenberg connection is given   by the two covariant derivatives
\begin{equation}
\nabla_{1}\phi(x)=-\frac{2i\pi x}{\theta}\phi(x)
\quad\mathrm{and}\quad
\nabla_{2}\phi(x)=\frac{d\phi(x)}{dx}.\label{connectionx}
\end{equation}
This connection minimize the Yang-Mills action \cite{connesrieffel}
and has a constant curvature given by
\begin{equation}
F_{12}=\left[\nabla_{1},\nabla_{2}\right]=\frac{2i\pi}{\theta}.
\end{equation}

In this example, with the euclidian metric
$g_{\mu\nu}=\delta_{\mu\nu}$, 
the action functional \eqref{actiontorus} reads
\begin{align}
\hspace{-1cm}
S[\phi,\overline{\phi}\,]&:=\int_{{\Bbb R}}dx\, 
\overline{\phi}(x)\left(
  -\frac{d^{2}}{dx^{2}}+\frac{4\pi^{2}}{\theta^{2}}
x^{2}\right)\phi(x)\,+\mu_0^{2}
\int_{{\Bbb R}}dx\,\overline{\phi}(x)\,\phi(x)\,
\cr
&\hspace{4cm}+\frac{\lambda}{2}\mathop{\sum}\limits_{m,n\in\Z}
\int_{{\Bbb R}}dx\,
\overline{\phi}(x{+}n{+}m\theta)\,\phi(x{+}n)\,
\overline{\phi}(x)\,\phi(x{+}m\theta).
\label{actionx}
\end{align}
To derive this expression from \eqref{scalarx}, we used the Poisson
re-summation formula in the sense of tempered distributions to write
$\sum e^{2i\pi nx}=\sum\delta(x+n)$.
The kinetic part of this action is 
simply the energy of an harmonic oscillator. The
interaction takes a non-local form in $x$-space because of the
summation over $m$ and $n$, but reduces to the an-harmonic oscillator
for $m=n=0$. We shall further comment on the non-local structure of
this interaction in section \ref{matmod}, once we have derived the
Feynman rules.

So far, we have seen that for the Heisenberg module there are two
notions of dimension. The first one, $d$, is the spectral dimension of
the noncommutative torus, whereas the second, $d/2$, is the dimension
of the representation space entering in
$\E_\SS(\R^{d/2})=\SS(\R^{d/2})$. In the sequel, in view of quantum
field application, we will see that the pertinent notion of dimension
is those of the noncommutative torus. In the following and unless
otherwise specified, $d$ will always denote the dimension of the
noncommutative torus, which moreover has to be even.
  
\subsubsection{Symmetries and field equation}

We are now going to review the classical discrete symmetries of the Model.
First note that this action is invariant under a
version of the Langmann-Szabo duality \cite{duality}. If we replace
$\phi$ by its Fourier transform
\begin{equation}
\eta(\xi)=\int_{{\Bbb R}}dx\, e^{-2i\pi x\xi}\,\phi(x),
\end{equation}
then the action is invariant, up to a rescaling of the different
parameters: 
\begin{equation}
S_{\lambda,\,\mu_0,\,\theta}[\phi,\overline{\phi}]=
{\textstyle \frac{1}{\theta^{2}}}
S_{\theta\lambda,\,\th^2\mu_0,\,1/\theta}[\eta,\overline{\eta}].\label{LSx}
\end{equation}
This duality is an essential tool in the study of the
renormalizability of the model.

Besides the Langmann-Szabo duality, there is another discrete symmetry
involving the Fourier transform. This symmetry is related to the action
of the modular group $SL(2,{\Bbb Z})$ on the modulus $\tau$ in the
upper half plane. Here $\tau$ 
parameterizes the constant matrix of determinant
one, entering in the kinetic term as
\begin{equation}
g=\frac{1}{\Im(\tau)}
\begin{pmatrix}
1&\Re(\tau)\cr
\Re(\tau)&|\tau|^{2}
\end{pmatrix}.
\end{equation}
Then, the action is invariant under simultaneous changes of the field and
the modulus under the generators $S$ and $T$ of the modular group,
\begin{equation}
\left\{
\begin{array}{lrcrcrcl}
S:&\phi(x)&\rightarrow&\frac{1}{\sqrt{\theta}}\int_{{\Bbb R}}d\xi\, 
e^{\frac{-2i\pi x\xi}{\theta}}\,\phi(\xi),&
\quad&\tau&\rightarrow&-\frac{1}{\tau},\\
T:&\phi(x)&\rightarrow&e^{\frac{-i\pi x^{2}}{\theta}}\,\phi(x),&\quad
&\tau&\rightarrow&\tau+1.
\end{array}
\right.
\end{equation}
For such a metric, the Langmann-Szabo duality must also be accompanied
by the action of $S$ on the modulus. The transformations given by $S$
and $T$ define outer automorphisms of the algebra of the noncommutative
torus and correspond to large diffeomorphsims of the torus. Other
outer automorphsisms are given by the translations defined in
\eqref{tor}, but they do not leave the action invariant because of the
non-trivial connection. Translation invariance can only be recovered
by including gauge fields.


We have displayed the construction in the two dimensional case with
the simplest Heisenberg module. More general modules can easily be
obtained in any even dimension  by simply tensoring this module by
itself $d/2$ times. In this case the action takes the same form as in
(\ref{actionx}) with $x$, $m$ and $n$ replaced by $d/2$-dimensional
vectors. The Langmann-Szabo duality takes the same form as before
except that the coupling constant $\lambda$ transforms as
\begin{equation}
\lambda\rightarrow \lambda\,\theta^{2-\frac{d}{2}}. 
\end{equation} 
In particular, the coupling constant is invariant in four dimensions.

\quad

It is also interesting to look at the classical field equation:
\begin{equation}
\frac{\delta S[\phi,\overline\phi]}{\delta\overline\phi}=0,\quad
\frac{\delta S[\phi,\overline\phi]}{\delta\phi}=0,
\end{equation}
where, as usual, the functional derivatives are defined in the weak
sense, with respect to the paring \eqref{scalarE}.
Because the action is ``symmetric'' in $\phi$ and $\overline\phi$, it
is sufficient to look at one of them. The first one reads
\begin{equation}
H\phi=-\lambda\, \phi(\phi,\phi)_{\A_\th}
=-\lambda\sum_{\ga\in\Ga}\langle\phi,\phi U_\ga\rangle_{L^2(\R)}\, 
\phi U_{-\ga},
\end{equation}
where
\begin{equation}
\label{oscharm}
H:=-\frac{d^2}{dx^2}+\frac{4\pi^2}{\th^2}x^2+\mu_0^2
\end{equation}
is the harmonic oscillator Hamiltonian with frequency $2\pi/\th$. 
In the explicit realization of
the module $\E_\SS=\SS(\R)$, the field equation can be rewritten
as
\begin{equation}
\big(-\frac{d^2}{dx^2}+\frac{4\pi^2}{\th^2}x^2+\mu_0^2\big)\phi(x)
=-\lambda\sum_{n,m\in\Z}\phi(x+n)\,
\overline\phi(x+n+m\th)\,\phi(x+m\th).
\end{equation}
A very important task would be to study the solutions of the
classical field equation.

\subsection{Holomorphic representation}

For our purposes it is convenient to work with an equivalent representation
of the Heisenberg modules in terms of holomorphic functions. In the
general case, these are constructed out of the space of square
integrable holomorphic functions on ${\Bbb C}$ with respect to
the scalar product
\begin{equation}
\langle \phi,\chi\rangle_{B}
:=\int d\mu(z,\overline{z})\,\overline{\phi}(\overline{z})\,\chi(z),
\end{equation}
with the Gau\ss ian measure, normalized as 
$d\mu(z,\overline{z}):=(\omega/\pi) e^{-\omega |z|^{2}}d\Re(z)\,d\Im(z)$, 
and  $\omega >0$ is an arbitrary parameter. 
We denote by ${\cal H}_{B}:=L^2_{hol}(\C,d\mu)$ the corresponding 
Hilbert space. It also admits a projective representation of ${\Bbb
  C}\simeq{\Bbb R}\times{\Bbb R}^{*}$ given by
\begin{equation}
\big(T_{v}\phi\big)(z)
:=e^{-\frac{\omega |v|^{2}}{2}-\omega \overline{v}z}
\phi(z+v), \quad v\in\C.\label{projB}
\end{equation}  
The operators $T_{v}$ are unitary with respect to the
scalar product $\langle.,.\rangle_B$ and obey to the multiplication rule
\begin{equation}
T_{v}T_{w}=
e^{\frac{\omega }{2}(\overline{v}w-\overline{w}v)}
T_{v+w}.
\end{equation}
Let us choose a square lattice in ${\Bbb C}$ parametrized by $l>0$,
\begin{equation}
\Gamma:=\left\{ \gamma=l( m+in)\quad\mathrm{with}\quad (m,n)\in{\Bbb Z}^{2}
\right\}.
\end{equation}
One can define a right action $U_\ga$ of ${\cal A}_{\theta}$, with
$\th=\frac{\omega l^2}{\pi}$, by restricting the action \eqref{projB}
to the lattice $\Ga$, i.e.\ $\phi U_\ga:=T_{\ga}\phi$, $\ga\in\Ga$. 
Here and in the
following, we made a slight abuse of notation by denoting by $U_\ga$
an element of $\A_\th$ as well as the operator acting on the right on
$\H_B$.
Note that in this framework, the dual lattice is given by
\begin{equation}
\widehat{\Gamma}:=
\big\{
  \widehat{\gamma}=\frac{\pi}{\omega  l}(m'+in')
\quad\mathrm{with}\quad (m',n')\in{\Bbb Z}^{2}
\big\},
\end{equation}
and gives rise to an action of $\A_{1/\th}$.
The scalar product with values in the algebra $\A_\th$ is defined as in 
\eqref{scalarA}, with
$\langle\cdot,\cdot\rangle_B$ denoting the Bargmann scalar product:
\begin{equation}
\left(\phi,\chi\right)_{{\cal A}_{\theta}}=
\sum_{\ga\in \Ga}\langle\phi,\chi U_\ga\rangle_B\,U_{-\ga}.\label{scalarB}
\end{equation}

The connection follows from the infinitesimal action of the
translation group on $\H_B$. It reads
\begin{equation}
\nabla_{1}\phi(z)=
\frac{\pi}{i\omega  lR}\left[\frac{d\phi}{dz}(z)+\omega  z\phi(z)\right]
\quad\mathrm{and}\quad
\nabla_{2}\phi(z)=
\frac{\pi}{\omega  lR}\left[\frac{d\phi}{dz}(z)-\omega  z\phi(z)\right],
\end{equation}
and of course they fulfil the Leibniz rule $\nabla_i(\phi
U_\ga)=(\nabla_i\phi)U_\ga +\phi(\delta_i U_\ga)$, $i=1,2$, where the
derivations of $\A_\th$ have been defined in \eqref{derivation}.
Note that we have introduced an extra parameter $R>0$ which has the
dimension of a length and has to be thought as the `radius' of the
noncommutative torus.
It is to be noted that this connection has constant curvature $
F_{12}=\left[\nabla_{1},\nabla_{2}\right]=\frac{2i\pi^{2}}{kl^{2}R^{2}}
$ and that $ \mathrm{Tr}_{{\cal
    A}_{\theta}}\left[F_{12}\right]=2i\pi\frac{\pi}{kl^{2}}$.  This
provides a topological invariant of the bundle, analogous to the first
Chern class. Usually, one includes the factor $\frac{\pi}{kl^{2}}$ in
the trace since it corresponds to the {\it dimension} of the bundle
\cite{garden}. Therefore, the topological invariant belongs to
$2i\pi{\Bbb Z}$, which holds for arbitrary projective modules and
arbitrary connections over a two-dimensional noncommutative torus.


{}From a physical point of view, the Bargmann module appears in the
analysis of the motion of an electron confined to the $x,y$ plane in
an external uniform magnetic field of strength $B$ along an orthogonal
axis. If we denote by $-e$ and $m$
the charge and the mass of the electron, the Hamiltonian is, in the
Landau gauge,
\begin{equation}
H_{\mathrm{ L}}=-\frac{\hbar^{2}}{2m}\left[
\left( \frac{\partial}{\partial x}-\frac{ieB}{2\hbar}y\right)^{2}
+\left( \frac{\partial}{\partial y}+\frac{ieB}{2\hbar}x\right)^{2}
\right].\label{hamiltonian}
\end{equation}
Translation invariance is realized through the magnetic translation
operators
\begin{equation}
\big(T_{a,b}\psi\big)(x,y)
:=e^{\frac{ieB}{2\hbar}(ay-bx)}\psi(x+a,y+b),\label{mag}
\end{equation}
which form a projective representation of the translation group,
\begin{equation}
T_{a,b}T_{a',b'}=e^{\frac{ieB}{2\hbar}(ab'-ba')}T_{a+a',b+b'}.
\end{equation}
Let us now assume that the magnetic field is strong enough so that the
electron is confined to the lowest Landau level, which is the ground
state of the Hamiltonian $H_{\mathrm{L}}$ with energy $\frac{\hbar eB}{2m}$.

The wave functions pertaining to the lowest Landau level are
conveniently written using a complex coordinate $z=x+iy$, 
\begin{equation}
\psi(z,\overline{z})=e^{-\frac{eB}{4\hbar}|z|^{2}}\phi(z),
\end{equation}
where $\phi$ is an arbitrary holomorphic function with square summable
Taylor coefficients. Thus, the lowest
Landau level is infinitely degenerated and its wave functions are in
one-to-one correspondence with analytic functions $\phi$ in the
Bargmann space ${\cal H}_{B}$ with
$\omega =\frac{eB}{2\hbar}$. Besides, 
a magnetic translation by $v=a+ib$ acts on $\phi$ via the action
of ${\Bbb C}$ on holomorphic functions \eqref{projB}. 

The noncommutative tori arise naturally when we consider the effect
of an atomic lattice in perturbation theory. Indeed, let us assume that
in addition to the magnetic field, the electron is submitted to a
potential $V$ created by a square lattice of spacing
$l$. Then, the potential is ${\Bbb Z}^{2}$-periodic,
$V(x+lm,y+nl)=V(x,y)$ for any $(m,n)\in{\Bbb Z}^{2}$. Furthermore, let us
make the assumption that the magnetic field is strong enough so that
one can consider the electron to be confined to the lowest
Landau level even in the presence of $V$. Therefore, the implementation
of the lattice translations imply that the lowest Landau level is
nothing but the Bargmann module ${\cal H}_{B}$ over the
noncommutative torus ${\cal
  A}_{\theta}$ with a parameter 
\begin{equation}
\theta=\frac{\omega  l^{2}}{\pi}=
\frac{eBl^{2}}{2\pi\hbar}=\frac{\Phi}{\Phi_{0}},\label{deftheta}
\end{equation} 
where $\Phi=Bl^{2}$ is the flux of $B$ through the unit cell of the lattice
and $\Phi_{0}=\frac{2\pi\hbar}{e}$ is the quantum flux. This
noncommutative torus may be seen as a {\it Noncommutative
  Brioullin zone} since it replaces the ordinary Brioullin zone in the
presence of the magnetic field. For this
interpretation to hold, it is necessary to take $R=\frac{2\pi}{l}$
which corresponds to the size of the Brioullin zone. Let us note that
the topological invariant given by $\Tr_{\A_\Th}(F_{12})$, 
is nothing but the Hall conductivity \cite{bellissard}.

If the electron remains confined to the lowest Landau level, then $V$
has to be projected on ${\cal H}_{B}$ and lattice invariance
translates into the statement that $V$ must commute
with the left action of ${\cal A}_{\theta}$. Therefore, $V$ is an
element of the algebra ${\cal A}_{1 /\theta}$ acting on the right. This
is a general  constraint on the action of $V$, solely derived from
symmetry considerations and independent of the precise form of $V$.

\quad

Obviously, the projective modules constructed out of the Schwartz space and the
Bargmann space are equivalent, the explicit equivalence being the unitary
Bargmann transform $B:\,{\cal H_{S}}\rightarrow {\cal
  H}_{\B}$ defined as
\begin{equation}
\big(B\chi\big)(z)=\left(\frac{\omega }{\pi}\right)^{1/2}\int_{{\Bbb R}}dx\,
e^{-\frac{\omega (x^{2}+z^{2})}{2}+\sqrt{2}\omega  zx}\chi(x),
\end{equation} 
and its inverse is
\begin{equation}
\big(B^{-1}\phi\big)(x)=\left(\frac{\omega }{\pi}\right)^{1/2}
\int_{{\Bbb C}}d\mu(z,\overline{z})\,
e^{-\frac{\omega (x^{2}+\overline{z}^{2})}{2}+\sqrt{2}\omega  
\overline{z}x}\phi(z),
\end{equation}
for $\chi\in L^2(\R)$ and $\phi\in\H_B$. 
This corresponds to the equivalence between the vertical
polarization on the phase space ${\Bbb R}\times {\Bbb R}^{*}$ and the
holomorphic one. Then, one defines the smooth
module $\E_B$ as the image of Schwartz space under the Bargmann transform.
   
Consider now the projective action $T_{a,b}$ of ${\Bbb R}\times {\Bbb
  R}^{*}$ on the Schwartz functions of the real line. Its image
through the Bargmann transform is an action of ${\Bbb C}$ by the
operators $T_{v}$ given in \eqref{projB} with 
$v=\frac{1}{\sqrt{2}}\left(a+\frac{ib}{\omega }\right)$.
This allows to establish a general correspondence between the two
projective modules, their scalar products and their connections. In
what follows, it is particularly convenient to set
$\omega =\frac{2\pi}{\theta}$, in such a way that the lattices $\Gamma$
and $\widehat{\Gamma}$ read, in the holomorphic formulation,
\begin{equation}
\Gamma=\Big\{\frac{\theta}{\sqrt{2}}\left(m+in\right)
\quad\mathrm{with}\quad(m,n)\in{\Bbb Z}^{2}\Big\},
\end{equation}
and 
\begin{equation}
\widehat{\Gamma}=\Big\{\frac{1}{\sqrt{2}}\left(m'+in'\right)
\quad\mathrm{with}\quad(m',n')\in{\Bbb Z}^{2}\Big\},
\end{equation}
It amounts to set $l=\frac{\theta}{\sqrt{2}}$ in the general square
lattice introduced at the beginning of this section. An arbitrary
lattice in ${\Bbb C}$ would correspond to an arbitrary lattice in ${\Bbb
R}\times{\Bbb R}^{*}$ using the inverse Bargmann transform. From now
on, we restrict ourselves to the preceding values of $\omega $ and $l$.

In this framework, the two operators representing the connection read
\begin{equation}
\nabla_{1}\phi(z)=
\frac{1}{i\sqrt{2}}\left[\frac{d\phi}{dz}(z)+\omega  z\phi(z)\right]
\quad\mathrm{and}\quad
\nabla_{2}\phi(z)=
\frac{1}{\sqrt{2}}\left[\frac{d\phi}{dz}(z)-\omega  z\phi(z)\right],
\label{connectionB}
\end{equation}
as follows from \eqref{connectionB} with $\omega =\frac{2\pi}{\theta}$,
$l=\frac{\theta}{\sqrt{2}}$ and $R=1$. The operator \eqref{oscharm} 
entering into the definition of the kinetic term is 
\begin{equation}
\label{trans}
H=-\left(\nabla_{1}\right)^{2}-\left(\nabla_{2}\right)^{2}+\mu_0^{2}=
2\omega \left(z\frac{d}{dz}+\frac{1}{2}\right)+\mu_0^{2},
\end{equation}
which is nothing but the image through the Bargmann transform of the
harmonic oscillator Hamiltonian $-\frac{d^{2}}{dx^{2}}+\omega^2 
x^{2}+\mu_0^{2}$. Note that there is a constant term in $H$ even if 
$\mu_0=0$. Thus,
even in a massless theory, there is {\it a priori} no infrared
divergences at the tree level. Of course, this picture may change at
higher  loop order because of the IR/UV mixing.

In the development of the quantum theory associated to these
projective modules, it will be interesting to have at our disposal higher
dimensional analogues of the previous construction. While
the general structure of projective modules over higher dimensional
noncommutative tori is extremely rich \cite{rieffel}, we can easily
construct simple examples in any even dimension $d$ by simply
taking the $\frac{d}{2}^{\mathrm{th}}$ tensor power of the previous
module. This corresponds to a Bargmann space of holomorphic functions
on ${\Bbb C}^{d/2}$. 
The corresponding lattices are simply obtained by replacing
the integers $m$, $n$, $m'$ and $n'$ by vectors in ${\Bbb
  Z^{\frac{d}{2}}}$. This
procedure leads to a noncommutative algebra ${\cal A}_{\Theta}$ with
a $d\times d$ matrix $\Theta$ which is a direct sum of $2\times 2$ 
antisymmetric matrices,
\begin{equation}
\Theta=\begin{pmatrix}
0&\theta&&&\cr-\theta&0&&&\cr&&\ddots&&
\cr&&&0&\theta\cr&&&-\theta&0
\end{pmatrix}.
\label{mostgene}
\end{equation}
To emphasis the difference with a generic noncommutative torus algebra
$\A_\Th$, we will denote the algebra corresponding to the specific
choice \eqref{mostgene} by $\A_\th$, like in the two-dimensional case.

\section{QFT and diagrammatics}

\subsection{Feynman rules}

\label{feynman}

\begin{fmffile}{fmfqft2-pmod}

  In this section, we restore the notation $\phi^\dagger$, instead of
  the complex conjugate notation $\overline\phi$, to denote an element
  of the dual module $\E^*$.  The reason is that we would like to
  sketch a representation free construction of Feynmann rules for a
  field theory on the Heisenberg modules corresponding to the choice
  \eqref{mostgene}. Otherwise specified, $\E$ can be either
  $\E_\SS(\R^{d/2})$ or $\E_B(\C^{d/2})$, or even another unitarily equivalent
  module.  We also emphasize that such a procedure does not rely on
  the particular structures of the module studied, so that it can be
  employed for the construction of any quantum field theory on a
  projective module.

  The action functional $S[\phi,\phi^{\dagger}]$ we have defined in
  the previous section can serve as a basis for the construction of a
  perturbative quantum field theory. This amounts to define and study
  the $(2N)$-point functions $G_{2N}$, given in terms of the
  functional integral
\begin{equation}
G_{2N}:=
\frac{\displaystyle \int [D\phi] [D\phi^{\dagger}]\,
\left( \phi\otimes\phi^{\dagger}\otimes\cdots
\otimes\phi\otimes\phi^{\dagger}\right)\,
e^{-S[\phi,\phi^{\dagger}]}}
{\displaystyle \int [D\phi] [D\phi^{\dagger}]\,e^{-S[\phi,\phi^{\dagger}]}},
\label{correlation}
\end{equation}
as a perturbation of a free Gau{\ss}ian functional integral. Because the
projective module can be realized as a dense sub-space of an 
Hilbert space of functions with
a kinetic term given by a second order differential operator, the
free field theory is always well-defined. We
will always assume in this section that the functional integral has
been regularized, for example by integrating only over a finite
dimensional subspace of the projective module. Later on, we shall
use other regularization techniques suitable for renormalization.

This means that once the theory is regularized, 
the Green's functions $G_{2N}$ define distributions 
on the projective tensor product
$\E^{\ox N}\ox{\E^*}^{\ox N}\simeq\SS(\R^{dN})$ (recall that
$\SS(\R^{d/2})$ is a nuclear space). The evaluation of $G_{2N}$ 
on $2N$ test functions
$\psi_i\in\E$, $\chi_i^\dagger\in\E^*$, $i=1,\cdots,N$, is given by the
following path integral formula:
\begin{align}
&G_{2N}\big(\chi^\dagger_1,\cdots,\chi^\dagger_n,\psi_1,\cdots,\psi_2\big)
=\\
&
\frac{\displaystyle \int [D\phi] [D\phi^{\dagger}]\,
\Tr_{\A_\th}\left[(\chi_1,\phi)_{\A_\th}\right]\,
\Tr_{\A_\th}\left[(\phi,\psi_1)_{\A_\th}\right]\cdots
\Tr_{\A_\th}\left[(\chi_n,\phi)_{\A_\th}\right]\,
\Tr_{\A_\th}\left[(\phi,\psi_n)_{\A_\th}\right]\,
e^{-S[\phi,\phi^{\dagger}]}}
{\displaystyle \int [D\phi] [D\phi^{\dagger}]\,
e^{-S[\phi,\phi^{\dagger}]}}.\nonumber
\label{correlation2}
\end{align}

It has to be noted that the field $\phi$ and its
conjugate $\phi^{\dagger}$ must occur an equal number of times, as
required by the invariance under global phase multiplication. Any
correlation function not fulfilling this condition vanishes
identically. Note that the field $\phi$ lives in a module ${\cal
E}$ and its conjugate $\phi^{\dagger}$ in the dual module ${\cal
E}^{*}$, so that
the correlation function with $2N$ fields is conveniently seen as a
linear map from ${\cal E}^{\otimes N}$ to itself.

As usual in QFT, all the correlations functions can be gathered into
the generating functional
\begin{equation}
Z[J,J^{\dagger}]=
\frac{\displaystyle \int [D\phi] [D\phi^{\dagger}]\
e^{-S[\phi,\phi^{\dagger}]+
\Tr_{\A_\th}\left[(J,\phi)_{\A_\th}\right]+
\Tr_{\A_\th}\left[(\phi,J)_{\A_\th}\right]}}
{\displaystyle \int [D\phi] [D\phi^{\dagger}]\,e^{-S[\phi,\phi^{\dagger}]}}.
\end{equation}
{}From $Z[J,J^{\dagger}]$ we define $W[J,J^{\dagger}]=\log
Z[J,J^{\dagger}]$ and $\Gamma[\phi,\varphi^{\dagger}]$ as the Legendre
transform of $W[J,J^{\dagger}]$,
\begin{equation*}
\Gamma[\varphi,\varphi^{\dagger}]=
\Tr_{\A_\th}\big[(\varphi,J)_{\A_\th}\big]+\Tr_{\A_\th}
\big[(J,\varphi)_{\A_\th}\big]-W[J,J^{\dagger}]
\quad\mbox{with}\qquad \varphi=\frac{\delta W}{\delta J^{\dagger}}
\quad\mbox{and}\quad\varphi^{\dagger}=\frac{\delta W}{\delta J}.
\end{equation*}
Whereas $Z$ is obtained as a sum over all Feynman diagrams, $W$ only
involves connected diagrams and $\Gamma$ only 1PI diagrams, i.e.\ diagrams
that remain connected if one cuts any internal line. In the sequel, we
shall always restrict our analysis to 1PI diagrams.

Since the  action functional can be written as
\begin{equation}
S[\phi,\phi^{\dagger}]=\mathrm{Tr}_{\A_\th}\Big[
(\phi,H\phi)_{\A_\th}+\frac{\lambda}{2}
\left(\phi,\phi\right)_{\A_\th}^{2}\Big],
\end{equation}
it is convenient to rewrite the quartic 
interaction using a  Hubbard-Stratonovitch transform as
\begin{equation}
e^{-\frac{\lambda}{2}\mathrm{Tr}_{{\cal A}_{\theta}}\big[
  \left(\phi,\phi\right)_{\A_\th}^{2}\big]}
=\int[DA]\,
e^{-\mathrm{Tr}_{{\cal A}_{\theta}}\big[\frac{\lambda A^{2}}{2}
+i\lambda A(\phi,\phi)_{\A_\th} \big]},
\end{equation}
where we integrate over all Hermitian elements $A^*=A\in{\cal
  A}_{\theta}$. From a physical perspective, this amounts to consider
$A$ as a random external field subjected to Gau\ss ian probability law.

As usual, a regularization is self-understood and we
have dropped an irrelevant normalization factor. Thus, the correlation
function reads
\begin{equation}
G_{2N}=\frac{\displaystyle \int [D\phi] [D\phi^{\dagger}] [DA]\,
\left( \phi\otimes\phi^{\dagger}\otimes\cdots
\otimes\phi\otimes\phi^{\dagger}\right)\,
e^{-\mathrm{Tr}_{{\cal A}_{\theta}}\big[(\phi,H\phi)_{\A_\th}
+\frac{\lambda A^{2}}{2}+i\lambda A(\phi,\phi)_{\A_\th}\big]}}
{\displaystyle 
\int [D\phi] [D\phi^{\dagger}][DA]\,e^{-\mathrm{Tr}_{{\cal A}_{\theta}}
\big[(\phi,H\phi)_{\A_\th}
+\frac{\lambda A^{2}}{2}+i\lambda A(\phi,\phi)_{\A_\th}\big]}}.
\label{correlationA}
\end{equation}
This allows us to derive the Feynman rules in a very simple way. There
is a trivial propagator for the $A$-field
\begin{align}
\parbox{23mm}{\begin{picture}(20,5)
\put(0,0){\begin{fmfgraph}(20,5)
\fmfleft{l}
\fmfright{r}
\fmf{photon}{l,r}
\fmf{phantom_arrow}{l,r}
\end{fmfgraph}}
\put(9,5){\mbox{\small$\gamma$}}
\end{picture}}= 
\frac{1}{\lambda}\;.
\end{align}

\noindent
Note that the field $A$ only appears in internal lines since we 
only compute correlations of $\phi$ and $\phi^{\dagger}$. We have also
used an explicit expansion over the noncommutative Fourier modes 
$A=\sum_{\gamma}A_{\gamma} U_{\gamma}$.
The fields $\phi$ and $\phi^{\dagger}$ propagate according to
\begin{align}
\parbox{23mm}{\begin{picture}(20,5)
\put(0,0){\begin{fmfgraph}(20,5)
\fmfleft{l}
\fmfright{r}
\fmf{fermion}{l,r}
\end{fmfgraph}}
\put(15,4){}
\put(2,4){}
\end{picture}}= 
H^{-1}\;,
\end{align}
\noindent
where $H$ is defined in \eqref{trans}.
Because the fields are complex, the propagator is not symmetric and an
orientation on the lines is necessary. We always consider
$\phi^{\dagger}$ as incoming and $\phi$ as outgoing, so that the
propagator makes sense as a linear map from ${\cal E}$ to itself.
Finally, the interaction vertex is
\begin{align}
\parbox{38mm}{\begin{picture}(35,15)
\put(0,0){\begin{fmfgraph}(35,20)
\fmfleft{l1,l2}
\fmfright{r1,r2}
\fmf{fermion}{r1,i1,l1}
\fmf{phantom}{l2,i2,r2}
\fmf{photon}{i1,i2}
\fmf{phantom_arrow,tension=0}{i2,i1}
\end{fmfgraph}}
\put(4,3){}
\put(28,3){}
\put(19,10){\mbox{\small$\gamma$}}
\end{picture}}= 
i\lambda U_\gamma.\label{vertexA}
\end{align}
\noindent
It is important to note that there is a cyclic ordering of the fields
$A$, $\phi^{\dagger}$ and $\phi$ at the vertices which is due to the
fact that the scalar product $(\phi,\phi)_{\A_\th}$ takes its values
in the algebra ${\cal A}_{\theta}$. The vertex \eqref{vertexA} always
has the $A$-field attached to the right when following the arrows of
the $\phi$-field. Exchanging $\phi$ and $\phi^{\dagger}$ would involve
the scalar product $(\phi,\phi)_{{\cal B}_\th}$ with values in the
commutant of ${\cal A}_{\theta}$ in $\L({\H_\E})$. (Recall that
$\H_\E$ is the Hilbert space obtained by completion of $\E$ with
respect to the norm subordonate to  the scalar product
$\langle.,.\rangle_\E=\Tr_{\A_\th}[(.,.)_{\A_\th}]$.)
We shall come back to this point when discussing duality. \\
It is worthwhile to notice that all the operators entering in the
Feynmann rules extend to bounded operators on $\H_\E$. Instead being
interpreted as a distribution, a typical regularized Feynman diagram
with $N$-incoming and $N$-outgoing lines will then be conveniently
seen as a linear map from $\H^{\ox N}$ to itself.

For instance, an arbitrary diagram with $2N$ external legs is obtained
as follows:

\begin{itemize}

\item
Draw $N$ oriented lines representing the propagation of the field
$\phi$. Each of these lines represents an operator from ${\H_\E}$
to itself, obtained by multiplying propagators $H^{-1}$ and interactions
$U_\gamma$.

\item
If they occur, draw closed oriented loops made of propagators $H^{-1}$ and
vertices $U_\gamma$ and take the traces of the corresponding
operators.

\item
Relate all interactions by wavy lines representing the propagation of
$A$ and sum over all the corresponding  $\gamma$.
  
\end{itemize}

It is important to know at this stage that the divergences we will
encounter may both come from the traces and the sums over $\gamma$'s.


Because of the orientation of the lines, there are no non-trivial
symmetry factors, as e.g.\ in QED.\\

The simplest diagram in the 2-point function ${\cal H}_\E
\rightarrow{\cal H}_\E$ is
\begin{align}
\parbox{50mm}{\begin{picture}(45,20)
\put(0,0){\begin{fmfgraph}(45,20)
\fmfbottom{l,r}
\fmf{fermion,tension=2}{r,i1}
\fmf{fermion}{i1,i2}
\fmf{fermion,tension=2}{i2,l}
\fmffreeze
\fmf{photon,right}{i1,i2}
\fmffreeze
\fmf{phantom_arrow,left}{i2,i1}
\end{fmfgraph}}
\put(2,5){}
\put(40,5){}
\put(28,4){}
\put(20,15){\mbox{\small$\gamma$}}
\put(14,4){}
\end{picture}}
=-\lambda\mathop{\sum}\limits_{\gamma}U_{\!-\gamma}\,H^{-1}\,U_{\gamma}.
\end{align}
Here we have made a little abuse of notation, using the same symbol
$U_\ga$ to denote an element of the noncommutative torus and the
operator acting (on the right) on $\H_\E$. Also, the trace which will
appear in the next diagram, is not the noncommutative torus trace but
the operator trace on $\H_\E$.
A diagram contributing to the 2-point function involving a closed loop
is
\begin{align}
\parbox{35mm}{\begin{picture}(25,30)
\put(0,0){\begin{fmfgraph}(25,30)
\fmfbottom{l,r}
\fmftop{t}
\fmf{fermion}{r,i1,l}
\fmffreeze
\fmf{photon,tension=3}{i1,i2}
\fmf{phantom_arrow,tension=0}{i2,i1}
\fmf{fermion,right}{i2,t}
\fmf{fermion,right}{t,i2}
\end{fmfgraph}}
\put(-1,6){}
\put(24,5){}
\put(18,13){}
\put(3,13){}
\put(14,8){\mbox{\small$\gamma$}}
\end{picture}} 
=-\lambda\sum_\ga\mbox{Tr}\left( U_{\!-\gamma}\,H^{-1}\,\right)U_{\gamma}.
\label{tp2}
\end{align}
The simplest  non-planar diagram contributing to the 2-point
function is
\begin{align}
\parbox{50mm}{\begin{picture}(50,20)
\put(0,0){\begin{fmfgraph}(45,20)
\fmfbottom{l,r}
\fmf{plain,tension=2}{r,i1}
\fmf{fermion}{i1,i2}
\fmf{plain}{i2,i3}
\fmf{fermion}{i3,i4}
\fmf{plain,tension=2}{i4,l}
\fmffreeze
\fmf{photon,right=.6,rubout=4}{i1,i3}
\fmf{photon,right}{i2,i4}
\fmffreeze
\fmf{phantom_arrow,left}{i4,i2}
\fmffreeze
\end{fmfgraph}}
\put(2,5){}
\put(40,5){}
\put(28,4){}
\put(18,16){\mbox{\small$\gamma_1$}}
\put(33,10){\mbox{\small$\gamma_2$}}
\put(14,4){}
\end{picture}}
=\lambda^{2}\mathop{\sum}\limits_{\gamma_{1},\gamma_{2}}
U_{\!-\gamma_{1}}\,H^{-1}\,U_{-\gamma_{2}}\,H^{-1}\,
U_{\gamma_{1}}\,H^{-1}\,U_{\gamma_{2}},
\end{align}
where $\gamma_2$ has the same orientation as $\gamma_1$.
Analogously, let us list a few diagrams contributing to the 4-point function
${\cal H}_\E\otimes{\cal H}_\E\rightarrow{\cal H}_\E\otimes {\cal
  H}_\E$, starting with the tree level contribution:
\begin{align}
\parbox{40mm}{\begin{picture}(35,20)
\put(0,0){\begin{fmfgraph}(35,20)
\fmfleft{l1,l2}
\fmfright{r1,r2}
\fmf{fermion}{r1,i1,l1}
\fmf{fermion}{l2,i2,r2}
\fmf{photon}{i1,i2}
\fmf{phantom_arrow,tension=0}{i2,i1}
\end{fmfgraph}}
\put(4,3){}
\put(28,3){}
\put(28,16){}
\put(4,15.5){}
\put(19,10){\mbox{\small$\gamma$}}
\end{picture}}= 
-\lambda\mathop{\sum}\limits_{\gamma}U_{\!-\gamma}\otimes U_{\gamma}.
\end{align}
This diagram simply represents the vertex once the $A$-field has been
integrated out.

The first non-trivial four-point diagram with a loop is
\begin{align}
\parbox{48mm}{\begin{picture}(50,20)
\put(0,0){\begin{fmfgraph}(45,20)
\fmfleft{l1,l2}
\fmfright{r1,r2}
\fmf{fermion,tension=2}{r1,i1}
\fmf{fermion,tension=1}{i1,i3}
\fmf{fermion,tension=2}{i3,l1}
\fmf{fermion,tension=2}{l2,i4}
\fmf{fermion,tension=1}{i4,i2}
\fmf{fermion,tension=2}{i2,r2}
\fmffreeze
\fmf{photon}{i1,i2}
\fmf{phantom_arrow,tension=0}{i2,i1}
\fmf{photon}{i3,i4}
\fmf{phantom_arrow,tension=0}{i3,i4}
\end{fmfgraph}}
\put(4,-18.5){}
\put(38,-18.5){}
\put(27,-19){}
\put(15,-18.5){}
\put(38,7){}
\put(4,7){}
\put(15,7){}
\put(27,7){}
\put(8,7){\mbox{\small$\gamma_2$}}
\put(34,7){\mbox{\small$\gamma_1$}}
\end{picture}}  
=\lambda^{2}\mathop{\sum}\limits_{\gamma_{1},\gamma_{2}}
\left(U_{-\gamma_{1}}\,H^{-1}\,U_{\gamma_{2}}\right)\otimes
\left(U_{\!-\gamma_{2}}\,H^{-1}\,U_{\gamma_{1}}\right).
\end{align}
This diagram is planar and a non-planar one, can be constructed by
exchanging $\gamma_{1}$ and $\gamma_{2}$ in the second tensor product,
\begin{align}
\parbox{48mm}{\begin{picture}(45,23)
\put(0,1){\begin{fmfgraph}(45,20)
\fmfleft{l1,l2}
\fmfright{r1,r2}
\fmf{fermion,tension=2}{r1,i1}
\fmf{fermion,tension=1}{i1,i3}
\fmf{fermion,tension=2}{i3,l1}
\fmf{fermion,tension=2}{l2,i4}
\fmf{fermion,tension=1}{i4,i2}
\fmf{fermion,tension=2}{i2,r2}
\fmffreeze
\fmf{photon}{i1,i5,i4}
\fmf{phantom_arrow,tension=0}{i5,i1}
\fmf{photon,rubout=5}{i3,i7}
\fmf{photon}{i2,i7}
\fmf{phantom}{i3,i6,i7,i2}
\fmf{phantom_arrow,tension=0}{i2,i6}
\end{fmfgraph}}
\put(4,-18.5){}
\put(38,-18.5){}
\put(27,-19){}
\put(15,-18.5){}
\put(38,7){}
\put(4,7){}
\put(15,7){}
\put(27,7){}
\put(29,6){\mbox{\small$\gamma_1$}}
\put(28,14){\mbox{\small$\gamma_2$}}
\end{picture}}
=\lambda^2\mathop{\sum}\limits_{\gamma_{1},\gamma_{2}}
\left(U_{-\gamma_{2}}\,H^{-1}\,U_{-\gamma_{1}}\right)\otimes
\left(U_{\gamma_{2}}\,H^{-1}\,U_{\gamma_{1}}\right).
\end{align}
With a trace, the simplest example is
\begin{align}
\parbox{40mm}{\begin{picture}(35,27)
\put(4,0){\begin{fmfgraph}(35,25)
\fmfleft{l1,l2}
\fmfright{r1,r2}
\fmf{fermion}{r1,i1,l1}
\fmf{fermion}{l2,i2,r2}
\fmffreeze
\fmf{photon,tension=3}{i1,i3}
\fmf{phantom_arrow,tension=0}{i3,i1}
\fmf{photon,tension=3}{i2,i4}
\fmf{phantom_arrow,tension=0}{i2,i4}
\fmf{fermion,right}{i3,i4}
\fmf{fermion,right}{i4,i3}
\end{fmfgraph}}
\put(4,-13){}
\put(28,-13){}
\put(10,-3){}
\put(22,-3){}
\put(21,7){}
\put(10,7){}
\put(28,17){}
\put(4,17){}
\put(23.5,3){\mbox{\small$\gamma_1$}}
\put(23.5,20){\mbox{\small$\gamma_2$}}
\end{picture}}
=\lambda^{2}\mathop{\sum}\limits_{\gamma_{1},\gamma_{2}}
\mathrm{Tr}\left( H^{-1}\,U_{\gamma_{2}}\,H^{-1}\,U_{-\gamma_{1}}
\right)\,U_{-\gamma_{2}}\otimes U_{\gamma_{1}}.
\end{align}
Finally, let us give a more complicated example, contributing to the
6-point function ${\cal H}_\E\otimes{\cal H}_\E\otimes{\cal H}_\E
\rightarrow{\cal H}_\E\otimes {\cal H}_\E\otimes{\cal H}_\E$,
\begin{align}
\parbox{40mm}{\begin{picture}(40,40)
\put(0,0){\begin{fmfgraph}(40,40)
\fmfstraight
\fmfbottomn{u}{8}
\fmftopn{o}{8}
\fmffreeze
\fmfleftn{l}{8}
\fmfrightn{r}{4}
\fmffreeze
\fmfforce{(.5w,.5h)}{l1}
\fmfforce{(.3w,.7h)}{l3}
\fmfforce{(.1w,.5h)}{l5}
\fmfforce{(.3w,.3h)}{l7}
\fmfforce{(.44w,.64h)}{l2}
\fmfforce{(.16w,.64h)}{l4}
\fmfforce{(.16w,.36h)}{l6}
\fmfforce{(.44w,.36h)}{l8}
\fmfforce{(.9w,.4h)}{r1}
\fmfforce{(.8w,.5h)}{r2}
\fmfforce{(.7w,.4h)}{r3}
\fmfforce{(.8w,.3h)}{r4}
\fmfshift{(-.2w,-.2h)}{o1}
\fmfshift{(-.2w,-.1h)}{o2}
\fmfshift{(-.2w,.0h)}{o3}
\fmfshift{(.20w,-.28h)}{o8}
\fmfshift{(.20w,-.21h)}{o7}
\fmfshift{(.20w,-.14h)}{o6}
\fmfshift{(.20w,-.07h)}{o5}
\fmfshift{(.20w,.0h)}{o4}
\fmffreeze
\fmf{plain}{o4,o8}
\fmf{plain}{o1,o3}
\fmf{plain}{u1,u8}
\fmf{fermion,right}{l1,l5}
\fmf{plain,right}{l5,l1}
\fmf{plain,right}{r1,r3}
\fmf{plain,right}{r3,r1}
\fmffreeze
\fmf{phantom_arrow,right}{r4,r2}
\fmf{photon}{u3,l7}
\fmf{photon}{o2,l4}
\fmf{photon}{u6,r4}
\fmf{photon,right,rubout=4}{u7,u5}
\fmf{photon}{l1,r3}
\fmf{phantom_arrow}{l1,r3}
\fmf{photon}{r2,o6}
\fmf{photon,rubout=4}{l2,o7}
\fmf{photon,rubout=4,left=1.5}{l8,l5}
\fmf{phantom_arrow}{o1,o2}
\fmf{phantom_arrow}{o4,o6}
\fmf{phantom_arrow}{u5,u3}
\end{fmfgraph}}
\put(35,3){\mbox{\scriptsize $\gamma_1$}}
\put(34,8){\mbox{\scriptsize $\gamma_2$}}
\put(35,22){\mbox{\scriptsize $\gamma_3$}}
\put(25,30){\mbox{\scriptsize $\gamma_4$}}
\put(25,20){\mbox{\scriptsize $\gamma_5$}}
\put(-3,12){\mbox{\scriptsize $\gamma_6$}}
\put(7,2){\mbox{\scriptsize $\gamma_7$}}
\put(-3,30){\mbox{\scriptsize $\gamma_8$}}
\end{picture}}
\end{align}

\begin{equation}
\begin{array}{rl}
=\lambda^8\sum\limits_{\gamma_{1},...,\gamma_{8}}&
\begin{array}{l}
U_{\gamma_8} \otimes \left(U_{\gamma_4}\,H^{-1}\,U_{\gamma_3}\right)\otimes 
\left(U_{\gamma_7}\,H^{-1}\,U_{\gamma_1}\,H^{-1}\,U_{\gamma_2}\,H^{-1}\,
U_{-\gamma_1}\right)\\
\times\mathrm{Tr}\left(H^{-1}\,U_{-\gamma_5}\,H^{-1}\,U_{\gamma_6}\,H^{-1}\,
U_{-\gamma_7}
\,H^{-1}\,U_{-\gamma_6}\,H^{-1}\,U_{-\gamma_8}\,H^{-1}\,U_{-\gamma_4}\right)\\
\times\mathrm{Tr}\left(H^{-1}\,U_{-\gamma_2}\,H^{-1}\,U_{\gamma_5}\,H^{-1}\,
U_{-\gamma_3}\right)
\end{array}
\end{array},
\end{equation}
where $\gamma_2,\gamma_3,\gamma_4,\gamma_7,\gamma_8$ leave the loops and
the internal $\gamma_1$ and $\gamma_6$ first leave and then arrive,
accordingly to the orientation.

\medskip

Let us note that on the definition of an 1PI diagram, 
we impose the irreducibility condition only for
the $\phi$-lines, not for the $A$-lines. In fact, only the 
$\phi$-lines are really internal lines. Thus we reserved the terminology to
the latter. The wavy $A$-lines are just a
convenient way to visualize the interaction and serve to indicate the
identifications of the operators $U_{\gamma}$ inserted in the
diagram.\\

For the simplest example of a Heisenberg module in $x$-space, we have
seen that the theory is invariant under the Langmann-Szabo
duality, see \eqref{LSx}. This is a general fact that follows from the
existence of two compatible scalar products on the projective module 
${\cal E}$, under fairly general conditions \cite{rieffel}. Indeed,
if we denote by ${\cal B}_{\theta}$ the endomorphism algebra 
$\mathrm{End}_{{\cal A}_{\theta}}({\cal E})$, then one defines a scalar
product $(\cdot,\cdot)_{{\cal B}_{\theta}}$ on $\E$, which is linear with
respect to the left action of ${\cal B}_{\theta}$ on the first
variable and anti-linear in the second one. The two scalar products are
compatible in the sense that
\begin{equation}
\phi\left(\chi,\psi\right)_{{\cal A}_{\theta}}=
\left(\phi,\chi\right)_{{\cal{B}_{\theta}}}\psi,
\end{equation}
for any $\phi,\chi,\psi\in{\cal E}$.  The algebra ${\cal B}_{\theta}$
is in general another noncommutative torus constructed with the dual
lattice. If we denote by $\mathrm{Tr}_{{\cal B}_{\theta}}$ its
normalized trace, then the interaction can be transformed as
\begin{align}
\mathrm{Tr}_{{\cal A}_{\theta}}\Big[
\left(\phi_{1},\phi_{2}\right)_{{\cal A}_{\theta}}
\left(\phi_{3},\phi_{4}\right)_{{\cal A}_{\theta}}\Big]&=
\mathrm{Tr}_{{\cal A}_{\theta}}\Big[
\big(\phi_{1},\phi_{2}
\left(\phi_{3},\phi_{4}\right)_{{\cal A}_{\theta}}\big)_{{\cal A}_{\theta}}
\Big]\nonumber \\
&=\mathrm{Tr}_{{\cal A}_{\theta}}\Big[
\big(\phi_{1},\left(\phi_{2}
,\phi_{3}\right)_{{\cal B}_{\theta}}\phi_{4}\big)_{{\cal A}_{\theta}}
\Big] \nonumber \\
&=\mathrm{Tr}_{{\cal B}_{\theta}}\Big[
\big(\left(\phi_{2},\phi_{3}\right)_{{\cal B}_{\theta}}
\phi_{4},\phi_{1}\big)_{{\cal B}_{\theta}}
\Big] \nonumber\\
&=\mathrm{Tr}_{{\cal B}_{\theta}}\Big[
\left(\phi_{4},\phi_{1}\right)_{{\cal B}_{\theta}}
\left(\phi_{2},\phi_{3}\right)_{{\cal B}_{\theta}}
\Big],
\label{duality}
\end{align}
where we have used $\mathrm{Tr}_{{\cal
      A}_{\theta}}\left[(\phi,\chi)_{{\cal A}_{\theta}}\right]=
\mathrm{Tr}_{{\cal
      B}_{\theta}}\left[(\chi,\phi)_{{\cal B}_{\theta}}\right]$.  We
  have also labeled explicitly all the fields in order to better
 visualize the manipulation we have made. 

At the level of the Feynman diagrams, especially for the (total)
vertex, this duality reads

\bigskip

\begin{equation}
\sum_{\ga\in\Ga}
\parbox{20mm}{\begin{picture}(15,15)
\put(0,0){\begin{fmfgraph}(15,15)
\fmfstraight
\fmftopn{o}{3}\fmfbottomn{u}{3}
\fmf{fermion}{u3,u2,u1}
\fmf{fermion}{o1,o2,o3}
\fmf{photon}{u2,o2}
\end{fmfgraph}}
\end{picture}}
\
\quad 
\longleftrightarrow
\quad
\sum_{i\hat\ga\in\widehat\Ga}
\parbox{20mm}{\begin{picture}(15,15)
\put(4,0){\begin{fmfgraph}(15,15)
\fmfstraight
\fmfleftn{l}{3}\fmfrightn{r}{3}
\fmf{fermion}{l3,l2,l1}
\fmf{fermion}{r1,r2,r3}
\fmf{photon}{l2,r2}
\end{fmfgraph}}
\end{picture}}.
\end{equation}
\bigskip

\noindent
Then, it is easy to see that this duality exchanges different types of
diagrams: for planar diagrams lines and bubbles are exchanged.
The dual of an arbitrary diagram is constructed in two
steps. First, replace any wavy line by a double line. Then, in each
double line, relate the two lines by a new wavy line. The dual theory
can also  be written using a Hubbard-Stratonovitch transform with an
auxiliary field $B\in{\cal B}_{\theta}$, but now the opposite
convention have to be used: following the arrows, the field $B$ leaves
on the right.

\begin{equation}
\begin{array}{ccc}
\parbox{40mm}{\begin{picture}(40,40)
\put(0,5){\begin{fmfgraph}(40,40)
\fmfstraight
\fmfbottomn{u}{4}
\fmf{plain}{u1,u2}
\fmf{plain}{u3,u4}
\fmf{fermion}{u3,u2}
\fmfleft{c1}
\fmfright{c2}
\fmffreeze
\fmfforce{(.5w,.5h)}{c1}
\fmfforce{(.5w,.5h)}{c2}
\fmffreeze
\fmfshift{(-.20w,-.1h)}{c1}
\fmfshift{(.20w,-.1h)}{c2}
\fmffreeze
\fmfleft{d1}
\fmfright{d2}
\fmftop{d3}
\fmftop{d4}
\fmffreeze
\fmfforce{(.35w,.54h)}{d1}
\fmfforce{(.65w,.54h)}{d2}
\fmfforce{(.35w,.26h)}{d3}
\fmfforce{(.65w,.26h)}{d4}
\fmffreeze
\fmf{fermion,right}{c1,c2}
\fmf{fermion,right}{c2,c1}
\fmf{photon}{u2,d3}
\fmf{photon}{u3,d4}
\fmf{photon,right=2}{d2,d1}
\end{fmfgraph}}
\end{picture}}
&\longleftrightarrow&
\parbox{40mm}{\begin{picture}(40,40)
\put(0,-2){\begin{fmfgraph}(40,40)
\fmfstraight
\fmftopn{u}{5}
\fmf{fermion}{u4,u3,u2}
\fmf{plain}{u1,u2}
\fmf{plain}{u4,u5}
\fmffreeze
\fmftop{c1,c2,c3,c4,c5}
\fmffreeze
\fmfforce{(.33w,.3h)}{c1}
\fmfforce{(.67w,.3h)}{c2}
\fmfforce{(.36w,.7h)}{c3}
\fmfforce{(.64w,.7h)}{c4}
\fmfforce{(.5w,.84h)}{c5}
\fmffreeze
\fmf{fermion,left}{c2,c1}
\fmf{fermion,left}{c1,c2}
\fmffreeze
\fmf{plain,left}{c3,c4}
\fmf{fermion,left}{c4,c3}
\fmffreeze
\fmf{photon}{u3,c5}
\fmf{photon,right=.35}{u2,c1}
\fmf{photon,left=.35}{u4,c2}
\end{fmfgraph}}
\end{picture}}\quad.
\end{array}
\end{equation}

For the simple Heisenberg module obtained from Schwartz functions on
${\Bbb R}$, the scalar product with values in ${\cal
  B}_{\theta}={\cal A}_{1/\theta}$ is simply
\begin{align}
\left(\chi,\phi\right)_{{\cal A}_{1/\theta}}&=
\frac{1}{\theta}\mathop{\sum}\limits_{\widehat{\gamma}\in\widehat{\Gamma}}
\langle\phi,U_{\widehat{\gamma}}\chi\rangle_{L^2(\R)}\,\,
U_{\!-\widehat{\gamma}}\\
&=\frac{1}{\theta}\mathop{\sum}\limits_{\widehat{\gamma}\in\widehat{\Gamma}}
\left(\int_{{\Bbb R}}dx \,\overline{\phi}(x)\,
e^{\frac{i\pi mn}{\theta}}\,\chi(x+m)\,e^{\frac{2i\pi
    nx}{\theta}}\right)\,
U_{\!-\widehat{\gamma}}.
\end{align}
One passes from one scalar product to the other one by Poisson
re-summation. In the study of the renormalizability, it will prove to
be helpful to use
the Poisson re-summation for some of the $\gamma$'s only. This
corresponds to a duality operation on only some of the vertices.\\

\subsection{Reduced theory}

To make contact with the standard analysis of divergences
of a field theory in the Moyal plane \cite{filk}, it is helpful to  write
the propagator as
\begin{equation}
\label{param}
H^{-1}=\int_{0}^{\infty}\!d\alpha\, e^{-\alpha H}.
\end{equation}
Therefore, the contribution of any diagram $G$ with $N$ internal 
lines and $M$ vertices is
expressed as an integral over $N$ Schwinger parameters and a sum over
$M$ lattice elements
\begin{equation}
\int d\alpha_{1}\cdots d\alpha_{N}
\mathop{\sum}\limits_{\gamma_{1},\dots,\gamma_{M}}\,
I_{G}(\alpha_{1},\dots,\alpha_{N};\gamma_{1},\dots,\gamma_{M}).
\end{equation}
We refer to the next section for a precise evaluation of this integral
in the Bargmann representation.

At a heuristic level, it is expected that all the divergences arise
from the regions of integration in which one or several Schwinger
parameters go to zero. Indeed, if all the Schwinger parameters are
different from zero, the traces in $I_{G}$ as well as the sums over
lattice elements $\gamma$ are convergent as tempered distributions.
We shall see in the next sections the details of this fact for all
one-loop diagrams.

Therefore, it is interesting to analyze
$I_{G}(\alpha_{1},\dots,\alpha_{N};\gamma_{1},\dots,\gamma_{M})$ when
all Schwinger parameters are set to zero. In this case, there are no
propagators anymore and only the insertions of $U_{\gamma}$ remain. For
any trace around a closed loop of internal lines, there is a
divergence proportional to the trace of the product of the operators
$U(\gamma)$ along the loop. If we renormalize the trace such that the
trace of the identity is $1$, this reduced vertex reads

\begin{equation}
\begin{array}{ccc}
\parbox{35mm}{\begin{picture}(30,30)
\put(0,0){\begin{fmfgraph}(30,30)
\fmfcurved
\fmfsurroundn{v}{6}
\fmfbottom{c}
\fmffreeze
\fmfforce{(.5w,.5h)}{c}
\fmffreeze
\fmfsurroundn{y}{6}
\fmfshift{(-0.3w,0h)}{y1}
\fmfshift{(0.3w,0h)}{y4}
\fmffreeze
\fmf{fermion,right}{y1,y4}
\fmf{fermion,right}{y4,y1}
\fmffreeze
\fmfshift{(-0.15w,-0.28h)}{y2}
\fmfshift{(-0.15w,0.28h)}{y6}
\fmfshift{(0.15w,-0.28h)}{y3}
\fmfshift{(0.15w,0.28h)}{y5}
\fmffreeze
\fmf{photon}{v1,y1}
\fmf{photon}{v2,y2}
\fmf{photon}{v3,y3}
\fmf{photon}{v5,y5}
\fmf{photon}{v6,y6}
\end{fmfgraph}}
\put(2,11){$\vdots$}
\put(2,16){$\vdots$}
\put(16,27){\mbox{$\gamma_3$}}
\put(26,17){\mbox{$\gamma_2$}}
\put(24,3){\mbox{$\gamma_1$}}
\put(2,25){\mbox{$\gamma_4$}}
\put(2,5){\mbox{$\gamma_n$}}
\end{picture}}&
\rightarrow&
\parbox{35mm}{\begin{picture}(25,30)
\put(0,0){\begin{fmfgraph}(25,30)
\fmfcurved
\fmfsurroundn{v}{6}
\fmfbottom{c}\fmfdot{c}
\fmffreeze
\fmfforce{(.5w,.5h)}{c}
\fmf{photon}{v1,c}
\fmf{photon}{v2,c}
\fmf{photon}{v3,c}
\fmf{photon}{v5,c}
\fmf{photon}{v6,c}
\end{fmfgraph}}
\put(5,11){$\vdots$}
\put(5,16){$\vdots$}
\put(13,27){\mbox{$\gamma_3$}}
\put(24,17){\mbox{$\gamma_2$}}
\put(20,3){\mbox{$\gamma_1$}}
\put(0,25){\mbox{$\gamma_4$}}
\put(0,5){\mbox{$\gamma_n$}}
\end{picture}}
\end{array}
\end{equation}
\begin{equation}
\mathop{\prod}\limits_{i<j}e^{i\pi\theta(\gamma_{i},\gamma_{j})}\,
\delta\left(\gamma_{1}+\cdots+\gamma_{M}\right),\nonumber
\end{equation}
where the ordering is given by the orientation of the loop.
By the same method, we can reduce an open line. It is important to
keep in mind that the reduced diagrams are also to be thought of as
ribbon diagrams, because the cyclic orders at the vertices matters:

\begin{equation}
\begin{array}{ccc}
\parbox{40mm}{\begin{picture}(40,20)
\put(0,5){\begin{fmfgraph}(40,15)
\fmfstraight
\fmftopn{o}{7}
\fmfbottomn{u}{7}
\fmf{plain}{o1,o7}
\fmf{phantom_arrow,tension=0}{o4,o2}
\fmf{photon}{u2,o2}
\fmf{photon}{u4,o4}
\fmf{photon}{u5,o5}
\fmf{photon}{u6,o6}
\end{fmfgraph}}
\put(11,5){$\dots$}
\put(32,2){\mbox{$\gamma_1$}}
\put(25,2){\mbox{$\gamma_2$}}
\put(19,2){\mbox{$\gamma_3$}}
\put(4,2){\mbox{$\gamma_M$}}
\end{picture}}&
\rightarrow&
\parbox{40mm}{\begin{picture}(40,20)
\put(0,5){\begin{fmfgraph}(40,15)
\fmfstraight
\fmftopn{o}{3}\fmfdot{o2}
\fmfbottomn{u}{7}
\fmf{plain}{o1,o3}
\fmf{phantom_arrow,tension=0}{o3,o2,o1}
\fmf{photon}{u2,o2}
\fmf{photon}{o2,u4}
\fmf{photon}{u5,o2}
\fmf{photon}{o2,u6}
\end{fmfgraph}}
\put(32,2){\mbox{$\gamma_1$}}
\put(25,2){\mbox{$\gamma_2$}}
\put(19,2){\mbox{$\gamma_3$}}
\put(4,2){\mbox{$\gamma_M$}}
\put(11,5){$\dots$}
\end{picture}}
\end{array}
\end{equation}
\begin{equation}
\mathop{\prod}\limits_{i<j}e^{i\pi\theta(\gamma_{i},\gamma_{j})}\,
U_{\gamma_{1}+\cdots+\gamma_{M}},\nonumber
\end{equation}
where the ordering is again given by the orientation of the line.

The reduced theory behaves as a very simple model with vertices given
by the previous two types of diagrams, propagators equal to
one and momenta corresponding to the lattice elements. External legs of
the reduced diagrams are given by the external
legs of the lines that have been reduced and the sum of the lattice
elements along the line corresponds to the external momenta of the
reduced diagram.  Momentum is conserved at the internal vertices and
only the loops of the reduced diagram involve a non-trivial
summation. Besides, any diagram carries a phase factor that is
readily computed using the techniques introduced by T. Filk
\cite{filk}.

It follows from this work that the phase associated with a planar
diagram (i.e.\ a diagram that can be drawn without crossings) only
depends on the external momenta of the reduced diagram. Therefore, the
diagram is independent of the loop momenta flowing into the reduced
diagram and the corresponding summation is divergent. For example, in
the ladder diagram contributing to the 4-point function, the summation reads 
\begin{align}
\parbox{30mm}{\begin{picture}(30,25)
\put(0,2){\begin{fmfgraph}(30,20)
\fmfstraight
\fmftopn{o}{3}\fmfdot{o2}
\fmfbottomn{u}{3}\fmfdot{u2}
\fmf{fermion}{o1,o2,o3}
\fmf{fermion}{u3,u2,u1}
\fmf{photon,right=.5}{u2,o2}
\fmf{photon,right=.5}{o2,u2}
\fmf{phantom_arrow,right=.5}{u2,o2}
\fmf{phantom_arrow,left=.5}{u2,o2}
\end{fmfgraph}}
\put(22,10){\mbox{\small$\gamma_2$}}
\put(4,10){\mbox{\small$\gamma_1$}}
\end{picture}}
=\mathop{\sum}\limits_{\gamma_{1},\gamma_{2}}
U_{\gamma_{1}}\,U_{\gamma_{2}}\otimes
U_{\!-\gamma_{1}}\,U_{\!-\gamma_{2}}=
\mathop{\sum}\limits_{\gamma}
U_{\gamma}\otimes U_{\!-\gamma}
\!\!\underbrace{\Big(\mathop{\sum}\limits_{\gamma_{1}} 1\Big)}
_{\mathrm{ constant\,\,divergence}},
\end{align}
with $\gamma=\gamma_{1}+\gamma_{2}$. The remaining sum over
$\gamma_{1}$ is obviously divergent and its divergent part is proportional
to the vertex  $\sum_{\gamma}U_{\gamma}\otimes U_{\!-\gamma}$ so that
one can infer that it is renormalizable.

The non-planar analogue of the previous diagram behaves very
differently. Its contribution to the 4-point function is
\begin{align}
\parbox{30mm}{\begin{picture}(30,25)
\put(0,2){\begin{fmfgraph}(30,20)
\fmfstraight
\fmftopn{o}{3}\fmfdot{o2}
\fmfbottomn{u}{3}\fmfdot{u2}
\fmf{fermion}{o1,o2,o3}
\fmf{fermion}{u3,u2,u1}
\fmffreeze
\fmftop{c1}
\fmffreeze
\fmfforce{(.5w,.5h)}{c1}
\fmftop{c2}
\fmffreeze
\fmfforce{(.5w,.5h)}{c2}
\fmffreeze
\fmfshift{(.05w,.05h)}{c1}
\fmfshift{(-.05w,-.05h)}{c2}
\fmffreeze
\fmftop{c}
\fmffreeze
\fmfforce{(.5w,.5h)}{c}
\fmf{photon,right=.5}{o2,c}
\fmf{photon,left=.5}{c,u2}
\fmf{phantom_arrow,right=.5}{u2,c}
\fmf{photon,left=.55}{u2,c2}
\fmf{phantom_arrow,left=.55}{u2,c2}
\fmf{photon,right=.55}{c1,o2}
\end{fmfgraph}}
\put(20,14){\mbox{\small$\gamma_2$}}
\put(8,14){\mbox{\small$\gamma_1$}}
\end{picture}}
=\mathop{\sum}\limits_{\gamma_{1},\gamma_{2}}
U_{\gamma_{1}}\,U_{\gamma_{2}}\otimes
U_{\!-\gamma_{2}}\,U_{\!-\gamma_{1}}=
\mathop{\sum}\limits_{\gamma}
U_{\gamma}\otimes U_{\!-\gamma}
\underbrace{\Big(\mathop{\sum}\limits_{\gamma_{1}} 
e^{2i\pi\theta(\gamma_{1},\gamma)}\Big)}
_{\gamma\!-\!\mathrm{dependent\,\,divergence}}.
\end{align}

\noindent

In the integral case, the phase factor is trivial and the integral
diverges for all $\gamma$ as in the commutative theory. If the phases
are all rational, then there is a divergence for those $\gamma$ such
that $U_{\gamma}$ is central, for other values of $\gamma$ there is an
exact cancellation. In the irrational case, there is obviously a
divergence when $\gamma=0$. We
shall see  in the section on 1-loop renormalization that this is the
only divergence when  $\theta$ satisfies a Diophantine condition.\\

All the properties of the noncommutative field theory we have
presented in this section are independent of the representation we
choose. As such, they are best derived in a general setting, without
using any specific representation for the operators $H$ and $U_{\gamma}$.
In the next two sections, we shall use the position space (or
equivalently, by Langmann-Szabo duality, the momentum space
representation) to uncover an analogy with matrix models and with the
Grosse-Wulkenhaar one \cite{Grosse:2004yu}, and the
holomorphic representation for an explicit evaluation of Feynman
diagrams and their divergences.

\subsection{Position space Feynman rules and relations with other models}

\subsubsection{Relation with matrix models}
\label{matmod}

Let us come back to the non-local structure
of the interaction in $x$-space. For the simplest module,
i.e.\ $\E_\SS=\SS(\R^2)$,  the interaction term given by
\eqref{actionx}, can be written as
\begin{equation}
\int_{{\Bbb R}^{4}}dx\,dy\,dz\,dt\,V(x,y,z,t)\,
\overline\phi(x)\phi(y)\overline\phi(z)\phi(t),
\end{equation}
with 
\begin{equation}
V(x,y,z,t)=\frac{\lambda}{2}\sum_{m,n\in{\Bbb Z}}
\delta(y-x-m\theta)\delta(z-x-m\theta-n)\delta(t-x-n).\label{vertexx}
\end{equation}
This leads to a vertex
%
\begin{equation}
\parbox{44mm}{\begin{picture}(44,35)
\put(4,4){\begin{fmfgraph}(30,25)
\fmfstraight
\fmfleftn{l}{3}\fmfrightn{r}{3}
\fmf{fermion}{l3,l2,l1}
\fmf{fermion}{r1,r2,r3}
\fmf{photon}{l2,r2}
\end{fmfgraph}}
\put(2,30){\mbox{\small$x$}}
\put(1,1){\mbox{\small$x\!+\!m\theta$}}
\put(30,30){\mbox{\small$x\!+\!n$}}
\put(26,1){\mbox{\small$x\!+\!n\!+\!m\theta$}}
\end{picture}}
\end{equation}

The non-local structure of this interaction given by the vertex
\eqref{vertexx} is intuitively understood
as follows. First let us notice that the interaction involves
products of the fields at the four points $x$, $x+m\theta+n$, $x+n$
and $x+m\theta$. These four points are identical in the quotient space
${\Bbb R}\big/({\Bbb Z}\!+\!\theta{\Bbb Z})$ so that the theory would be
local if it could be formulated on such a quotient. This is possible
if $\theta$ is rational but otherwise the quotient space is badly
behaved from a topological viewpoint, namely because $\Z+\th\Z$ is
dense in $\R$ when $\th$ is irrational. Such a quotient space is related
to Kronecker's foliation and is fruitfully understood using the
powerful techniques presented in \cite{connes}. Here we shall content 
ourselves with a very rough analysis by disregarding the topological 
difficulties so that we simply write any real number as
$x=[x]+k\theta+l$, where $[x]$ is the class of $x$ in
the quotient and $k$ and $l$ are elements of ${\Bbb Z}$. Accordingly,
the interaction vertex can be written as
%
\begin{equation}
\parbox{70mm}{\begin{picture}(70,35)
\put(12,4){\begin{fmfgraph}(30,25)
\fmfstraight
\fmfleftn{l}{3}\fmfrightn{r}{3}
\fmf{fermion}{l3,l2,l1}
\fmf{fermion}{r1,r2,r3}
\fmf{photon}{l2,r2}
\end{fmfgraph}}
\put(3,31){\mbox{\small$[x]\!+\!k\theta\!+\!l$}}
\put(0,0){\mbox{\small$[x]\!+\!(m\!+\!k)\theta\!+\!l$}}
\put(32,31){\mbox{\small$[x]\!+\!k\theta\!+\!l\!+\!n$}}
\put(30,0){\mbox{\small$[x]\!+\!(m\!+\!k)\theta\!+\!(l\!+\!n)$}}
\end{picture}}
\end{equation}
It appears that all the fields are evaluated at a point which corresponds
to the same equivalence class $[x]$ whereas the interaction
mixes the integers $k$ and $l$ in a particular pattern
reminiscent of matrix models. Indeed in a matrix model involving
$(p\x q)$-rectangular matrices with an interaction of the type
\begin{equation}
V(M,M^{\dagger})=\frac{\lambda}{2}\mathrm{Tr}\big[M^{\dagger}MM^{\dagger}
M\big]
=\frac\lambda2 \sum_{0\leq i_1,i_2\leq q-1\atop 0\leq j_1,j_2\leq p-1}
(M^{\dagger})_{i_1j_1}M_{j_1i_2}(M^{\dagger})_{i_1j_2}M_{j_2i_1},
\end{equation}
the interaction vertex is

\bigskip

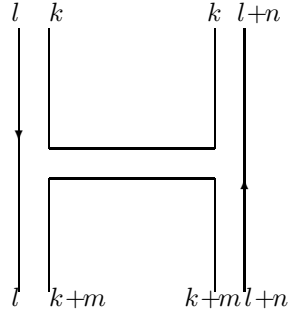
\begin{center}

\parbox{40mm}{\begin{picture}(40,40)
\put(33,3){\vector(0,1){15}}
\put(33,18){\line(0,1){20}}
\put(3,38){\vector(0,-1){15}}
\put(3,23){\line(0,-1){20}}
\put(7,38){\line(0,-1){16}}
\put(7,22){\line(1,0){22}}
\put(29,38){\line(0,-1){16}}
\put(7,3){\line(0,1){15}}
\put(7,18){\line(1,0){22}}
\put(29,3){\line(0,1){15}}
\put(2,39){\mbox{\small$l$}}
\put(7,39){\mbox{\small$k$}}
\put(2,1){\mbox{\small$l$}}
\put(7,1){\mbox{\small$k\!+\!\!m$}}
\put(28,39){\mbox{\small$k$}}
\put(32,39){\mbox{\small$l\!+\!\!n$}}
\put(33,1){\mbox{\small$l\!+\!\!n$}}
\put(25,1){\mbox{\small$k\!+\!\!m$}}
\end{picture}}

\end{center}

\bigskip

The integers $k$, $l$, $k+m$ and $l+n$ correspond to matrix indices
$i_1$, $j_1$, $i_2$ and $j_2$.  In the framework of rectangular matrix models,
the duality \eqref{duality} exchanges $M$ with its adjoint
$M^{\dagger}$ \cite{Cicuta:1986tn}.

If $\theta=\frac{p}{q}$, with $p$ and $q$ two relatively prime
positive integers, then the relation with rectangular matrix models
can be established as follows. Define the vector bundle $F\to\T^2$
over a commutative 2-torus of radius $1/q$, whose fibers are $(p\x q)$
complex matrices, in terms of its module of smooth sections
$\Ga^\infty(F)$. The latter is defined as the set of $(p\x
q)$ matrix-valued smooth functions on $[0,1/q]\x[0,1/q]$, which are
periodic in the second variable but only quasi-periodic in the first:
\begin{equation}
\left\{
\begin{array}{r@{\;}c@{\;}l}
M(x,y+\frac{1}{q})&=&M(x,y),\\[1ex]
M(x+\frac{1}{q},y)&=&\Omega_{p}^{a}(qy)M(x,y)\Omega_{q}^{-b}(-qy),
\label{boundary}
\end{array}
\right. 
\end{equation}
where $a$ and $b$ are two integers such that $aq+bp=1$ and
$\Omega_{N}(y)$ is the $N\times N$ matrix defined by
\begin{equation}
\Omega_{N}(y)=
\begin{pmatrix}
0&1&&\cr
&\ddots&\ddots&\cr
&&&1\cr
e^{2i\pi y}&&&0
\end{pmatrix}.
\end{equation}

Now, given $\phi\in\SS(\R)$ one associates an element of
$\Ga^\infty(F)$ via the map $\rho$

\begin{equation}
\rho:\SS(\R)\to\Ga^\infty(F),\quad \phi(x)\mapsto
M_{ij}(x,y)=
\sum_{n\in{\Bbb Z}}\phi\Big(x+\frac{iq+jp+npq}{q}\Big)e^{-2i\pi
  nqy},
\label{matrix}
\end{equation}  
with $i\in\left\{0,1,\dots, p-1\right\}$ and $j\in\left\{ 0,1,\dots,
  q-1\right\}$. 
Using the identity
\begin{equation}
[\Omega^{n}_{N}(y)]_{ij}=
\sum_{k}\delta_{j,i+n-kN}e^{2i\pi ky},
\end{equation}
where only one term in the sum is non-zero,
it is easy to prove that $M$ satisfies the boundary conditions
\eqref{boundary}. 
Conversely, one can define a function $\phi\in\SS(\R)$ form a 
matrix $M\in\Ga^\infty(F)$ using the map $\rho^*$ (the notation will
be justified soon):
\begin{equation}
\rho^*:\Ga^\infty(F)\to\SS(\R),\quad M(x,y)\mapsto
\phi(x)=q\int_{0}^{\frac{1}{q}}dy\, M_{00}(x,y).
\label{function}
\end{equation}
{}From the boundary conditions \eqref{boundary}, it is straightforward
to see that the latter is well defined as an element of $\SS(\R)$.

The map $\rho$
establishes an isomorphism between $\SS(\R)$ and $\Ga^\infty(F)$, 
whose inverse is $\rho^*$. Moreover, $\rho$ and $\rho^*$ 
extend to unitary operators at the
level of $L^2$-completions. To this aim, let $L^2(\T^2,F)$ be the
completion of $\Ga^\infty(F)$ with respect to the norm $\|.\|_F$ induced from
the scalar product
$\displaystyle \langle M,N\rangle_F
=q\int_0^{1/q}dx\int_0^{1/q}dy\Tr[M^\dagger(x,y) N(x,y)]$.
\begin{lem}
For $\th=\frac p q$, with $p$ and $q$ relatively prime positive
integer, the map $\rho$ 
is a unitary operator from $L^2(\R)$ to
$L^2(\T^2,F)$, whose adjoint is $\rho^*$.
\label{lemlem}
\end{lem}
\begin{proof}
First note that the maps \eqref{matrix} $\rho:\SS(\R)\to\Ga^\infty(F)$
and \eqref{function} $\rho^*:\Ga^\infty(F)\to\SS(\R)$ are inverse each
other. Indeed, for any $\phi\in\SS(\R)$, we have
\begin{align*}
\big(\rho^*\rho\phi\big)(x)=q\int_0^{1/q}dy\big(\rho\phi\big)_{00}(x,y)
&=q\int_0^{1/q}dy\sum_{n\in\Z}\phi(x+np)\,e^{-2i\pi nqy}\\
&=\sum_{n\in\Z}\delta_{n,0}\,\phi(x+np)=\phi(x).
\end{align*}

Conversely, for any $M\in\Ga^\infty(F)$, we find
\begin{align*}
\big(\rho\rho^*M\big)_{ij}(x,y)&=
\sum_{n\in\Z}\big(\rho^*M\big)\big(x+\frac{iq+jp+npq}{q}\big)\, 
e^{-2i\pi nqy}\\
&=
q\sum_{n\in\Z}\int_0^{1/q}dz\,M_{00}\big(x+\frac{iq+jp+npq}{q},z\big)\, 
e^{-2i\pi nqy}.
\end{align*}
Using the boundary conditions \eqref{boundary} and the
Poisson re-summation formula (in the sense of tempered distributions),
the former expression reads
$$
q\sum_{n\in\Z}\int_0^{1/q}dz\,M_{ij}(x,z)\,
e^{-2i\pi nq(y-z)}=
q\sum_{n\in\Z}\int_{n/q}^{(n+1)/q}dz\,M_{ij}(x,z)\,
\delta(qz-qy)=M_{ij}(x,y).
$$

Now for $\phi,\chi\in\SS(\R)$, denote by 
$M,N\in\Ga^\infty(F)$ the corresponding 
matrices. Let us also introduce the bijective map 
\begin{equation}
\begin{array}{lrcl}
\Xi:
&\left\{0,1,\dots,p-1\right\}\times\left\{0,1,\dots,p-1\right\}\x\Z&
\rightarrow&{\Bbb Z}\\
&(i,j,n)&\mapsto&iq+jp+npq.
\end{array}
\end{equation}
Using this map, one has
\begin{align*}
\langle M,N\rangle_F&=q\int_{0}^{\frac{1}{q}}dx\int_{0}^{\frac{1}{q}}dy\,
\mathrm{Tr}\big[M^{\dagger}(x,y)N(x,y)\big]\\
&=
\mathop{\sum}\limits_{\Xi(i,j,n)\in{\Bbb Z}}\int_{0}^{\frac{1}{q}}dx\,
\overline{\phi}\Big(x+\frac{\Xi(i,j,n)}{q}\Big)
\chi\Big(x+\frac{\Xi(i,j,n)}{q}\Big)\\
&=
\int_{-\infty}^{+\infty}dx\, \overline{\phi}(x)\chi(x)
=\langle\phi,\chi\rangle_{L^2(\R)}.
\end{align*}
Thus $\|\rho\phi\|_F=\|\phi\|_{L^2(\R)}$ and
$\|\rho^*M\|_{L^2(\R)}=\|M\|_F$, for any  $\phi\in\SS(\R)$ and
$M\in\Ga^\infty(F)$. 
By density, this proves that $\rho$ and $\rho^*$ extend to isometries
on $L^2(\R)$ and $L^2(\T^2,F)$ and that they are adjoint each
other. This concludes the proof since one has already checked that
$\rho^*\rho=1_{L^2(\R)}$, $\rho\rho^*=1_{L^2(\T^2,F)}$.
\end{proof}

It is worthwhile to notice that the map $\rho$ possesses more
structures. In particular, it preserves the module structure, where
$\Ga^\infty(F)$ is equipped with a left action of the bundle algebra of
$(p\x p)$-matrices over $\T^2$, with boundary conditions :

\begin{equation}
\left\{
\begin{array}{r@{\;}c@{\;}l}
A(x,y+\frac{1}{q})&=&A(x,y),\\[1ex]
A(x+\frac{1}{q},y)&=&\Omega_{p}^{a}(qy)A(x,y)\Omega_{p}^{a}(qy)^\dagger,
\label{boundary2}
\end{array}
\right. .
\end{equation}
In particular,
the two covariant derivatives defining the connection
\eqref{connectionx} on the Schwartz space translate into
\begin{equation}
\nabla_{1}=\frac{\partial}{\partial y}-\frac{2i\pi p
  x}{q}+A\cdot+\cdot B\quad\mathrm{and}\quad
\nabla_{2}=\frac{\partial}{\partial x},
\end{equation}
on the rectangular matrices. In this definition, $A$ and $B$ are two
square $p\times p$ and $q\times q$ constant diagonal matrices acting by
left and right multiplication.  Their diagonal elements are
\begin{equation}
A_{kk}=-2i\pi p\, k\quad\mathrm{and}\quad B_{ll}=-\frac{2i\pi }{p}\, l.
\end{equation}
Therefore, at rational $\theta$, the Heisenberg module simply
describes a bundle of rectangular matrices over an ordinary torus. These
relations generalize the one given in \cite{kirilov}, which
corresponds to the case $p=q=1$.\\

Consider now a rectangular matrix model, with matrix-valued functions 
satisfying the boundary conditions \eqref{boundary} and an action
given by
\begin{align}
&S[M,M^{\dagger}]=\\
&\quad\int_{0}^{\frac{1}{q}}dx\int_{0}^{\frac{1}{q}}dy\,
\mathrm{Tr}\big[\nabla_{\mu}M^{\dagger}(x,y)\nabla^{\mu}M(x,y)
+\mu_0^{2}M^\dagger(x,y)M(x,y)
+\frac{\lambda}{2}\big(M^\dagger (x,y)M(x,y)\big)^{2}
\big].\nonumber
\end{align}
If we express the matrix $M$ in terms of $\phi$ via the map $\rho$, 
the matrix model
action agrees with the Heisenberg module action \eqref{actionx} up to
a factor $q$,
\begin{equation}
qS[M,M^{\dagger}]=S[\phi,\overline{\phi}].
\end{equation} 

The equality for the quadratic part of the action is already
contained in the Lemma \ref{lemlem}.
The interaction term can be treated along the same lines. The matrix
model interaction reads
\begin{equation}
\frac{q\lambda}{2}
\sum_{i,j,k,l}\int_{0}^{\frac{1}{q}}dx\int_{0}^{\frac{1}{q}}dy\,
\overline{M}_{ij}(x,y)\,M_{ik}(x,y)\,\overline{M}_{lk}(x,y)\,M_{lj}(x,y).
\end{equation}
Expressing $M$ in terms of $\phi$ and integrating over $y$ yields
\begin{equation}
\begin{array}{rl}
\frac{\lambda}{2}\mathop{\sum}\limits_{N,N',N''}
\mathop{\sum}\limits_{i,j,k,l}
\displaystyle{\int}_{0}^{\frac{1}{q}}dx&
\overline{\phi}\left(x+\frac{iq+jp+Npq}{q}\right)
\phi\left(x+\frac{iq+kp+N'pq}{q}\right)\cr
&\times\,\overline{\phi}\left(x+\frac{lq+kp+(N'+N''-N)pq}{q}\right)
\phi\left(x+\frac{lq+jp+N''pq}{q}\right).
\end{array}
\end{equation}
After a shift of the integration variable by $\Xi(i,j,N)$ and a change
of summation indices  
\begin{equation}
\left\{
\begin{array}{rcl}
N'-jp-N&\rightarrow&N',\cr
N''-iq-N&\rightarrow&N'',
\end{array}
\right.
\end{equation}
the interaction reads
\begin{equation}
\begin{array}{rl}
\frac{\lambda}{2}\mathop{\displaystyle{\sum}}\limits_{\Xi(i,j,N)}
\mathop{\displaystyle{\sum}}\limits_{N',N''}
\mathop{\displaystyle{\sum}}\limits_{k,l}
\displaystyle{\int}_{\Xi(i,j,N)}^{\frac{\Xi(i,j,N)+1}{q}}dx&
\overline{\phi}\left(x\right)
\phi\left(x+\frac{kp+(N')pq}{q}\right)\cr
&\times\,\overline{\phi}\left(x+\frac{lq+kp+(N'+N'')pq}{q}\right)
\phi\left(x+\frac{lq+N''pq}{q}\right).
\end{array}
\end{equation}
Finally, the identification with the vertex in $x$-space
\begin{equation}
\frac{\lambda}{2}\int_{+\infty}^{-\infty}dx\,\mathop{\sum}\limits_{m,n\in{\Bbb
  Z}} \overline{\phi}\left(x\right)\phi\Big(x+m\frac{p}{q}\Big)
\overline{\phi}\Big(x+m\frac{p}{q}+n\Big)\phi\left(x+n\right),
\end{equation}
follows from the euclidian division of $m$ by $q$ and of $n$ by $p$,
\begin{equation}
\left\{
\begin{array}{rcl}
m&=&k+qN',\cr
n&=&l+pN''.
\end{array}
\right. 
\end{equation}

Let us note that the action in terms of the Schwartz function $\phi$
only involves the ratio $\frac{p}{q}$ and does not depend separately
on $p$ and $q$. Therefore, it is suited to the study of the limit
case 
\begin{equation}
p\rightarrow\infty \quad \mathrm{and} \quad
q\rightarrow\infty,\quad \mathrm{with} \quad
\frac{p}{q}\rightarrow\theta,
\end{equation}
where $\theta$ is a fixed positive real number. Thus, one has
\begin{equation}
\lim_{p,q\to\infty\atop p/q\to\th}
\int[DM][DM^{\dagger}]e^{-qS[M,M^{\dagger}]}{\cal O}\big(M,M^{\dagger}\big)
=\int[D\phi][D\overline{\phi}]e^{-S[\phi,\overline\phi]}
{\cal O}
\left(\phi,\overline{\phi}\right).
\end{equation}
In this equation, one has to keep in mind that the matrix-valued function $M$
satisfies the boundary condition \eqref{boundary}, which means that
its period is $1/q$, 
along the two directions $x$ and $y$, which goes to zero as $q\to\infty$. In
this respect, the quantum field theory over the Heisenberg module
corresponds to a twisted reduced matrix model, the size of the
underlying space going to zero. In the next sections, we shall see
that the renormalization properties of these models involves some
number-theoretical properties of $\theta$ for non-planar diagrams.

\subsubsection{Relation with the Grosse-Wulkenhaar model}

Recall that in \cite{Grosse:2004yu, RVW} it is proven that the
duality-covariant $\phi_4^4$-theory on the Moyal
hyper-plane is renormalizable
to all orders in perturbation theory. In the exact self-dual case
($\Omega=1$ in the language of \cite{Grosse:2004yu}), the theory is
defined by the following action functional
\begin{equation}
\label{eq:classaction}
S[\phi]:=\frac12\int d^4x\, \phi(x)\big(\tri+X^2+\mu_0^2\big)\phi(x)
+\frac{\lambda}{2} \int d^4x\,(\phi\star\phi\star\phi\star\phi)(x),
\end{equation}
where $\tri=-\pa^\mu\pa_\mu$ is the Laplacian and $X^2=X^\mu X_\mu$,
with $X^\mu= 2 (\theta^{-1})^{\mu\nu} x_\nu$, where
$\theta_{\mu\nu}=-\theta_{\nu\mu}\in \mathbb{R}$ are the components of
the deformation matrix which defines the Moyal product $\star$, and 
$x^\mu$ denotes the ordinary Cartesian coordinate on $\R^4$.

It is worthwhile to notice that the kinetic term of that theory is
exactly the same as in the model we study here: it is given by an
harmonic oscillator Hamiltonian. But the origin of this term is quite
different. For the action \eqref{eq:classaction}, the term $X^2$ has
been ``added by hand'' to cure the UV/IR-mixing problem. More
specifically, such a term is required by the renormalization flow. But to
see it, one has to formulate the theory on the matrix base (see
\cite{Grosse:2004yu}). Moreover the presence of such term allows to
restore the Langmann-Szabo duality, which holds only for the
interaction term without the quadratic potential.  For the Heisenberg
module, the harmonic oscillator potential has a clear geometric origin
since it comes from the connection \eqref{connectionx}. In particular,
whereas the frequency of the oscillator is a free parameter in
\cite{RVW} (and has to be renormalized), this is not the case in our
model where the frequency is inverse proportional to the deformation
parameter. Furthermore, we will see in the next section, that the
value of this parameter is preserved by the renormalization flow.
However, in spite of the presence of the confining potential, the
 $\phi_4^4$-theory on the Heisenberg module will still suffer
from an UV/IR-mixing. Fortunately, in this case the UV/IR-entanglement
is easy to cure (at least at one-loop) by the adjunction of another
interaction term.

To investigate the analogies and differences of the interactions of the
two models, it is useful to rewrite the vertex \eqref{vertexx} as
\begin{equation}
V(x,y,z,t)=\frac{\lambda}{2^{2d}\theta^{2d}}\sum_{m,n\in{\Bbb Z^d}}
e^{\frac{2i\pi }{\theta}m(x-y)}\,\delta(x-y+z-t)\,
\label{vertexx2}
e^{2i\pi n(x-t)}.
\end{equation}
This formula can be obtained from \eqref{vertexx} 
after a Poisson re-summation on the index $m$. 
Here we have considered the module $\SS(\R^d)$ over a
$2d$-dimensional noncommutative torus, with deformation matrix of the
form \eqref{mostgene}. On the other hand, the interaction
vertex associated to \eqref{eq:classaction} in $d$-dimensions is given by
\begin{equation}
V_{\star}(x,y,z,t)=\frac{\lambda}{\pi^d\th^d}
e^{2i\theta^{-1}(x,y)}\,\delta(x-y+z-t)\,
e^{2i\theta^{-1}(z,t)},
\label{vertexm}
\end{equation}
where $\theta^{-1}$ is the antisymmetric bilinear form associated to the
inverse deformation matrix $\theta_{\mu\nu}$. This leads to a
correspondence between the field theory on the Heisenberg
module in dimension $2d$ (the dimension of the underlying
noncommutative torus) and the field
theory on the Moyal plane of dimension $d$. 
This correspondence relates the phases in the
interactions as
\begin{equation}
e^{2i\theta^{-1}(x,y)}\leftrightarrow
\sum_{m\in{\Bbb Z}^d}
e^{\frac{2i\pi m}{\theta}(y-x)}
\quad\mathrm{and}\quad
e^{2i\theta^{-1}(z,t)}\leftrightarrow
\sum_{n\in{\Bbb Z}^d}
e^{2i\pi n(x-t)}.
\end{equation}
The dissymmetry between the two phases has a deep impact on the power
counting of the two theories: The correct analogue of the commutative
$\phi_4^4$-theory on the noncommutative hyperplane is made out of the
Moyal product on $\R^4$, whereas in the non-trivial module case, it
has to be formulated on $\SS(\R^2)$, not on $\SS(\R^4)$.  Indeed, it
is the dimension of the noncommtative torus, not of the representation
space, that matters. This lead to the following question: How can it
be that two theories with the same propagator in different dimensions
behave the same? The answer comes precisely from dissymmetry between
the vertices \eqref{vertexx2} and \eqref{vertexm}.

To see how the vertex affects the evaluation of the Feynman diagrams,
let us have a look at the one-loop planar two-point function,
associated to a quartic interaction, for the two models. In the case
of the projective module over a $(2d)$-noncommutative torus, if we
evaluate the diagram \eqref{tp2} in external fields
$\vf_1,\vf_2\in\SS(\R^d)$, we obtain up to a numerical factor
\begin{equation}
\Tr\big((\vf_1,\vf_2)_{\A_\th}\,H^{-1}\big)=
\sum_{\ga\in\Ga}\Tr\big(U_{-\ga}\,H^{-1}\big)\,
\langle\vf_1,\vf_2 U_\ga\rangle_{L^2(\R^d)}.
\end{equation}
Recall that in $d$-dimensions, the spectrum of $H^{-1}=(\tri+X^2)^{-1}$ is
$(d/2+|n|)^{-1}$, $n\in\N^d$. For $\ga=0$ (we will see that the terms with
$\ga\ne 0$ give rise to finite contributions), the divergence is
logarithmic for $d=1$, quadratic for $d=2$ and so on. The behavior of
this diagram for  $d=2$ is reminiscent to a just-renormalizable
$\phi_4^4$ theory.\\
Alternatively, one can study the UV-divergences using a parametric
representation \eqref{param} for the propagator and looking at the
small-$\beta$ behavior. With the help of the Mehler formula, one can
express the kernel of $H^{-1}$ as
\begin{equation}
\label{eq:propagator}
H^{-1}(x,y)=\frac{\omega^2}{4\pi^2}\int_0^\infty d\b\,
\frac{1}{\sinh^{d/2}(2\omega \b)}
e^{-\frac{\omega}{4}\big(\coth(\omega \b)|x-y|^2
+\tanh(\omega \b)|x+y|^2\big)},
\end{equation}
where $\omega$ is the frequency of the oscillator. If we regularize
the theory by setting 
\begin{equation}
\label{eq:regprop}
H^{-1}\to H_\epsilon^{-1}=\int_\epsilon^\infty d\beta\,e^{-\beta H},
\end{equation}
then one gets up to an irrelevant constant
\begin{equation}
\Tr\big(H_\epsilon^{-1}\big)=\int_{\R^d}dx\,H_\epsilon^{-1}(x,x)
=\int_\epsilon^\infty d\beta\,\Big(\frac{\cosh(\omega\beta)}
{\sinh(\omega\beta)\sinh(2\omega\beta)}\Big)^{d/2}.
\end{equation}
This yields the same conclusion: the divergence  (in $\eps^{-1/2}$)
is logarithmic for $d=1$, quadratic for $d=2$.\\
Now, the same diagram for the duality-covariant model on the Moyal
plane reads
\begin{equation}
\label{pln}
\Tr\big(L(\vf_1\star\vf_2)\,H^{-1}\big),
\end{equation}
where $L(\vf)$ denotes the operator of left Moyal-multiplication by
$\vf$ (a look at \cite{UVIR} can help the unfamiliar reader to derive
such formula). We show in the appendix \ref{dualcov} that the leading
divergence of \eqref{pln} is given by
\begin{equation}
\int_{\R^d}dx\,\vf_1\star\vf_2(x)\,H^{-1}_\epsilon(x,x).
\end{equation}
The last expression equals
\begin{equation}
\int_{\R^d}dx\,\vf_1\star\vf_2(x)\,
\int_\epsilon^\infty d\b\,
\frac{1}{\sinh^{d/2}(2\omega \b)}
e^{-\omega\tanh(\omega \b)|x|^2}.
\end{equation}
Due to the presence of the fields $\vf_1,\vf_2$ inside the
$x$-integral, the former can be estimated by
\begin{equation}
\int_{\R^d}dx\,\big|\vf_1\star\vf_2 \big|(x)\,
\int_\epsilon^\infty d\b\,
\frac{1}{\sinh^{d/2}(2\omega \b)}.
\end{equation}
Since in the Moyal plane case only even dimensions are allowed (to
respect the symplectic pairs structure of the Moyal product), the
divergence of the planar two-point diagram \eqref{pln} is (only)
logarithmic for $d=2$ and (as expected) quadratic for $d=4$.

This is a strong evidence that our field theory on $\SS(\R^d)$ doesn't
behave accordingly to the dimension $d$ of the representation space,
but accordingly to the dimension $2d$ of the underlying noncommutative torus!

\subsection{Feynman diagrams in the holomorphic representation}


{}From now on, we shall work in the holomorphic representation for a
Bargmann module in two dimensions, obtained by 
tensoring twice the module of the previous type, i.e.\ corresponding
to a four-dimensional deformation matrix of the form
\eqref{mostgene}. In the explicit 
computation of Feynman diagrams, we have to evaluate traces and
kernels of
products of operators of the type $H^{-1}$ and $U_{\gamma}$ in the
holomorphic representation. To this aim, recall that an operator $A$ is
said to be an operator with distribution kernel $A(z,\overline{z}')$
if its action on $\phi\in{\cal H}_B$ can be written as
\begin{equation}
A\phi(z)=
\int_{{\Bbb C}^{2}} d\mu(z',\overline{z}')
A(z,\overline{z}')\phi(z').
\end{equation}
For instance, the identity is an operator with kernel
$I(z,\overline{z}')=e^{\omega \overline{z}'z}$; the reproducing kernel
of the Bargmann space.  By convention, our
kernels are holomorphic in the in first variable
(associated to an incoming field on the diagram) and anti-holomorphic
in the second one (associated to an outgoing field). 

{}Now and for all, we will fix a regularization scheme. To this aim, we
define the regularized propagator $H^{-1}_\epsilon$ as follows
\begin{equation}
H^{-1}_\epsilon
:=\int_{\epsilon}^{\infty}d\beta\,e^{-\beta H}=H^{-1}e^{-\epsilon H}.
\end{equation} 
As soon as $\epsilon>0$, the operator $H^{-1}_{\epsilon}$ is a
strictly positive compact operator with fast decreasing
eigenvalues, and in particular trace-class. 
This operator admits a kernel given by 
\begin{equation}
H^{-1}_{\epsilon}(z_{2},\overline{z}_{1})=\frac{1}{2}\int_{\epsilon}^{\infty}
d\beta\, e^{-\frac{\beta m^{2}}{2}}I(z_{2},e^{-\beta\omega}\overline{z}_{1}).
\label{eq:propagatorbarg}
\end{equation}
As such, it provides a natural regularization of the
Feynman integrals.
Indeed, if we come back to $\E_\SS(\R^2)$ by the
inverse Bargmann transform, it is easy to see 
that $H^{-1}_\epsilon(x,y)$ is a function
of Schwartz class in both variables (have a look at \eqref{eq:propagator} 
versus \eqref{eq:regprop}). Since the vertex \eqref{vertexx}
is a tempered distribution, this means that all the contractions with
the regularized propagator are now well defined. Using the Bargmann
transform back, it is immediate to see that the same holds for the
module $\E_B(\C^2)$.

The regularized propagator, given by the
expression \eqref{eq:propagatorbarg}, is graphically represented by a
single oriented line:
\begin{align}
\parbox{23mm}{\begin{picture}(20,5)
\put(0,0){\begin{fmfgraph}(20,5)
\fmfleft{l}
\fmfright{r}
\fmf{fermion}{l,r}
\end{fmfgraph}}
\put(15,4){\mbox{\small$\overline{z}_2$}}
\put(2,4){\mbox{\small$z_{1}$}}
\end{picture}}= 
H_\eps^{-1}(z_1,\overline{z}_{2})\;.
\end{align}

\noindent
Using the definition \eqref{projB}, it is easy to see that
the operator $U_{\gamma}$ also admits a kernel, given by
\begin{equation}
U_{\gamma}(z_{1},\overline{z_{2}})=
\mathrm{e}^{-\frac{\omega}{2}|\gamma|^2
-\omega\overline{\gamma} z_{1}} \,I(z_{1}+\gamma,\overline{z_{2}})
=\mathrm{e}^{-\frac{\omega}{2}|\gamma|^2+\omega\overline{z_{2}}\gamma}\,
I(z_{1},\overline{z_{2}}-\overline{\gamma})\;.
\end{equation}


\noindent 
Thus in the holomorphic representation, the (total or quadri-valent) 
vertex reads
\begin{align}
V(z_1,\overline{z}_2,z_3,\overline{z}_4)&=\frac\lambda2\sum_\ga
U_\ga(z_1,\overline{z}_2)\,U_{-\ga}(z_3,\overline{z}_4)\nonumber\\
&=\frac{\lambda}{2}\sum_\gamma 
\mathrm{e}^{-\omega|\gamma|^2}\,
\mathrm{e}^{\omega(\overline{z_2}-\overline{z_4})\gamma}\,
I(z_1,\overline{z_2}-\overline\gamma)\,
I(z_3,\overline{z_4}+\overline\gamma)\nonumber\\
&=\frac{\lambda}{2}\sum_\gamma \mathrm{e}^{-\omega|\gamma|^2}\,
I(\gamma,\overline{z_2}-\overline{z_4})\,
I(z_1,\overline{z_2}-\overline\gamma)\,
I(z_3,\overline{z_4}+\overline\gamma)
\end{align}
and is represented by
\begin{align}
\parbox{40mm}{\begin{picture}(35,20)
\put(0,0){\begin{fmfgraph}(35,20)
\fmfleft{l1,l2}
\fmfright{r1,r2}
\fmf{fermion}{r1,i1,l1}
\fmf{fermion}{l2,i2,r2}
\fmf{photon}{i1,i2}
\fmf{phantom_arrow,tension=0}{i2,i1}
\end{fmfgraph}}
\put(4,3){\mbox{\small$\overline{z_2}$}}
\put(28,3){\mbox{\small$z_1$}}
\put(30,16){\mbox{\small$\overline{z_4}$}}
\put(4,15.5){\mbox{\small$z_3$}}
\put(19,10){\mbox{\small$\gamma$}}
\end{picture}}= 
V(z_1,\overline{z}_2,z_3,\overline{z_4})\;.
\label{vertex}
\end{align}

When drawing the Feynman graphs it is important to notice that
propagator and vertex are oriented. When following the scalar field
arrow, the wavy $\gamma$-line always leaves the vertex to the
right. Alternatively, to stress the link with the Hubbard-Stratonovich
transformation, we may split the vertex (\ref{vertex}) into two
tri-valent vertices
\begin{align}
\parbox{38mm}{\begin{picture}(35,15)
\put(0,0){\begin{fmfgraph}(35,20)
\fmfleft{l1,l2}
\fmfright{r1,r2}
\fmf{fermion}{r1,i1,l1}
\fmf{phantom}{l2,i2,r2}
\fmf{photon}{i1,i2}
\fmf{phantom_arrow,tension=0}{i2,i1}
\end{fmfgraph}}
\put(4,3){\mbox{\small$\overline{z_2}$}}
\put(28,3){\mbox{\small$z_1$}}
\put(19,10){\mbox{\small$\gamma$}}
\end{picture}}= i\lambda
U_\gamma(z_1,\overline{z_2}),
\label{tri-vertex}
\end{align}
and a propagator for the auxiliary field
\begin{align}
\parbox{23mm}{\begin{picture}(20,5)
\put(0,0){\begin{fmfgraph}(20,5)
\fmfleft{l}
\fmfright{r}
\fmf{photon}{l,r}
\fmf{phantom_arrow}{l,r}
\end{fmfgraph}}
\put(9,5){\mbox{\small$\gamma$}}
\end{picture}}= 
\frac{1}{\lambda}\;.
\end{align}

 

\subsection{Parametric representation}

\label{parametric}

A careful analysis of the involved terms shows that
a general (regularized) graph always has the following structure
\begin{equation*}
\Gamma_n[z_1,\overline{z_2}...,z_{2N-1},\overline{z_{2N}}]=
\sum_{\gamma_1,...,\gamma_V}\int_\epsilon^\infty\prod_{k=1}^{L}
(d\beta_k\;\mathrm{e}^{-\frac{\mu_0^2}{2}\beta_k})\;
\frac{1}{U_\Gamma(\beta)}
\;\mathrm{e}^{-\gamma_i\,\mathcal{Q}_{ij}(\beta)\,\bar{\gamma}_j+
\mathcal{B}_i(z,\beta)\,\bar{\gamma}_i+
\gamma_i\,\overline{\mathcal{B}}_i(\bar{z},\beta)},
\end{equation*}
up to a symmetry factor due to it's multiple appearance and a
term purely depending on $z_1,...,\overline{z}_{2N}$.
The notation means that having chosen an arbitrary graph, $L$
indicates the number of internal scalar field line integrations and
$V$ is the number of $\gamma$-lines. Missing indices for $\beta$ or
$z$ is just short for taking all. 
We now give a complete set
of rules to construct the matrix $\mathcal{Q}_{ij}(\beta)$ and
the linear terms $\mathcal{B}_i(z,\beta)$, 
$\overline{\mathcal{B}}_i(\bar{z},\beta)$ respectively. 
These can be used
to directly write down the result, skipping Gau{\ss}ian integrations
or tracing out loops.
\medskip

Draw a graph and properly orient the scalar lines as well as 
the $\gamma$-lines. For each internal line, there is 
a $\beta$-integration accompanied by a factor 
$\exp({-\frac{\mu_0^2}{2}\beta})$. For each $\gamma$ there is a
sum. To construct the diagonal $\mathcal{Q}_{ii}$, one has to
take care of five different types of $\gamma$-lines:
\begin{enumerate}
\item Each $\gamma_i$ which connects two different external 
lines gives 
$\mathcal{Q}_{ii}=\omega$. 
\item A $\gamma_i$ connecting a line with a bubble leads to 
\begin{equation}
\mathcal{Q}_{ii}=\frac{\omega}{(1-\mathrm{e}^{-\omega\sum\beta_k})},
\end{equation} 
where the sum counts the internal integrations of the associated
bubble.
\item 
Each $\gamma_i$ that connects two loops $L_1$ and $L_2$ generates
a term 
\begin{equation}
\mathcal{Q}_{ii}=\omega\frac{(1-\mathrm{e}^{-\omega\sum\beta_k})}
{(1-\mathrm{e}^{-\omega\sum_{L_1}\beta_k})
(1-\mathrm{e}^{-\omega\sum_{L_2}\beta_k})},
\end{equation}
where $\sum_{L_1}$, respectively $\sum_{L_2}$,
counts the internal integrations of the associated loop and
the $\beta$-sum in the numerator counts them all.
\item An internal $\gamma_i$ on an external line gives
\begin{equation}
\mathcal{Q}_{ii}=\omega(1-\mathrm{e}^{-\omega\sum\beta_k}), 
\end{equation}
where the sum counts the $\beta$-integrations appearing on 
the part of the external line which is enclosed by that $\gamma_i$.
\item 
Every $\gamma_i$, which represents an internal
line inside a bubble generates
\begin{equation}
\mathcal{Q}_{ii}=
\omega\frac{(1-\mathrm{e}^{-\omega\sum_{H1}\beta_k})
(1-\mathrm{e}^{-\omega\sum_{H2}\beta_k})}{(1-\mathrm{e}^{-\omega\sum\beta_k})},
\end{equation}
where the sum in the denominator counts all $\beta$-integrations of the
bubble and $\sum_{H_1}$ $(\sum_{H_2})$ counts the $\beta$-
integrations of the first (second) half of the bubble, each defined
by the points the gamma is attached to.
\end{enumerate}
For the off diagonal elements $\mathcal{Q}_{ij}$, there are
essentially two different cases to take care of.
\begin{enumerate}
\item Lines. For a $\gamma_i$ which is attached to a chosen line, 
there are three possibilities. The $\gamma_i$ may leave or arrive,  
or it may be 'internal'. Following the orientation of the line,
denote by $\gamma_i^{(1)}$ the first position the $\gamma_i$ is
attached to and by $\gamma_i^{(2)}$ the (possible) second position.
Of course, the latter might be empty. 
Further, define $\mathrm{sign}(\gamma_i^{(a)})$ to be
$+1$ if it is arriving and  $-1$ otherwise.
The same convention holds for a $\gamma_j$ with $i\ne j$. As a last
ingredient, define $L_{ij}^{(a,b)}$ to be the sum over internal
integration variables $\beta$ enclosed by the positions 
$\gamma_i^{(a)}$ and $\gamma_j^{(b)}$ on the line.
Then, the final contribution is
\begin{equation}
\mathcal{Q}_{ij}=\omega\sum_{a,b,a<b}\mathrm{sign}(\gamma_i^{(a)})
\;\mathrm{sign}(\gamma_j^{(b)})\;\mathrm{e}^{-\omega\,L_{ij}^{(a,b)}}.
\end{equation}
With slight abuse of notation, the ordering $a<b$ is given
by the orientation of the line.
\item Loops. If there is a $\gamma_i$ and a $\gamma_j$ attached
to a chosen loop, the same conventions as in 1) can be used. Here,
$L_{ij}^{(a,b)}$ is defined to be the sum over internal integration
variables on the shortest loop segment connecting $\gamma_i^{(a)}$
and $\gamma_j^{(b)}$ following the orientation of the loop.
One finds
\begin{equation}
\mathcal{Q}_{ij}=\frac{\omega}{(1-\mathrm{e}^{-\omega\sum\beta_k})}
\sum_{a,b}\mathrm{sign}(\gamma_i^{(a)})
\;\mathrm{sign}(\gamma_j^{(b)})\;\mathrm{e}^{-\omega\,L_{ij}^{(a,b)}}.
\end{equation}
In this case, note that there is no ordering in the sum over $a,b$,
but again, the convention is to take only $a=2$ ($b=2$) if the
associated $\gamma$ really is internal.
\end{enumerate}
The general procedure to construct the matrix $\mathcal{Q}_{ij}$ is
then to regard line by line and loop by loop all $\gamma$'s that are
attached to it and to apply the rules given above. For the linear
terms $\mathcal{B}_i(z,\beta)$ and 
$\overline{\mathcal{B}}_j(\bar{z},\beta)$ it is sufficient to restrict
oneself to a line by line analysis.
\begin{enumerate}
\item For the first one, for every line decorated with an incoming
  $z_k$ and outgoing $\overline{z}_l$ denote by $\gamma_i^{(1)}$ and
  $\gamma_i^{(2)}$, as before, the positions a $\gamma_i$ might be
  attached to. Define $L_i^{(a)}$ to be the sum of internal
  integration variables on the line segment between $z_k$ and
  $\gamma_i^{(a)}$ following the orientation.  Note that $L_i^{(a)}$
  can be the empty sum.  One has
\begin{equation}
\mathcal{B}_i=-\omega\, z_k\;\sum_a \mathrm{sign}(\gamma_i^{(a)})\;
\mathrm{e}^{-\omega\,L_i^{(a)}},
\end{equation}
and the sum only runs from $a=1$ to $a=2$ if there is a second
attachment point.
\item The second term is constructed completely analogously. It is
\begin{equation}
\overline{\mathcal{B}}_i=\omega\,\overline{z}_l\;\sum_a \mathrm{sign}
(\gamma_i^{(a)})\;\mathrm{e}^{-\omega\,L_i^{(a)}},
\end{equation}
with $L_i^{(a)}$ being the sum of $\beta$'s on the oriented 
line segment between $\gamma_i^{(a)}$ and $\overline{z}_l$.
\end{enumerate}
The pre-factor $U_\Gamma(\beta)$ is a product of terms of the form
$(1-\mathrm{e}^{-\omega\sum\beta_k})^{d/2}$. Namely, for each
scalar loop $L$ it associates such a factor, that is, 
for $d=4$ dimensions
\begin{equation}
U_\Gamma(\beta)=\prod_L\;(1-\mathrm{e}^{-\omega\sum_L\beta_k})^{2}.
\end{equation}
The sum $\sum_L \beta_k$ stands again for the sum over the
internal integration variables defined on the corresponding loop.
\medskip

As an example, consider the first 4-point graph 
$\Gamma^{(1)}_{4a}[z_1,\overline{z_2},z_3,\overline{z_4}]$ 
(given in (\ref{G4a-graph})). For $\mathcal{Q}_{ij}$ and $\mathcal{B}_i$,
$\overline{\mathcal{B}}_i$ one finds
\begin{equation}
\mathcal{Q}_{ij}=\frac{\omega}{(1-\mathrm{e}^{-\omega(\beta_1+\beta_2)})}
\left( \begin{array}{cc}
1 & -\mathrm{e}^{-\omega\beta_2}\\
-\mathrm{e}^{-\omega\beta_1} & 1
\end{array}\right),
\end{equation}
and
\begin{equation}
\mathcal{B}_i=\omega\;\left( \begin{array}{c} 
-z_1\\z_3
\end{array}\right),\quad
\overline{\mathcal{B}}_i=\omega\;\left( \begin{array}{c} 
\overline{z_2}\\-\overline{z_4}
\end{array}\right),
\end{equation}
as well as
\begin{equation}
U_{\Gamma^{(1)}_{4a}}=(1-\mathrm{e}^{-\omega(\beta_1+\beta_2)})^2.
\end{equation}

To complete the construction, finally include for each external line
decorated with incoming $z_k$ and outgoing $\overline{z}_l$ a term
$\exp\{z_k\overline{z}_l\mathrm{e}^{-\omega\sum_i\beta_i}\}$ with
the sum given by the (possibly empty) sum over all integrations 
variables on that line.

\section{One-loop renormalization}

We start this section by showing that for a $\phi^4$-theory 
on the simplest Heisenberg module $\E_B(\C^2)$ over $\T^4_\th$, only the two- 
and four-point functions are divergent in the one-loop
approximation.

\begin{prop}
At one loop, for $N>2$,  the $2N$-point Green's functions
 are finite as tempered distributions.
\end{prop}
\begin{proof}
In the one-loop approximation, the 1PI quantum effective action in
external fields $\vf_1\in\E$, $\overline{\vf_2}\in\E^*$ is given by
\begin{equation}
\Ga^{(1)}[\vf_1,\overline{\vf_2}]=\frac12 \ln\det\big(H^{-1}(H+\lambda\,
B_{\vf_1,\vf_2})\big),
\end{equation}
where $H$ is the harmonic oscillator Hamiltonian in two dimensions,
$\lambda$ is the coupling constant and $B_{\vf_1,\overline{\vf_2}})$ is a bounded
operator on
$\H_\E$, depending only on $\vf_1\in\E$, $\overline{\vf_2}\in\E^*$. 
($\E$ can be either $\E_\SS(\R^2)$ or $\E_\B(\C^2)$.)
Here the definition of $\Ga^{(1)}[\vf_1,\overline{\vf_2}]$ includes the
normalization of the partition function, yielding the $H^{-1}$
pre-factor in the determinant. To see that only the two- and four-point
functions need regularization, we perform a Dyson expansion. It
formally reads:
\begin{equation}
\ln\det\big(H^{-1}(H+\lambda B_{\vf_1,\overline{\vf_2}}))\big)=\Tr\ln\big(1 
+\lambda\,H^{-1}\,B_{\vf_1,\overline{\vf_2}})\big)
=\sum_{n=1}^\infty(-)^{n+1}\frac{\lambda^n}{n}\,
\Tr\big((H^{-1}\,B_{\vf_1,\overline{\vf_2}}))^n\big).
\label{dyson}
\end{equation}
But in $d=2$ dimensions, the spectrum of $H^{-1}$ is
$\{(1+n_1+n_2)^{-1},\,n_1,n_2\in\N\}$, so that $H^{-1}$ belongs to
$\L^p(\H_\E)$ (the $p$-th Schatten class) for $p>2$. Thus only the two
first terms of \eqref{dyson} are divergent. Moreover, the truncated series
(starting at $n=3$) is absolutely convergent for reasonable small
values of (the absolute value of) the coupling constant:
\begin{align*}
\Big|\sum_{n=3}^\infty(-)^{n+1}\frac{\lambda^n}n\,
\Tr\big((H^{-1}\,B_{\vf_1,\overline{\vf_2}}))^n\big)\Big|&\leq
\sum_{n=3}^\infty\|B_{\vf_1,\overline{\vf_2}})\|^n\, 
\frac{|\lambda|^n}{n}\,\|H^{-1}\|^{n-3}\,
\|H^{-3}\|_1\\
&\leq \sum_{n=3}^\infty 
\|B_{\vf_1,\overline{\vf_2}})\|^n\,\frac{|\lambda|^n}{n}.
\end{align*}
\end{proof}
\subsection{2-point sector}

At one loop we have the following contributions for the two-point
function: 
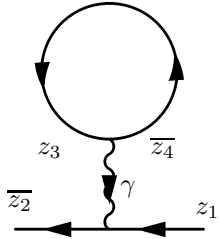
\begin{align}
\parbox{35mm}{\begin{picture}(25,30)
\put(0,3){\begin{fmfgraph}(25,30)
\fmfbottom{l,r}
\fmftop{t}
\fmf{fermion}{r,i1,l}
\fmffreeze
\fmf{photon,tension=3}{i1,i2}
\fmf{phantom_arrow,tension=0}{i2,i1}
\fmf{fermion,right}{i2,t}
\fmf{fermion,right}{t,i2}
\end{fmfgraph}}
\put(-1,6){\mbox{\small$\overline{z_2}$}}
\put(24,5){\mbox{\small$z_1$}}
\put(18,13){\mbox{\small$\overline{z_4}$}}
\put(3,13){\mbox{\small$z_3$}}
\put(14,8){\mbox{\small$\gamma$}}
\end{picture}} 
\Gamma^{(1)}_{2a}[z_1,\overline{z_2}] &= 
2 \int d\mu(z_3,\overline{z_3})\,d\mu(z_4,\overline{z_4}) 
 V(z_1,\overline{z_2},z_3,\overline{z_4})\,
H^{-1}_\eps(z_4,\overline{z_3})\;.
\end{align}
The graph of this type appears twice, once with
loop connecting $\overline{z_4}\to z_3$ (which is shown) and once 
with loop connecting $\overline{z_2}\to z_1$.
Replacing $\gamma \mapsto -\gamma$ shows that both contributions are 
identical.

The $z_4$-integration is directly eliminated using one reproducing
kernel of the vertex \eqref{vertex}, and we are left with
\begin{align}
\Gamma^{(1)}_{2a}[z_1,\overline{z_2}] 
&= \frac{\lambda}{2} \,I(z_1,\overline{z_2})
\int_\eps^\infty \!\!d\beta \sum_\ga
\int d\mu(z,\bar z)\,\mathrm{e}^{-\frac{\b \mu_0^2}{2}}
\mathrm{e}^{-\omega(|\gamma|^2-|z|^2e^{-\b\omega}+\bar{z}\ga
  e^{-\b\omega}-\bar\ga z+\bar\ga z_1-\bar{z}_2\ga)}\;.
\end{align}
Now the Gau\ss ian $z$-integral can be evaluated to give
\begin{align}
\Gamma^{(1)}_{2a}[z_1,\overline{z_2}] 
= \frac{\lambda}{2}\, I(z_1,\overline{z_2})
\int_\eps^\infty \!\!d\beta \sum_\ga
\,\frac{\mathrm{e}^{-\frac{\b \mu_0^2}{2}}
\mathrm{e}^{-\omega(\frac{|\ga|^2}{1-\mathrm{e}^{-\beta \omega}}
+\bar{\gamma}z_1 -\overline{z_2}\ga)}}{
(1-\mathrm{e}^{-\beta \omega})^2}\;.
\label{mass-ren-1}
\end{align}
These two operations, use of reproducing kernels and Gau\ss ian
integration for loops, can be tedious for more complicated
graphs. This is the reason why we provide the general rules for the parametric
representation of an arbitrary Feynman graph in subsection
\ref{parametric}.

In the last expression \eqref{mass-ren-1}, only the term $\ga=0$ is
divergent. This divergence is local and quadratic in the cut-off
($\Lambda=\epsilon^{-1/2}$),
yielding a mass counter-term:
\begin{align}
\Gamma^{(1)}_{2a,div}[z_1,\overline{z_2}] 
= \frac{\lambda}{2}\, I(z_1,\overline{z_2})
\int_\eps^\infty \!\!d\beta 
\,\frac{\mathrm{e}^{-\frac{\b \mu_0^2}{2}}}{
(1-\mathrm{e}^{-\beta \omega})^2}\;.
\label{mass-counterterm-1}
\end{align}
The (integrated) sum over $\gamma \neq 0$ is convergent as a
distribution on $\E_B(\C^2)$, in the limit $\eps\to0$. (Recall that as
a Fr\'echet space, $\E_B(\C^2)$ is defined as the image of
$\E_\SS(\R^2)=\SS(\R^2)$ under the Bargmann transform.) 
Indeed, we have
for any $\vf_1,\vf_2\in\E_B(\C^2)$
\begin{align}
&\Gamma^{(1)}_{2a,conv}[\overline{\varphi_1},\varphi_2] 
\nonumber
\\*
&\quad= \int
d\mu(z_1,\overline{z_1})\,d\mu(z_2,\overline{z_2})\,\overline{\vf_1}
(\overline{z_1})
\,\vf_2(z_2)\,
 I(z_1,\overline{z_2})\int_0^\infty \!\!d\beta \sum_{\ga\neq 0}
\,\frac{\mathrm{e}^{-\frac{\b \mu_0^2}{2}}
\mathrm{e}^{-\omega(\frac{|\ga|^2}{1-\mathrm{e}^{-\beta \omega}}
+\bar{\gamma}z_1 -\overline{z_2}\ga)}}{
(1-\mathrm{e}^{-\beta \omega})^2}
\nonumber
\\
&\quad= \sum_{\ga\neq 0}\langle\vf_1,\vf_2U_\ga\rangle_\B
\int_0^\infty \!\!d\beta 
\,\frac{\mathrm{e}^{-\frac{\b \mu_0^2}{2}}
\mathrm{e}^{-\omega\frac{|\ga|^2}{2}
\frac{1+\mathrm{e}^{-\beta \omega}}{1-\mathrm{e}^{-\beta \omega}}
}}{(1-\mathrm{e}^{-\beta \omega})^2}\;.
\end{align}
The $\beta$-integral is bounded from above since $|\gamma|^2 \geq
\frac{\theta^2}{2}$. Because
$\langle\vf_1,\vf_2U_\ga\rangle_\B$ is by construction a
Schwartz sequence, the sum is finite.  More precise estimations can be
obtained using
\begin{align}
\frac{1-e^{-x_0}}{x_0} x \leq 
1-e^{-x} \leq x \qquad \text{for} \qquad 0\leq x \leq x_0\;.
\label{main-inequality}
\end{align}

The second contribution to the one-loop two-point function is the graph
\begin{align}
\parbox{50mm}{\begin{picture}(45,20)
\put(0,2){\begin{fmfgraph}(45,20)
\fmfbottom{l,r}
\fmf{fermion,tension=2}{r,i1}
\fmf{fermion}{i1,i2}
\fmf{fermion,tension=2}{i2,l}
\fmffreeze
\fmf{photon,right}{i1,i2}
\fmffreeze
\fmf{phantom_arrow,left}{i2,i1}
\end{fmfgraph}}
\put(2,5){\mbox{\small$\overline{z_2}$}}
\put(40,5){\mbox{\small$z_1$}}
\put(28,4){\mbox{\small$\overline{z_4}$}}
\put(20,15){\mbox{\small$\gamma$}}
\put(14,4){\mbox{\small$z_3$}}
\end{picture}} 
\Gamma^{(1)}_{2b}[z_1,\overline{z_2}] &= 2
\int d\mu(z_3,\overline{z_3})\,d\mu(z_4,\overline{z_3}) 
 V(z_1,\overline{z_4},z_3,\overline{z_2})\,H^{-1}_\eps(z_4,\overline{z_3})\;.
\end{align}
There are again two graphs of this type which after a change of
variables give the same contribution. 
Using \eqref{vertex} and  \eqref{eq:propagatorbarg}, we find
\begin{align}
\Gamma^{(1)}_{2b}[z_1,\overline{z_2}] 
&= \frac{\lambda}{2}\,I(z_1,\overline{z_2}) 
\sum_\ga \int_\eps^\infty\!\!d\beta \;
\mathrm{e}^{-\frac{\beta \mu_0^2}{2}}\mathrm{e}^{ -\omega(
\bar{\gamma}(1-\mathrm{e}^{-\beta\omega}) \gamma
+\bar{\gamma}(1-\mathrm{e}^{-\beta\omega})z_1
+\overline{z_2}(1-\mathrm{e}^{-\beta\omega})\gamma 
+ \overline{z_2}(1-\mathrm{e}^{-\beta\omega})z_1)}\;.
\label{Gamma12b-1}
\end{align}

A small $\beta$ spoils the Gau\ss ian decay for large $|\gamma|$. 
This can be made transparent by a Poisson re-summation.
For that, the convention on the Fourier transform we use is
$\hat{f}(\xi)= \int e^{-2i\pi x\xi}f(x)dx$. The formula we obtain with
this convention is
\begin{equation}
\label{eq:Poisson}
\sum_\ga \mathrm{e}^{-\omega a|\ga|^2}\,\mathrm{e}^{-\omega(\bar b
  \ga+\bar\ga c)}=(a\th)^{-2}\sum_{\ga^*}\mathrm{e}^{\frac{\omega}{a}(\bar
  b+i\bar\ga^*)(c+i\ga^*)}, \sepword{for all} a\in\R^+_*,\,\,b,c\in\C^2,
\end{equation}
with the frequency $\omega=\frac{2\pi}{\th}$ and with
$\ga=\frac{\th}{\sqrt{2}}(n+im)\in\Ga\subset\C^2$ one has
$\ga^*=\frac{1}{\sqrt{2}}(n+im)\in\widehat\Ga\subset\C^2$. 
In order to simplify the notations, here we use the notation $\ga^*$
instead of $\widehat\ga$ to denote an element of the dual lattice.\\ 
Using this formula, we can continue the computation of 
(\ref{Gamma12b-1}):
\begin{align}
\Gamma^{(1)}_{2b}[z_1,\overline{z_2}] 
&=\frac{\lambda}{2\th^2}\,I(z_1,\overline{z_2}) 
\sum_{\gamma^*} \int_\eps^\infty\!\!d\beta \;
\frac{\mathrm{e}^{-\frac{\beta \mu_0^2}{2}}\mathrm {e}^{-\omega( 
\frac{|\ga^*|^2}{(1-\mathrm{e}^{-\beta\omega})}
-\mathrm{i}\overline{z_2}\gamma^* 
-\mathrm{i}\overline{\gamma^*}z_1)}}{
(1-\mathrm{e}^{-\beta\omega})^2}\;.
\label{mass-ren-2}
\end{align}
Thus, the diagram (\ref{mass-ren-2}) is, up to a normalization factor
$\theta^{-2}$ and the exchange $\ga\leftrightarrow i\ga^*$, the same
as \eqref{mass-ren-1}: only the term with $\ga^*=0$ is divergent, with
a local quadratic divergence.

We have proven the following
\begin{prop}
For any $\th\in\R$, the divergent part of the two-point 1PI Green
function associated with the action \eqref{actiontorus}, 
in its one-loop approximation,  is given by
\begin{equation*}
\Ga^{(1)}_{2,div}[\overline{\vf_1},\vf_2]=\frac{\lambda}{2\omega}
\,\Big(1+\frac{1}{\th^2}\Big)\,\Tr_{\A_\th}\big[(\vf_1,\vf_2)_{\A_\th}\big]
\,\Big(\frac{1}{\epsilon \omega} 
+\Big(1-\frac{\mu_0^2}{2\omega}\Big) \ln\frac 1 \epsilon \Big).
\end{equation*}
\end{prop}

\subsection{4-point sector: Planar graphs}

Next, we look at four-leg graphs. The first possibility is 
\begin{align}
\parbox{40mm}{\begin{picture}(35,17)
\put(0,-10){\begin{fmfgraph}(35,25)
\fmfleft{l1,l2}
\fmfright{r1,r2}
\fmf{fermion}{r1,i1,l1}
\fmf{fermion}{l2,i2,r2}
\fmffreeze
\fmf{photon,tension=3}{i1,i3}
\fmf{phantom_arrow,tension=0}{i3,i1}
\fmf{photon,tension=3}{i2,i4}
\fmf{phantom_arrow,tension=0}{i2,i4}
\fmf{fermion,right}{i3,i4}
\fmf{fermion,right}{i4,i3}
\end{fmfgraph}}
\put(4,-13){\mbox{\small$\overline{z_2}$}}
\put(28,-13){\mbox{\small$z_1$}}
\put(10,-3){\mbox{\small$z_1'$}}
\put(22,-3){\mbox{\small$\overline{z_2}'$}}
\put(21,7){\mbox{\small$z_3'$}}
\put(10,7){\mbox{\small$\overline{z_4}'$}}
\put(28,17){\mbox{\small$\overline{z_4}$}}
\put(4,17){\mbox{\small$z_3$}}
\put(18.5,-6){\mbox{\small$\gamma_1$}}
\put(18.5,11){\mbox{\small$\gamma_2$}}
\put(7,3){\mbox{\small$\beta_1$}}
\put(25,3){\mbox{\small$\beta_2$}}
\end{picture}}
&\Gamma^{(1)}_{4a}[z_1,\overline{z_2},z_3,\overline{z_4}]
\nonumber
\\*[-4ex] 
&=- 2 \int d\mu(z_1',\overline{z_1}')\,
 d\mu(z_2',\overline{z_2}')\, d\mu(z_3',\overline{z_3}')\, 
d\mu(z_4',\overline{z_4}')\,\nonumber
\\* 
& \quad \times 
V(z_1,\overline{z_2},z_1',\overline{z_2}')\,
V(z_3',\overline{z_4}',z_3,\overline{z_4})\,
H^{-1}_\eps(z_2',\overline{z_3}')\,
H^{-1}_\eps(z_4',\overline{z_1}')\;.
\label{G4a-graph}
\end{align}

\quad

This graph appears four times in the expansion, and because it
contains two vertices there is a combinatorial factor $\frac{1}{2!}$.
The minus sign is from the Legendre transformation of 1PI functions.
Three of the four integrations (e.g.\ over $z_2',z_3',z_4'$) can be
trivially done using the reproducing kernels in the vertex. Thus we
have
\begin{align}
\Gamma^{(1)}_{4a}[z_1,\overline{z_2},z_3,\overline{z_4}] 
&= -\frac{\lambda^2}{8} \int_\eps^\infty \!\!d\beta_1d\beta_2
\sum_{\gamma_1,\gamma_2}
\mathrm{e}^{-\frac{\mu_0^2}{2}(\beta_1+\beta_2)}\mathrm{e}^{-\omega(
|\gamma_1|^2+|\gamma_2|^2
-\overline{z_2}\gamma_1-\overline{z_2}z_1+\overline{\gamma_1}z_1
+\overline{z_4}\gamma_2-\overline{z_4}z_3-\overline{\gamma_2}z_3)}
\nonumber
\\
&\quad\times 
\int d\mu(z,\overline z) \;
\mathrm{e}^{-\omega(-\bar{z} \mathrm{e}^{-\beta_1 \omega} \gamma_2
+\bar{z} \mathrm{e}^{-(\beta_1+\beta_2) \omega}\gamma_1
-\overline{\gamma_2} \mathrm{e}^{-\beta_2 \omega} \gamma_1
-\bar{z}e^{-(\beta_1+\beta_2) \omega}z+
\overline{\gamma_2} \mathrm{e}^{-\beta_2 \omega}z -\overline{\gamma_1}z)}.
\nonumber
\end{align}

We  integrate over $z$, put $\gamma_2=\gamma_1+\gamma$, and obtain after
some reorganizations
\begin{align}
&\Gamma^{(1)}_{4a}[z_1,\overline{z_2},z_3,\overline{z_4}] 
\nonumber
\\*
&= -\frac{\lambda^2}{8}\sum_{\ga_1,\ga}\mathrm{e}^{-\omega|\ga_1|^2}\,
I(\ga_1,\overline{z_2}-\overline{z_4})\,
I(z_1,\overline{z_2}-\overline{\ga_1})\,I(z_3,\overline{z_4}+\overline{\ga_1})
\int_\eps^\infty \frac{d\beta_1d\beta_2\;
\mathrm{e}^{-\frac{\mu_0^2}{2}(\beta_1+\beta_2)}}{
(1-\mathrm{e}^{-(\beta_1+\beta_2)\omega})^2}
\nonumber
\\*
&\quad\times 
\mathrm{e}^{-\omega(\overline{z_4}\gamma-\overline{\gamma}z_3
+\frac{|\ga|^2}{
1-\mathrm{e}^{-(\beta_1+\beta_2)\omega}}
+\overline{\gamma_1}\frac{1-\mathrm{e}^{-\beta_1\omega}}{
1-\mathrm{e}^{-(\beta_1+\beta_2)\omega}}\gamma
+\overline{\gamma}
\frac{1-\mathrm{e}^{-\beta_2\omega}}{
1-\mathrm{e}^{-(\beta_1+\beta_2)\omega}}\gamma_1
+|\ga_1|^2\frac{(1-\mathrm{e}^{-\beta_1\omega})
(1-\mathrm{e}^{-\beta_2\omega})}{
1-\mathrm{e}^{-(\beta_1+\beta_2)\omega}})}\;.
\label{Gamma-4a}
\end{align}
For $\ga=0$ one obtains
\begin{align}
\Gamma^{(1)}_{4a,0}[z_1,\overline{z_2},z_3,\overline{z_4}] 
& = -\frac{\lambda}{4}\,V(z_1,\overline{z_2},z_3,\overline{z_4})
\int_\eps^\infty \frac{d\beta_1d\beta_2\;
\mathrm{e}^{-\frac{\mu_0^2}{2}(\beta_1+\beta_2)}}{
(1-\mathrm{e}^{-(\beta_1+\beta_2)\omega})^2}
\label{Gamma4a-1}
\\
&+\frac{\lambda^2}{8}\sum_{\ga_1}\mathrm{e}^{-\omega|\ga_1|^2}\,
I(\ga_1,\overline{z_2}-\overline{z_4})\,I(z_1,\overline{z_2}-\overline{\ga_1})
\,I(z_3,\overline{z_4}+\overline{\ga_1})
\nonumber
\\*
&\times 
\int_\eps^\infty \frac{d\beta_1d\beta_2\;
\mathrm{e}^{-\frac{\mu_0^2}{2}(\beta_1+\beta_2)}}{
(1-\mathrm{e}^{-(\beta_1+\beta_2)\omega})^2}
\Big(1-\mathrm{e}^{-\omega|\ga_1|^2\frac{(1-\mathrm{e}^{-\beta_1\omega})
(1-\mathrm{e}^{-\beta_2\omega})}{
1-\mathrm{e}^{-(\beta_1+\beta_2)\omega}}}\Big)\;.
\label{Gamma4a-2}
\end{align}
The line (\ref{Gamma4a-1}) gives the logarithmically divergent
renormalization of the coupling constant. It remains to show that the
remaining parts are convergent. In (\ref{Gamma4a-2}), if we add the
external fields $\vf_1,\vf_2,\vf_3,\vf_4\in\E_B(\C^2)$ 
and if use the inequality (\ref{main-inequality}), we obtain:
\begin{align*}
&\Big|\sum_{\ga_1}\langle\vf_1,\vf_2U_{\ga_1}\rangle_B\,
\langle\vf_3,\vf_4U_{-\ga_1}\rangle_B\,
\int_0^\infty \frac{d\beta_1d\beta_2\mathrm{e}^
{-\frac{\mu_0^2}{2}(\beta_1+\beta_2)}}{
(1-\mathrm{e}^{-(\beta_1+\beta_2)\omega})^2}
\Big(1-\mathrm{e}^{-\omega|\ga_1|^2\frac{(1-\mathrm{e}^{-\beta_1\omega})
(1-\mathrm{e}^{-\beta_2\omega})}{
1-\mathrm{e}^{-(\beta_1+\beta_2)\omega}}}\Big)\Big|\nonumber
\\*
&\leq \omega\sum_{\ga_1}\big|\langle\vf_1,\vf_2U_{\ga_1}
\rangle_B\big|\,
\big|\langle\vf_3,\vf_4U_{-\ga_1}\rangle_B|\,|\gamma_1|^2
\int_0^\infty d\beta_1\,d\beta_2\;
\mathrm{e}^{-\frac{\mu_0^2}{2}(\beta_1+\beta_2)}
\frac{(1-\mathrm{e}^{-\beta_1\omega})(1-\mathrm{e}^{-\beta_2\omega})
}{(1-\mathrm{e}^{-(\beta_1+\beta_2)\omega})^3}\;,
\end{align*}
which is finite since $\langle\vf_i,\vf_jU_{\ga_1}\rangle_B$ 
is a Schwartz sequence and because the $\b$-integral is now harmless.

It remains to estimate in (\ref{Gamma-4a}) the remaining sum over
$\gamma \neq 0$. Similar manipulations show that
\begin{align*}
&\Gamma^{(1)}_{4a,\neq 0}[\overline{\varphi_1},\varphi_2,
\overline{\vf_3},\vf_4] 
= -\frac{\lambda^2}{8}\sum_{\ga_1,\ga\neq 0}\langle\vf_1,\vf_2
U_{-\ga_1}\rangle_B \,\langle\vf_3,\vf_4
U_{\ga_1+\ga}\rangle_B \,
\int_0^\infty \frac{d\beta_1d\beta_2\;
\mathrm{e}^{-\frac{\mu_0^2}{2}(\beta_1+\beta_2)}}{
(1-\mathrm{e}^{-(\beta_1+\beta_2)\omega})^2}
\\
&\qquad\times 
\mathrm{e}^{-\frac{\omega}{2}(|\ga|^2\frac{1+\mathrm{e}
^{-(\beta_1+\beta_2)\omega}}{
1-\mathrm{e}^{-(\beta_1+\beta_2)\omega}}
+\overline{\gamma_1}\ga\frac{1-2\mathrm{e}^{-\beta_1\omega}
+\mathrm{e}^{-(\beta_1+\beta_2)\omega}}{
1-\mathrm{e}^{-(\beta_1+\beta_2)\omega}}
+\overline{\gamma}\gamma_1
\frac{1-2\mathrm{e}^{-\beta_2\omega}
+\mathrm{e}^{-(\beta_1+\beta_2)\omega}}{
1-\mathrm{e}^{-(\beta_1+\beta_2)\omega}}
+2|\ga_1|^2\frac{(1-\mathrm{e}^{-\beta_1\omega})
(1-\mathrm{e}^{-\beta_2\omega})}{
1-\mathrm{e}^{-(\beta_1+\beta_2)\omega}})}\;.
\end{align*}
Thus,
\begin{align*}
&\Big|\Gamma^{(1)}_{4a,\neq 0}[\overline{\varphi_1},\varphi_2,
\overline{\vf_3},\vf_4]\Big| 
\leq\frac{\lambda^2}{8}\sum_{\ga_1,\ga\ne0}\big|\langle\vf_1,\vf_2
U_{-\ga_1}\rangle_B \,\langle\vf_3,\vf_4
U_{\ga_1+\ga}\rangle_B\big| \,
\int_0^\infty \frac{d\beta_1d\beta_2\;
\mathrm{e}^{-\frac{\mu_0^2}{2}(\beta_1+\beta_2)}}{
(1-\mathrm{e}^{-(\beta_1+\beta_2)\omega})^2}
\nonumber
\\*
&\hspace{6cm}\times 
\mathrm{e}^{-\frac{\omega}{2}(|\ga|^2
\frac{1+\mathrm{e}^{-(\beta_1+\beta_2)\omega}}{
1-\mathrm{e}^{-(\beta_1+\beta_2)\omega}}
+(\overline{\ga_1}\ga+\overline{\ga}\ga_1+
2|\ga_1|^2)\frac{(1-\mathrm{e}^{-\beta_1\omega})
(1-\mathrm{e}^{-\beta_2\omega})}{
1-\mathrm{e}^{-(\beta_1+\beta_2)\omega}})}
\nonumber
\\*
&\quad\leq\frac{\lambda^2}{8}\sum_{\ga_1,\ga\ne0}\big|\langle\vf_1,\vf_2
U_{-\ga_1}\rangle_B \,\langle\vf_3,\vf_4
U_{\ga_1+\ga}\rangle_B\big| \,
\int_0^\infty \frac{d\beta_1d\beta_2\;
\mathrm{e}^{-\frac{\mu_0^2}{2}(\beta_1+\beta_2)}}{
(1-\mathrm{e}^{-(\beta_1+\beta_2)\omega})^2}
\mathrm{e}^{-\frac{\omega\theta^2}{4}\frac{\mathrm{e}^{-\beta_1\omega}
+\mathrm{e}^{-\beta_2\omega}}{
1-\mathrm{e}^{-(\beta_1+\beta_2)\omega}}}\;,
\end{align*}
which is finite.

The next graph is
\begin{align}
\parbox{48mm}{\begin{picture}(45,10)
\put(0,-15){\begin{fmfgraph}(45,20)
\fmfleft{l1,l2}
\fmfright{r1,r2}
\fmf{fermion,tension=2}{r1,i1}
\fmf{fermion,tension=1}{i1,i3}
\fmf{fermion,tension=2}{i3,l1}
\fmf{fermion,tension=2}{l2,i4}
\fmf{fermion,tension=1}{i4,i2}
\fmf{fermion,tension=2}{i2,r2}
\fmffreeze
\fmf{photon}{i1,i2}
\fmf{phantom_arrow,tension=0}{i2,i1}
\fmf{photon}{i3,i4}
\fmf{phantom_arrow,tension=0}{i3,i4}
\end{fmfgraph}}
\put(4,-18.5){\mbox{\small$\overline{z_2}$}}
\put(38,-18.5){\mbox{\small$z_1$}}
\put(27,-19){\mbox{\small$\overline{z_2'}$}}
\put(15,-18.5){\mbox{\small$z_1'$}}
\put(38,7){\mbox{\small$\overline{z_4}$}}
\put(4,7){\mbox{\small$z_3$}}
\put(15,7){\mbox{\small$\overline{z_4'}$}}
\put(27,7){\mbox{\small$z_3'$}}
\put(18.5,-13){\mbox{\small$\beta_1$}}
\put(18.5,1){\mbox{\small$\beta_2$}}
\put(6,-6){\mbox{\small$\gamma_2$}}
\put(35,-6){\mbox{\small$\gamma_1$}}
\end{picture}}  
&\Gamma^{(1)}_{4b}[z_1,\overline{z_2},z_3,\overline{z_4}] 
\nonumber
\\
&=-2 \int d\mu(z_1',\overline{z_1}')\,
 d\mu(z_2',\overline{z_2}')\, d\mu(z_3',\overline{z_3}')
\, d\mu(z_4',\overline{z_4}')\,
\nonumber\\ 
&
\times V(z_1,\overline{z_2}',z_3',\overline{z_4})
\,V(z_1',\overline{z_2},z_3,\overline{z_4}')
\,H^{-1}_\eps(z_2',\overline{z_1}')\,
H^{-1}_\eps(z_4',\overline{z_2}')\;.
\label{Gamma4b}
\end{align}

\quad

This graph appears with the same symmetry factor as the first one. 
All $z'$-integrations can be trivially done using the reproducing kernels
in the vertices, giving
\begin{align*}
&\Gamma_{4b}^{(1)}[z_1,\overline{z_2},z_3,\overline{z_4}] 
= -\frac{\lambda^2}{8} \sum_{\gamma_1,\gamma_2} 
   \int_\eps^\infty d\beta_1 d\beta_2\;
 \mathrm{e}^{-\frac{\mu_0^2}{2}(\beta_1+\beta_2)}
\mathrm{e}^{ -\omega( |\gamma_2|^2
+|\gamma_1|^2
+\overline{z_2} \gamma_2 
+\overline{\gamma_1} z_1 
+\overline{z_4}\gamma_1
+\overline{\gamma_2}z_3 
) }
\\*
&\qquad\times \mathrm{e}^{\omega(
\overline{z_2} \mathrm{e}^{-\beta_1 \omega} \gamma_1
+ \overline{\gamma_2}\mathrm{e}^{-\beta_1 \omega} \gamma_1
+ \overline{z_2}\mathrm{e}^{-\beta_1 \omega} z_1 
+ \overline{\gamma_2} \mathrm{e}^{-\beta_1 \omega} z_1 
+ \overline{z_4} \mathrm{e}^{-\beta_2\omega} \gamma_2
+ \overline{\gamma_1} \mathrm{e}^{-\beta_2\omega} \gamma_2
+ \overline{z_4} \mathrm{e}^{-\beta_2\omega} z_3
+ \overline{\gamma_1} \mathrm{e}^{-\beta_2\omega} z_3) }.
\end{align*}

There are two ways to proceed. We could set $\gamma_1=\gamma_2+\gamma$
and perform a Poisson re-summation \eqref{eq:Poisson} in the loop
variable $\gamma_2$. The second possibility, which underlines the
duality with the previous graph, consists in a Poisson re-summation
in both $\ga_1$ and $\ga_2$:
\begin{align*}
\Gamma_{4b}^{(1)}[z_1,\overline{z_2},z_3,\overline{z_4}] 
&= -\frac{\lambda^2}{8\th^4}\, \mathrm{e}^{\omega(\overline{z_4}z_1+
\overline{z_2}z_3)}
\sum_{\ga_1^*,\ga_2^*}\int_\eps^\infty
d\b_1\,d\b_2\,
\frac{\mathrm{e}^{-\frac{\mu_0^2}{2}(\b_1+\b_2)\omega}}{(1-\mathrm{e}^
{-(\b_1+\b_2)\omega})^{2}}
\mathrm{e}^{-\omega\frac{|\ga_1^*|^2+|\ga_2^*|^2}
{1-\mathrm{e}^{-(\b_1+\b_2)\omega}}}\\
&\qquad\x\mathrm{e}^{\omega(
\mathrm{i}\overline{\ga_1^*}z_1+
\overline{z_4}\mathrm{i}\ga_1^*
+\overline{z_2}\mathrm{i}\ga_2^*+\mathrm{i}\overline{\ga_2^*}z_3+
\frac{\mathrm{i}\overline{\ga_1^*}\mathrm{e}^{-\b_2\omega}
\mathrm{i}\ga_2^*+\mathrm{i}\overline{\ga_2^*}\mathrm{e}^{-\b_1\omega}
\mathrm{i}\ga_1^*}{1-\mathrm{e}^{-(\b_1+\b_2)\omega}})}.
\end{align*}

If we plug now $\ga_2^*=-\ga_1^*+\ga^*$, we see that this graph leads to the
same expression as the previous one \eqref{Gamma-4a}, up to a
constant factor and the exchange
$\ga_1\to \mathrm{i}\ga_1^*$, $\ga\to-\mathrm{i}\ga^*$ and 
$z_2 \leftrightarrow z_4$: 
\begin{align}
&\Gamma^{(1)}_{4b}[z_1,\overline{z_2},z_3,\overline{z_4}] 
\nonumber
\\*
&= -\frac{\lambda^2}{8\th^4}\sum_{\ga_1^*,\ga^*}
\mathrm{e}^{-\omega|\ga_1^*|^2}\,
I(\mathrm{i}\ga_1^*,\overline{z_4}-\overline{z_2})\,
I(z_1,\overline{z_4}-\overline{\mathrm{i}\ga_1^*})\,
I(z_3,\overline{z_2}+\overline{\mathrm{i}\ga_1^*})
\int_\eps^\infty \frac{d\beta_1d\beta_2\;
\mathrm{e}^{-\frac{\mu_0^2}{2}(\beta_1+\beta_2)}}{
(1-\mathrm{e}^{-(\beta_1+\beta_2)\omega})^2}
\nonumber
\\*
&\times 
\mathrm{e}^{-\omega(-\overline{z_2}\mathrm{i}\gamma^*
-\mathrm{i}\overline{\gamma^*}z_3
+\frac{|\ga^*|^2}{
1-\mathrm{e}^{-(\beta_1+\beta_2)\omega}}
+\mathrm{i}\overline{\gamma_1^*}\frac{1-\mathrm{e}^{-\beta_2\omega}}{
1-\mathrm{e}^{-(\beta_1+\beta_2)\omega}}\mathrm{i}\gamma^*
+\mathrm{i}\overline{\gamma^*}
\frac{1-\mathrm{e}^{-\beta_1\omega}}{
1-\mathrm{e}^{-(\beta_1+\beta_2)\omega}}\mathrm{i}\gamma_1^*
+|\ga_1^*|^2\frac{(1-\mathrm{e}^{-\beta_1\omega})
(1-\mathrm{e}^{-\beta_2\omega})}{
1-\mathrm{e}^{-(\beta_1+\beta_2)\omega}})}\;.
\label{Gamma-4b}
\end{align}

The divergent part is thus given by 
\begin{align}
&\Gamma^{(1)}_{4b,div}[z_1,\overline{z_2},z_3,\overline{z_4}] 
\nonumber
\\*
&= -\frac{\lambda^2}{8\th^4}\sum_{\ga_1^*}
\mathrm{e}^{-\omega|\ga_1^*|^2}\,
I(\mathrm{i}\ga_1^*,\overline{z_4}-\overline{z_2})\,
I(z_1,\overline{z_4}-\overline{\mathrm{i}\ga_1^*})\,
I(z_3,\overline{z_2}+\overline{\mathrm{i}\ga_1^*})
\int_\eps^\infty \frac{d\beta_1d\beta_2\;
\mathrm{e}^{-\frac{\mu_0^2}{2}(\beta_1+\beta_2)}}{
(1-\mathrm{e}^{-(\beta_1+\beta_2)\omega})^2}\;,
\label{Gamma-4bdiv}
\end{align}
which gives after a (final) Poisson re-summation \eqref{eq:Poisson}
another (local, logarithmically divergent) contribution for the
coupling constant renormalization: 
\begin{align}
&\Gamma^{(1)}_{4b,div}[z_1,\overline{z_2},z_3,\overline{z_4}] 
= -\frac{\lambda}{4\th^2}\,V(z_1,\overline{z_2},z_3,\overline{z_4})\,
\int_\eps^\infty \frac{d\beta_1d\beta_2\;
\mathrm{e}^{-\frac{\mu_0^2}{2}(\beta_1+\beta_2)}}{
(1-\mathrm{e}^{-(\beta_1+\beta_2)\omega})^2}\;.
\label{Gamma-4bdiv2}
\end{align}

We can summarize this estimations in the following
\begin{prop}
For any $\th\in\R$, the divergent part of the four-point planar 1PI
Green function associated with 
the action \eqref{actiontorus}, in its one-loop approximation, is given by
\begin{equation*}
\Ga^{(1)}_{4,planar,div}[\overline{\vf_1},\vf_2,\overline{\vf_3},\vf_3]
=-\frac{\lambda^2}{8\omega^2}
\,\Big(1+\frac{1}{\th^2}\Big)\,\Tr_{\A_\th}\big[(\vf_1,\vf_2)_{\A_\th}
\,(\vf_3,\vf_4)_{\A_\th}\big]
\,\ln\frac 1 \epsilon .
\end{equation*}
\end{prop}

\subsection{4-point sector: Non-planar graphs}

Up to now, we have found only local divergences which can be absorbed
in a mass and coupling constant renormalization. This is not the end
of the story. We will see that the next two four-point one-loop graphs
develop non-local divergences, in the sense that the counter-term
needed is not present in the classical action \eqref{actiontorus}.
Indeed, the next graphs are topologically non-planar and their
singularities are of the same kind as the singularity of the
non-planar sector of a scalar field theory on the noncommutative torus
or even on certain isospectral deformations \cite{UVIR}. This is the
discrete version of the so-called the UV/IR-entanglement phenomenon:
these new singularities are restricted to the zero-mode $\gamma=0$ of
the vertex. They should not appear in models where the zero-mode can
be consistently removed such as for $U(1)$-Yang-Mills theory on the
noncommutative torus \cite{Krajewski:1999ja}.

As we will see, the behavior of these non-planar graphs for the
Heisenberg module is highly sensitive to the number-theoretical aspect
of the deformation parameter $\th$. There are two different cases to
consider. When $\th$ is rational, we know that the model is a
commutative one and the singularities are local. For irrational $\th$,
in order to control the sums for non-vanishing $\ga$, we need to
impose a Diophantine type condition. We first give definitions.

\begin{defn}
\label{Dioph}
An irrational number $\th$ is said to satisfy a 
{\it Diophantine condition} if for
all $n\in\Z\setminus\{0\}$ there exist two constants $C,\delta>0$
such that
\begin{equation}
\label{dioph}
\|n\th\|_\T=\inf_{m\in\Z} |n\th-m|\geq C|n|^{-(1+\delta)}\;.
\end{equation}
We have the possibility to weaken this condition:  
An irrational number $\th$ is said to satisfy a {\it weak Diophantine 
condition} if for
all $n\in\Z\setminus\{0\}$ there exist two constants $C_1,C_2>0$
such that
\begin{equation}
\|n\th\|_\T=\inf_{m\in\Z} |n\th-m|\geq C_1\mathrm{e}^{ -C_2|n|}\;.
\label{weakdioph}
\end{equation}
\end{defn}
Those definitions extends for tuples of irrational number
$(\th_1,\cdots,\th_d)$, where the exponent $1+\delta$ in \eqref{dioph}
has to be replaced by $d+\delta$. 

A Diophantine condition is a way to measure ``how far from rationals''
is an irrational number. More precisely, it means that the convergence
of the continuous fractional expansion is quite slow.  Note that the
(ordinary) Diophantine condition is equivalent to the requirement that
the inverse torus-norm of $n\th$ is a tempered distribution on
$\SS(\Z\setminus\{0\})$. This is a natural requirement because of the
distributional nature of QFT and since the basic formulation of our
model relies on Schwartz sequences. Note also that irrational numbers
satisfying a Diophantine condition are not exceptional since they are
of full Lebesgue measure.  However, since at one-loop the non-planar
sector is only logarithmically divergent, this condition can be
weakened: We are able to control the convergence even under the
weakest version of the Diophantine condition \eqref{weakdioph}.

\quad

The first non-planar graph to compute is given by
\begin{align}
\parbox{62mm}{\begin{picture}(50,10)
\put(0,-20){\begin{fmfgraph}(50,25)
\fmfleft{l1,l2}
\fmfright{r1,r2}
\fmf{fermion,tension=2}{l2,i1}
\fmf{fermion,tension=1}{i1,i2}
\fmf{fermion,tension=1}{i2,i3}
\fmf{fermion,tension=2}{i3,r2}
\fmf{fermion,tension=1}{r1,i4,l1}
\fmffreeze
\fmf{photon}{i2,i6,i4}
\fmf{phantom_arrow}{i6,i4}
\fmffreeze
\fmf{phantom}{l2,a1,a2,i6}
\fmffreeze
\fmf{photon,left=0.3}{a2,i1}
\fmf{phantom_arrow,left=0.3}{a2,i1}
\fmf{photon,right=0.6,rubout=5}{a2,i3}
\end{fmfgraph}}
\put(8,-18){\mbox{\small$\overline{z_2}$}}
\put(38,-18){\mbox{\small$z_1$}}
\put(3,7){\mbox{\small$z_3$}}
\put(11,7){\mbox{\small$\overline{z_4'}$}}
\put(20,7){\mbox{\small$z_1'$}}
\put(26,7){\mbox{\small$\overline{z_2'}$}}
\put(36,7){\mbox{\small$z_3'$}}
\put(43,7){\mbox{\small$\overline{z_4}$}}
\put(30,1){\mbox{\small$\beta_1$}}
\put(17,1){\mbox{\small$\beta_2$}}
\put(11,-7){\mbox{\small$\gamma_2$}}
\put(20,-14){\mbox{\small$\gamma_1$}}
\end{picture}}  
&\Gamma^{(1)}_{4c}[z_1,\overline{z_2},z_3,\overline{z_4}] 
\nonumber
\\*
&=
-4\int d\mu(z_1',\overline{z_1}')\,
 d\mu(z_2',\overline{z_2}')\, d\mu(z_3',\overline{z_3}')\, 
d\mu(z_4',\overline{z_4}')\,\nonumber\\ 
&V(z_1,\overline{z_2},z_1',\overline{z_2}')\,
V(z_3,\overline{z_4}',z_3',\overline{z_4})
\,H^{-1}_\eps(z_2',\overline{z_3}')\,
H^{-1}_\eps(z_4',\overline{z_1}')\;.
\label{G4c-graph}
\end{align}
This graph appears eight times in the expansion, leading to twice 
the symmetry factor of the previous graphs. We have 
\begin{align}
  &\Gamma^{(1)}_{4c}[z_1,\overline{z_2},z_3,\overline{z_4}] 
= -\frac{\lambda^2}{4} \sum_{\gamma_1,\gamma_2} 
\int_\eps^\infty d\beta_1d\beta_2\;
\mathrm{e}^{-\frac{\mu_0^2}{2}(\beta_1+\beta_2) }
\mathrm{e}^{-\omega(
|\gamma_1|^2 +|\gamma_2|^2-\overline{z_2}\ga_1-(\overline{z_2}
-\overline{\ga_1})z_1
+\overline{z_4} \gamma_2) }
\nonumber
\\*
& \qquad\times \mathrm{e}^{\omega(-(\overline{z_4}
+\overline{\gamma_2})\mathrm{e}^{-\beta_1\omega}
\gamma_1
+ ((\overline{z_4}+\overline{\gamma_2})\mathrm{e}^{-(\beta_1+\beta_2)\omega}
+\overline{\gamma_1}\mathrm{e}^{-\beta_2 \omega}) \gamma_2 
+((\overline{z_4}+\overline{\gamma_2})\mathrm{e}^{-(\beta_1+\beta_2)\omega}
+\overline{\gamma_1}\mathrm{e}^{-\beta_2 \omega}-\overline{\gamma_2}) z_3)}
\;.
\end{align}
We perform a Poisson re-summation in the loop wavy line index (in $\gamma_2$):
\begin{align}
\label{eq:NP2}
 &\Gamma^{(1)}_{4c}[z_1,\overline{z_2},z_3,\overline{z_4}] 
= -\frac{\lambda^2}{4\th^2} \sum_{\gamma_1,\gamma_2^*} 
\mathrm{e}^{\omega(\overline{z_2}z_1+\overline{z_4}z_3+
\mathrm{i}\overline{\ga^*_2}z_3
+\overline{z_4}i\ga_2^*-\overline{\ga_1}z_1+\overline{z_2}\ga_1)}
\int_\eps^\infty d\beta_1d\beta_2
\frac{\mathrm{e}^{-\frac{\mu_0^2}{2}(\beta_1+\beta_2)}}{
(1-\mathrm{e}^{-(\b_1+\b_2)\omega})^2}\nonumber
\\*
&\hspace{6cm}
\times \mathrm{e}^{-\frac{\omega}{1-\mathrm{e}^{-(\b_1+\b_2)\omega}}(
|\ga_1|^2+|\ga_2^*|^2+\overline{\ga_1}i\ga_2^*\mathrm{e}^{-\b_2\omega}
-\mathrm{i}\overline{\ga_2^*}\ga_1\mathrm{e}^{-\b_1\omega})}\;.
\end{align}

In order to see what happens in the different arithmetical cases,
it is perhaps more enlightening to add external field
$\vf_1,\vf_2,\vf_3,\vf_4\in\E_B$. After some rearrangements, we obtain
\begin{align}
\label{eq:NP2a}
\Gamma^{(1)}_{4c}[\overline{\vf_1},\vf_2,\overline{\vf_3},\vf_4] 
&= -\frac{\lambda^2}{4\th^2}\sum_{\ga_1,\ga_2^*}
\langle\vf_1,\vf_2U_{\ga_1}\rangle_B\,
\langle\vf_3,\vf_4U_{\mathrm{i}\ga_2^*}\rangle_B\,
\int_\eps^\infty d\beta_1d\beta_2
\frac{\mathrm{e}^{-\frac{\mu_0^2}{2}(\beta_1+\beta_2)}}{
(1-\mathrm{e}^{-(\b_1+\b_2)\omega})^2}\nonumber\\
&\quad \times 
\mathrm{e}^{-\frac{\omega}{2}
\frac{1+\mathrm{e}^{-(\b_1+\b_2)\omega}}{1-\mathrm{e}^{-(\b_1+\b_2)\omega}}(
|\ga_1|^2+|\ga_2^*|^2)}\,
\mathrm{e}^{-\frac{\omega}{1-\mathrm{e}^{-(\b_1+\b_2)\omega}}(
\overline{\ga_1}\mathrm{i}\ga_2^*\mathrm{e}^{-\b_2\omega}
-\mathrm{i}\overline{\ga_2^*}\ga_1\mathrm{e}^{-\b_1\omega})}\;.
\end{align}

Now, the number-theoretical aspect of the deformation parameter $\theta$
comes into the play. Since 
$\gamma^*=\frac{1}{\sqrt{2}}(m+\mathrm{i}n)$, for $\theta\in\N$ 
(commutative case) we can set $\gamma_1=-\mathrm{i}\gamma_2^*$ in
\eqref{eq:NP2} (the sum $\gamma_1\neq -\mathrm{i}\ga_2^*$ gives a finite
contribution as it can be shown by similar estimates as those used
along this section) and obtain
\begin{equation*}
-\frac{\lambda^2}{4\th^2} \sum_{\ga_1}
\langle\vf_1,\vf_2U_{\ga_1}\rangle_B\,
\langle\vf_3,\vf_4U_{-\ga_1}\rangle_B\,
\int_\eps^\infty d\beta_1d\beta_2
\frac{\mathrm{e}^{-\frac{\mu_0^2}{2}(\beta_1+\beta_2) }}{
(1-\mathrm{e}^{-(\b_1+\b_2)\omega})^2}
\mathrm{e}^{-\omega|\ga_1|^2\frac{(1-\mathrm{e}^{-\b_1\omega})
(1-\mathrm{e}^{-\b_2\omega})}{1-\mathrm{e}^{-(\b_1+\b_2)\omega}}}\;.
\end{equation*}
Thus, using the routine inequalities, we can extract
the divergent part of
$\Gamma^{(1)}_{4d}[z_1,\overline{z_2},z_3,\overline{z_4}]$:
\begin{equation}
\Gamma^{(1)}_{4c,div}[z_1,\overline{z_2},z_3,\overline{z_4}]=
-\frac{\lambda}{2\th^2}V(z_1,\overline{z_2},z_3,\overline{z_4}) 
\int_\eps^\infty d\beta_1d\beta_2
\frac{\mathrm{e}^{-\frac{\mu_0^2}{2}(\beta_1+\beta_2) }}{
(1-\mathrm{e}^{-(\b_1+\b_2)\omega})^2}\;,
\end{equation}
so in the commutative case, the divergence is local as expected.

\quad

When $\th$ is rational, the divergence is local as well. Indeed, for
$\th=p/q$ we have $\gamma_1=\frac{p}{q\sqrt{2}}(m_1+\mathrm{i}n_1)$
and $-\mathrm{i}\gamma_2^*=\frac{1}{\sqrt{2}}(n_2-\mathrm{i}m_2)$.
Thus, if we restrict the sum to the set of $(m_1.n_1,m_2,n_2)\in\Z^8$,
such that $pn_1=-qm_2$ and $pm_1=qn_2$ (the rest is a finite sum that
corresponds to a finite multiplicity and
yields a finite contribution) and if we add external fields and use 
the transformation \eqref{matrix}, we find for  the divergent part: 
\begin{equation}
\label{divrational}
-\frac{q^2\lambda^2}{4p^2}
\int_0^{\frac1q}dx\int_0^{\frac1q}dy\Tr\big[M^\dagger M\big](x,y)\,
\Tr\big[M^\dagger M\big](x,y)\,
\int_\eps^\infty d\beta_1d\beta_2
\frac{\mathrm{e}^{-\frac{\mu_0^2}{2}(\beta_1+\beta_2) }}{
(1-\mathrm{e}^{-(\b_1+\b_2)\omega})^2}.
\end{equation}
\quad
Whereas unfamiliar in the standard formulation of a field theory
on a vector bundle with matricial fibers, such a term in the original
action is perfectly acceptable.

When $\theta$ is irrational, plus satisfying a Diophantine condition
(Definition \ref{Dioph}) to control the sums $\ga_1,\ga_2^*\ne0$, we
see in \eqref{eq:NP2} that only the mode $\ga_1=0$, $\ga_2^*=0$ will
contribute to the divergent part. There are three regions to consider:

\begin{tabular}{rl}
(I) \quad& $\gamma_1=\gamma_2^*=0$, \\
(II) \quad& $\gamma_1=0$, $\gamma_2^*\neq 0$,
\\
(III) \quad& $\gamma_1 \neq 0$, any $\gamma_2^*$. 
\end{tabular}
\\
In case (I), we obtain
\begin{equation}
\Gamma^{(1)}_{4c,\theta \notin \mathbb{Q},\text{(I)}}
[\overline{\vf_1},\vf_2,\overline{\vf_3},\vf_4]
= -\frac{\lambda^2}{4\th^2}
\langle\vf_1,\vf_2\rangle_B\,
\langle\vf_3,\vf_4\rangle_B\,
\int_\eps^\infty d\beta_1d\beta_2
\frac{\mathrm{e}^{-\frac{\mu_0^2}{2}(\beta_1+\beta_2) }}{
(1-\mathrm{e}^{-(\b_1+\b_2)\omega})^2}\;.
\end{equation}
It is worthwhile to notice that this divergence is non-local
and requires a counter-term of the form
\begin{equation}
\label{prodtrace}
\Tr_{\A_\th}\big[(\vf,\vf)_{\A_\th}\big]\,
\Tr_{\A_\th}\big[(\vf,\vf)_{\A_\th}\big]\;.
\end{equation}
This is the discrete version of the UV/IR-entanglement, reminiscent to
toric-noncommutative spaces \cite{UVIR}.

The estimation in region (II) is analogous to an estimation we
performed for planar graphs. It remains to estimate the region (III),
where the Diophantine condition will be used. We obtain in the limit
$\eps\to0$:
\begin{align}
 &\Big|\Gamma^{(1)}_{4c,\theta \notin \mathbb{Q},\text{(III)}}
[\overline{\vf_1},\vf_2,\overline{\vf_3},\vf_4]\Big|
\nonumber
\\
&\leq\frac{\lambda^2}{4\th^2}\sum_{\ga_1\ne0,\ga_2^*}
\big|\langle\vf_1,\vf_2U_{\ga_1}\rangle_B\big|\,
\big|\langle\vf_3,\vf_4U_{\mathrm{i}\ga_2^*}\rangle_B\big|\,
\nonumber
\\
&\qquad\times
\int_0^\infty d\beta_1d\beta_2
\frac{\mathrm{e}^{-\frac{\mu_0^2}{2}(\beta_1+\beta_2) }}{
(1-\mathrm{e}^{-(\b_1+\b_2)\omega})^2}
\mathrm{e}^{-\frac{\omega}{2}
\frac{1+\mathrm{e}^{-(\b_1+\b_2)\omega}}{1-\mathrm{e}^{-(\b_1+\b_2)\omega}}(
|\ga_1|^2+|\ga_2^*|^2)}
\mathrm{e}^{-\frac{\omega}{2}\frac{\mathrm{e}^{-\b_1}+\mathrm{e}^{-\b_2}}
{1-\mathrm{e}^{-(\b_1+\b_2)\omega}}(
\overline{\ga_1}\mathrm{i}\ga_2^*-\mathrm{i}\overline{\ga_2^*}\ga_1)}
\nonumber
\\
&=
\frac{\lambda^2}{4\th^2}\sum_{\ga_1\ne0,\ga_2^*}
\big|\langle\vf_1,\vf_2U_{\ga_1}\rangle_B\big|\,
\big|\langle\vf_3,\vf_4U_{\mathrm{i}\ga_2^*}\rangle_B\big|\,
\nonumber
\\
&\qquad \times 
\int_0^\infty d\beta_1d\beta_2
\frac{\mathrm{e}^{-\frac{\mu_0^2}{2}(\beta_1+\beta_2)}}{
(1-\mathrm{e}^{-(\b_1+\b_2)\omega})^2}
\mathrm{e}^{-\frac{\omega}{2}
\frac{\mathrm{e}^{-\b_1\omega}+\mathrm{e}^{-\b_2\omega}
}{1-\mathrm{e}^{-(\b_1+\b_2)\omega}}
|\ga_1+\mathrm{i}\ga_2^*|^2}
\mathrm{e}^{-\frac{\omega}{2}\frac{(1-\mathrm{e}^{-\b_1})
(1-\mathrm{e}^{-\b_2})}
{1-\mathrm{e}^{-(\b_1+\b_2)\omega}}(|\ga_1|^2+|\ga_2^*|^2)}
\nonumber
\\
&\leq
\frac{\lambda^2}{4\th^2}\sum_{\ga_1\ne0,\ga_2^*}
\big|\langle\vf_1,\vf_2U_{\ga_1}\rangle_B\big|\,
\big|\langle\vf_3,\vf_4U_{\mathrm{i}\ga_2^*}\rangle_B\big|\,
\nonumber
\\
&\qquad \times \int_0^\infty d\beta_1d\beta_2
\frac{\mathrm{e}^{-\frac{\mu_0^2}{2}(\beta_1+\beta_2) }}{
(1-\mathrm{e}^{-(\b_1+\b_2)\omega})^2}
\mathrm{e}^{-\frac{\omega}{2}
\frac{\mathrm{e}^{-\b_1\omega}+\mathrm{e}^{-\b_2\omega}}{
1-\mathrm{e}^{-(\b_1+\b_2)\omega}}
|\ga_1+\mathrm{i}\ga_2^*|^2}\;.
\end{align}
Using now the (regular) Diophantine condition according to Definition
\ref{Dioph}, and disregarding the $\b$-integrals from $1$ to $+\infty$
(which gives a finite contribution say $C_1$), we are left with
\begin{align}
\label{ucla}
 &\Big|\Gamma^{(1)}_{4c,\ga_1\neq0,\text{(III)}}
[\overline{\vf_1},\vf_2,\overline{\vf_3},\vf_4]\Big|
\nonumber
\\
&\leq C_1+ \frac{\lambda^2}{4\th^2}\sum_{\ga_2^*}
\big|\langle\vf_3,\vf_4U_{\mathrm{i}\ga_2^*}\rangle_B\big|\,
\sum_{\ga_1\neq 0}
\big|\langle\vf_1,\vf_2U_{\ga_1}\rangle_B\big|\,
\int_0^1 d\beta_1d\beta_2\,
\frac{\sup_{\ga^*_2}\Big\{
\mathrm{e}^{-\frac{\omega}{2}
\frac{\mathrm{e}^{-\b_1\omega}+\mathrm{e}^{-\b_2\omega}}{
1-\mathrm{e}^{-(\b_1+\b_2)\omega}}
|\ga_1+\mathrm{i}\ga_2^*|^2}\Big\}
}{(1-\mathrm{e}^{-(\b_1+\b_2)\omega})^2}
\nonumber
\\
&\leq C_1+C_2\,\sum_{\ga_1\ne0}
\big|\langle\vf_1,\vf_2U_{\ga_1}\rangle_B\big|\,
\int_0^1 d\beta_1d\beta_2\,
\frac{\mathrm{e}^{-\frac{\omega}{2}
\frac{\mathrm{e}^{-\b_1\omega}+\mathrm{e}^{-\b_2\omega}}{
1-\mathrm{e}^{-(\b_1+\b_2)\omega}}
\inf_{\ga^*_2}|\ga_1+\mathrm{i}\ga_2^*|^2}
}{(1-\mathrm{e}^{-(\b_1+\b_2)\omega})^2}
\nonumber
\\
& \leq C_1+C_2\,\sum_{\ga_1\ne0}
\big|\langle\vf_1,\vf_2U_{\ga_1}\rangle_B\big|\,
\int_0^1 d\beta_1d\beta_2\,
\frac{\mathrm{e}^{-\frac{C\omega}{2}
\frac{\mathrm{e}^{-\b_1\omega}+\mathrm{e}^{-\b_2\omega}}{
1-\mathrm{e}^{-(\b_1+\b_2)\omega}}
|\ga_1|^{-2(4+\delta)}}}{(1-\mathrm{e}^{-(\b_1+\b_2)\omega})^2}
\nonumber
\\
&\leq C_1+\sum_{\ga_1\ne0}
\big|\langle\vf_1,\vf_2U_{\ga_1}\rangle_B\big|\,
(C_3 + C_4 \ln |\ga_1|) \;,
\end{align}
which is finite because
$\langle\vf_i,\vf_jU_{\ga}\rangle_B$ is a Schwartz
sequence.

It should be clear that the same conclusion holds with the weak
Diophantine condition \eqref{weakdioph} instead.  The reason for that
works is that this graph is logarithmically divergent only.  However,
it is unclear for us if this condition will suffice at higher loops.
There is a general argument \cite{Chepelev:2000hm} that oriented just
renormalizable $\phi^4$-models do not have quadratically divergent
non-planar graphs. If this argument applies to our case, the weak
Diophantine condition might suffice in general.

\quad

The second non-planar graph we have to compute is
\begin{align}
\parbox{48mm}{\begin{picture}(45,8)
\put(0,-15){\begin{fmfgraph}(45,20)
\fmfleft{l1,l2}
\fmfright{r1,r2}
\fmf{fermion,tension=2}{r1,i1}
\fmf{fermion,tension=1}{i1,i3}
\fmf{fermion,tension=2}{i3,l1}
\fmf{fermion,tension=2}{l2,i4}
\fmf{fermion,tension=1}{i4,i2}
\fmf{fermion,tension=2}{i2,r2}
\fmffreeze
\fmf{photon}{i1,i5,i4}
\fmf{phantom_arrow,tension=0}{i5,i1}
\fmf{photon,rubout=5}{i3,i7}
\fmf{photon}{i2,i7}
\fmf{phantom}{i3,i6,i7,i2}
\fmf{phantom_arrow,tension=0}{i2,i6}
\end{fmfgraph}}
\put(4,-18.5){\mbox{\small$\overline{z_2}$}}
\put(38,-18.5){\mbox{\small$z_1$}}
\put(27,-19){\mbox{\small$\overline{z_2'}$}}
\put(15,-18.5){\mbox{\small$z_1'$}}
\put(38,7){\mbox{\small$\overline{z_4}$}}
\put(4,7){\mbox{\small$z_3$}}
\put(15,7){\mbox{\small$\overline{z_4'}$}}
\put(27,7){\mbox{\small$z_3'$}}
\put(18.5,-13){\mbox{\small$\beta_1$}}
\put(19.5,1){\mbox{\small$\beta_2$}}
\put(29,-9){\mbox{\small$\gamma_1$}}
\put(28,-2){\mbox{\small$\gamma_2$}}
\end{picture}}
&\Gamma^{(1)}_{4d}[z_1,\overline{z_2},z_3,\overline{z_4}] 
\nonumber
\\*[-1ex]
&=-2\int d\mu(z_1',\overline{z_1}')\,
 d\mu(z_2',\overline{z_2}')\, d\mu(z_3',\overline{z_3}')\, 
d\mu(z_4',\overline{z_4}')\,\nonumber\\* 
&\times 
V(z_1,\overline{z_2}',z_3,\overline{z_4}')\,
V(z_3',\overline{z_4},z_1',\overline{z_2})
\,H^{-1}_\eps(z_2',\overline{z_1}')\,
H^{-1}_\eps(z_4',\overline{z_3}')\;.
\label{Gamma4d}
\end{align}

\quad

We have
\begin{align*}
&\Gamma^{(1)}_{4d}[z_1,\overline{z_2},z_3,\overline{z_4}] 
=-\frac{\lambda^2}{8} \sum_{\gamma_1,\gamma_2}
\int_\eps^\infty d\beta_1 d\beta_2\, \mathrm{e}^{
-\frac{\mu_0^2}{2} (\beta_1+\beta_2)  }
\mathrm{e}^{
-\omega(|\gamma_1|^2+|\gamma_2|^2) }
\\
&\qquad\qquad \times
\mathrm{e}^{\omega(\overline{z_2}\gamma_2
+(\overline{z_2}-\overline{\gamma_2})\mathrm{e}^{-\beta_1 \omega} 
\gamma_1
+ (\overline{z_2}-\overline{\gamma_2})\mathrm{e}^{-\beta_1 \omega}z_1
-\overline{\gamma_1}z_1
-\overline{z_4}\gamma_2
-(\overline{z_4}+\overline{\gamma_2})\mathrm{e}^{-\beta_2 \omega} \gamma_1
+ (\overline{z_4}+\overline{\gamma_2})\mathrm{e}^{-\beta_2 \omega}z_3
+\overline{\gamma_1}z_3)}\;.
\end{align*}
We set $\gamma_2=-\gamma_1+\gamma$. The resulting Gau\ss ian decay in
$\gamma_1$ is suppressed for small $\beta_i$, which is best expressed
by a Poisson re-summation in $\gamma_1$:
\begin{align}
&\Gamma^{(1)}_{4d}[z_1,\overline{z_2},z_3,\overline{z_4}] 
\nonumber
\\
&=-\frac{\lambda^2}{8\theta^2} \sum_{\gamma_1^*,\gamma}
\int_\eps^\infty \frac{d\beta_1 d\beta_2\, \mathrm{e}^{
-\frac{\mu_0^2}{2} (\beta_1+\beta_2)  }}{
(2-\mathrm{e}^{-\beta_1 \omega}-\mathrm{e}^{-\beta_2 \omega})^2}
\mathrm{e}^{\omega(
(\overline{z_2}-\overline{z_4})\gamma
-|\gamma|^2
+ (\overline{z_2}-\overline{\gamma})\mathrm{e}^{-\beta_1 \omega}z_1
+ (\overline{z_4}+\overline{\gamma})\mathrm{e}^{-\beta_2 \omega}z_3)}
\nonumber
\\
& \times
\mathrm{e}^{\frac{\omega}{
(2-\mathrm{e}^{-\beta_1 \omega}-\mathrm{e}^{-\beta_2 \omega})} 
(\overline{\gamma}
(1-\mathrm{e}^{-\beta_1 \omega} -\mathrm{e}^{-\beta_2 \omega} )
-\overline{z_2} (1-\mathrm{e}^{-\beta_1 \omega} )
+\overline{z_4} (1-\mathrm{e}^{-\beta_2 \omega})
-\mathrm{i}\overline{\gamma_1^*})
(\gamma 
-(1-\mathrm{e}^{-\beta_1 \omega})z_1
+(1-\mathrm{e}^{-\beta_2 \omega})z_3
- \mathrm{i} \gamma_1^*)
}
\nonumber
\\
&=-\frac{\lambda^2}{8\theta^2} \sum_{\gamma_1^*,\gamma}
\int_\eps^\infty \frac{d\beta_1 d\beta_2\, \mathrm{e}^{
-\frac{\mu_0^2}{2} (\beta_1+\beta_2)  }}{
(2-\mathrm{e}^{-\beta_1 \omega}-\mathrm{e}^{-\beta_2 \omega})^2}
\mathrm{e}^{\omega\big(
(\overline{z_2}-\overline{z_4})\gamma
+ \overline{\gamma}(z_3-z_1)
+ \overline{z_2}z_1
+ \overline{z_4}z_3
-\overline{\gamma} \mathrm{i} \gamma_1^*
\big)}
\nonumber
\\
& \times
\mathrm{e}^{-\frac{\omega}{
(2-\mathrm{e}^{-\beta_1 \omega}-\mathrm{e}^{-\beta_2 \omega})} 
\big( |\gamma-\mathrm{i}\gamma_1^*|^2
+ (\overline{\gamma-\mathrm{i}\gamma_1^*})
((1-\mathrm{e}^{-\beta_2 \omega})z_3
-(1-\mathrm{e}^{-\beta_1 \omega})z_1)
+(\overline{z_2} (1-\mathrm{e}^{-\beta_1 \omega} )
-\overline{z_4} (1-\mathrm{e}^{-\beta_2 \omega}))
(\gamma - \mathrm{i} \gamma_1^*)\big)}
\nonumber
\\
& \times
\mathrm{e}^{-\frac{\omega(1-\mathrm{e}^{-\beta_1 \omega} )
(1-\mathrm{e}^{-\beta_2 \omega} )}{
(2-\mathrm{e}^{-\beta_1 \omega}-\mathrm{e}^{-\beta_2 \omega})} 
(\overline{z_2} + \overline{z_4})( z_1+z_3)  }
\;.
\label{G4c}
\end{align}

When $\theta$ is irrational, plus satisfying a Diophantine condition
(Definition \ref{Dioph}) to control the sums $\gamma,\gamma_1^*\neq 0$, we see
that only the mode $\gamma=0$, $\gamma_1^*=0$ will contribute to the
diverging part. Adding external fields and integrating over the
$z_i$-positions with the help of the reproducing
kernels, we obtain in this case
\begin{align}
&\hspace{-0.3cm}\Gamma^{(1)}_{4d,\gamma=\gamma_1^*=0}
[\overline{\varphi_1},\varphi_2,
\overline{\varphi_3},\varphi_4] 
 =-\frac{\lambda^2}{8\theta^2} \int d\mu(z_1,\overline{z_1})\,
d\mu(z_3,\overline{z_3})
\int_\eps^\infty \frac{d\beta_1 d\beta_2\, \mathrm{e}^{
-\frac{\mu_0^2}{2} (\beta_1+\beta_2)  }}{
(2-\mathrm{e}^{-\beta_1 \omega}-\mathrm{e}^{-\beta_2 \omega})^2}\;
\nonumber\\
&\overline{\varphi_1}(\overline{z_1}) 
\varphi_2\big(z_1- \tfrac{\omega(1-\mathrm{e}^{-\beta_1 \omega} )
(1-\mathrm{e}^{-\beta_2 \omega} )}{
(2-\mathrm{e}^{-\beta_1 \omega}-\mathrm{e}^{-\beta_2 \omega})} 
(z_1{+}z_3)\big)
\overline{\varphi_3}(\overline{z_3}) 
\varphi_4\big(z_3- \tfrac{\omega(1-\mathrm{e}^{-\beta_1 \omega} )
(1-\mathrm{e}^{-\beta_2 \omega} )}{
(2-\mathrm{e}^{-\beta_1 \omega}-\mathrm{e}^{-\beta_2 \omega})} 
(z_1{+}z_3)\big)\;.
\label{G4c0}
\end{align}
Now, using Taylor's theorem in the form $\varphi(z+y)=\varphi(z)+y 
\int_0^1 d\xi \;(\partial\varphi)(z+\xi y)$, it is easy to see that the
derivatives of $\varphi$ lead to a convergent
$\beta$-integral. Therefore, the divergent part is given by
\begin{align} 
\Gamma^{(1)}_{4d,div}[\overline{\varphi_1},\varphi_2,
\overline{\varphi_3},\varphi_4] 
&=-\frac{\lambda^2}{8 \theta^2} \,
\langle \varphi_1,\varphi_2\rangle_{\mathcal{B}}\,
\langle \varphi_3,\varphi_3\rangle_{\mathcal{B}}
\int_\eps^\infty \frac{d\beta_1 d\beta_2\, \mathrm{e}^{
-\frac{\mu_0^2}{2} (\beta_1+\beta_2)  }}{
(2-\mathrm{e}^{-\beta_1\omega}
-\mathrm{e}^{-\beta_2\omega})^2}\;.
\end{align}
Again, this divergence is quite problematic at this stage since it
corresponds to a counterterm of the form \eqref{prodtrace},
which was not present in the
classical action we have chosen \eqref{actiontorus}.

For $\theta\in\N$, the divergent part of the graph \eqref{Gamma4d} is
given by the mode $\gamma_1^*=-\mathrm{i}\gamma$. Using again Taylor's
theorem we obtain
\begin{align} 
\Gamma^{(1)}_{4d,\theta\in\N,div}[z_1,\overline{z_2},z_3,\overline{z_4}]
&=-\frac{\lambda}{4 \theta^2} V(z_1,\overline{z_2},z_3,\overline{z_4})
\int_\eps^\infty \frac{d\beta_1 d\beta_2\, \mathrm{e}^{
-\frac{\mu_0^2}{2} (\beta_1+\beta_2)  }}{
(2-\mathrm{e}^{-\beta_1\omega}
-\mathrm{e}^{-\beta_2\omega})^2}\;.
\end{align}
In contrast to the irrational case, this is now a local
counterterm. For $\th$ rational, the same discussion than in the
previous graph applies, and we are left with a divergence of the form
\eqref{divrational}.

It remains to show that in the irrational Diophantian case, 
the remaining sum over $(\gamma,\gamma_1^*)\neq (0,0)$ leads to a convergent 
integral. With the help of reproducing kernels we find for (\ref{G4c})
\begin{align*}
&\Gamma^{(1)}_{4d,\theta\notin\mathbb{Q},conv}[
\overline{\varphi_1},\varphi_2,
\overline{\varphi_3},\varphi_4]
=-\frac{\lambda^2}{8\theta^2} \sum_{(\gamma_1^*,\gamma)\neq (0,0)}
\int d\mu(z_1,\overline{z_1})\,d\mu(z_3,\overline{z_3}) 
\int_\eps^\infty \frac{d\beta_1 d\beta_2\, \mathrm{e}^{
-\frac{\mu_0^2}{2} (\beta_1+\beta_2)  }}{
(2-\mathrm{e}^{-\beta_1 \omega}-\mathrm{e}^{-\beta_2 \omega})^2}
\\
& \qquad\times \overline{\varphi_1}(\overline{z_1})
\varphi_2\big(z_1+\gamma
-\tfrac{(1-\mathrm{e}^{-\beta_1 \omega})}{
(2-\mathrm{e}^{-\beta_1 \omega}-\mathrm{e}^{-\beta_2 \omega})}
(\gamma-\mathrm{i}\gamma_1^*)
- \tfrac{(1-\mathrm{e}^{-\beta_1 \omega} )
(1-\mathrm{e}^{-\beta_2 \omega} )}{
(2-\mathrm{e}^{-\beta_1 \omega}-\mathrm{e}^{-\beta_2 \omega})} 
( z_1{+}z_3)\big)
\\
&\qquad \times \overline{\varphi_3}(\overline{z_3})
\varphi_4\big(z_3-\gamma
+\tfrac{(1-\mathrm{e}^{-\beta_2 \omega})}{
(2-\mathrm{e}^{-\beta_1 \omega}-\mathrm{e}^{-\beta_2 \omega})}
(\gamma-\mathrm{i}\gamma_1^*)
- \tfrac{(1-\mathrm{e}^{-\beta_1 \omega} )
(1-\mathrm{e}^{-\beta_2 \omega} )}{
(2-\mathrm{e}^{-\beta_1 \omega}-\mathrm{e}^{-\beta_2 \omega})} 
( z_1{+}z_3)\big)
\\
&\qquad \times 
\mathrm{e}^{\omega\big(
\overline{\gamma}(z_3-z_1)
-\overline{\gamma} \mathrm{i} \gamma_1^*
\big)}
\mathrm{e}^{-\frac{\omega}{
(2-\mathrm{e}^{-\beta_1 \omega}-\mathrm{e}^{-\beta_2 \omega})} 
\big( |\gamma-\mathrm{i}\gamma_1^*|^2
+ (\overline{\gamma-\mathrm{i}\gamma_1^*})
((1-\mathrm{e}^{-\beta_2 \omega})z_3
-(1-\mathrm{e}^{-\beta_1 \omega})z_1)\big)}\;.
\end{align*}
We re-express $\gamma,\gamma_1^*$-dependence of the field positions in terms 
of the unitaries $U$:
\begin{align*}
&\Gamma^{(1)}_{4d,\theta\notin\mathbb{Q},conv}[
\overline{\varphi_1},\varphi_2,
\overline{\varphi_3},\varphi_4]
 =-\frac{\lambda^2}{8\theta^2} \sum_{(\gamma_1^*,\gamma)\neq (0,0)}
\int d\mu(z_1,\overline{z_1})\,d\mu(z_3,\overline{z_3}) 
\int_\eps^\infty \frac{d\beta_1 d\beta_2\, \mathrm{e}^{
-\frac{\mu_0^2}{2} (\beta_1+\beta_2)  }}{
(2-\mathrm{e}^{-\beta_1 \omega}-\mathrm{e}^{-\beta_2 \omega})^2}
\nonumber
\\
&\qquad\times 
\overline{\varphi_1}(\overline{z_1})\;
\Big(\varphi_2 U_{\gamma} U_{-\tfrac{
(1-\mathrm{e}^{-\beta_1 \omega})(\gamma-\mathrm{i}\gamma_1^*)}{
(2-\mathrm{e}^{-\beta_1 \omega}-\mathrm{e}^{-\beta_2 \omega})}}
\Big)\big(z_1- \tfrac{(1-\mathrm{e}^{-\beta_1 \omega} )
(1-\mathrm{e}^{-\beta_2 \omega} )}{
(2-\mathrm{e}^{-\beta_1 \omega}-\mathrm{e}^{-\beta_2 \omega})} 
( z_1{+}z_3)\big)\;
\nonumber
\\
&\qquad\times \overline{\varphi_3}(\overline{z_3})\;
\Big(\varphi_4 U_{-\gamma} U_{\tfrac{
(1-\mathrm{e}^{-\beta_2 \omega})(\gamma-\mathrm{i}\gamma_1^*)
}{(2-\mathrm{e}^{-\beta_1 \omega}-\mathrm{e}^{-\beta_2 \omega})}
}\Big)\big(z_3- \tfrac{(1-\mathrm{e}^{-\beta_1 \omega} )
(1-\mathrm{e}^{-\beta_2 \omega} )}{
(2-\mathrm{e}^{-\beta_1 \omega}-\mathrm{e}^{-\beta_2 \omega})} 
( z_1{+}z_3)\big)
\nonumber
\\
& \qquad\times 
\mathrm{e}^{-\frac{\omega}{
(2-\mathrm{e}^{-\beta_1 \omega}-\mathrm{e}^{-\beta_2 \omega})} 
|\gamma-\mathrm{i}\gamma_1^*|^2
(1-\frac{(1-\mathrm{e}^{-\beta_1 \omega})^2
+(1-\mathrm{e}^{-\beta_2 \omega})^2
}{2(2-\mathrm{e}^{-\beta_1 \omega}-\mathrm{e}^{-\beta_2 \omega})})}
\\
&\qquad\times 
\mathrm{e}^{-\frac{\omega
(1-\mathrm{e}^{-\beta_1 \omega})
(1-\mathrm{e}^{-\beta_2 \omega})
(\mathrm{e}^{-\beta_1 \omega}-\mathrm{e}^{-\beta_2 \omega})
}{(2-\mathrm{e}^{-\beta_1 \omega}-\mathrm{e}^{-\beta_2 \omega})^2} 
(\overline{\gamma-\mathrm{i}\gamma_1^*})(z_1+z_3)}
\;.
\end{align*}
Now we absorb the last line in new unitaries, at the price of 
$(\gamma-\mathrm{i}\gamma_1^*)$-dependent positions:
\begin{align*}
&\Gamma^{(1)}_{4d,\theta\notin\mathbb{Q},conv}[
\overline{\varphi_1},\varphi_2,
\overline{\varphi_3},\varphi_4]
 =-\frac{\lambda^2}{8\theta^2} \sum_{(\gamma_1^*,\gamma)\neq (0,0)}
\int d\mu(z_1,\overline{z_1})\,d\mu(z_3,\overline{z_2}) 
\int_\eps^\infty \frac{d\beta_1 d\beta_2\, \mathrm{e}^{
-\frac{\mu_0^2}{2} (\beta_1+\beta_2)  }}{
(2-\mathrm{e}^{-\beta_1 \omega}-\mathrm{e}^{-\beta_2 \omega})^2}
\nonumber
\\
&\times 
\overline{\varphi_1}(\overline{z_1})\;
\left(\varphi_2 U_{-\gamma} U_{\tfrac{
(1-\mathrm{e}^{-\beta_1 \omega})(\gamma-\mathrm{i}\gamma_1^*)}{
(2-\mathrm{e}^{-\beta_1 \omega}-\mathrm{e}^{-\beta_2 \omega})}}
U_{ \tfrac{(1-\mathrm{e}^{-\beta_1 \omega} )
(1-\mathrm{e}^{-\beta_2 \omega} )(\mathrm{e}^{-\beta_1 \omega} 
-\mathrm{e}^{-\beta_2 \omega} )}{
2(2-\mathrm{e}^{-\beta_1 \omega}-\mathrm{e}^{-\beta_2 \omega})
(\mathrm{e}^{-\beta_2 \omega} (1-\mathrm{e}^{-\beta_1 \omega} )
+\mathrm{e}^{-\beta_1 \omega} (1-\mathrm{e}^{-\beta_2 \omega} ))
} (\gamma-\mathrm{i}\gamma_1^*)}
\right)
\\
&
\Big(z_1- \tfrac{(1-\mathrm{e}^{-\beta_1 \omega} )
(1-\mathrm{e}^{-\beta_2 \omega} )}{
(2-\mathrm{e}^{-\beta_1 \omega}-\mathrm{e}^{-\beta_2 \omega})} 
( z_1{+}z_3)
- \tfrac{(1-\mathrm{e}^{-\beta_1 \omega} )
(1-\mathrm{e}^{-\beta_2 \omega} )(\mathrm{e}^{-\beta_1 \omega} 
-\mathrm{e}^{-\beta_2 \omega} )}{
2(2-\mathrm{e}^{-\beta_1 \omega}-\mathrm{e}^{-\beta_2 \omega})
(\mathrm{e}^{-\beta_2 \omega} (1-\mathrm{e}^{-\beta_1 \omega} )
+\mathrm{e}^{-\beta_1 \omega} (1-\mathrm{e}^{-\beta_2 \omega} ))
} (\gamma-\mathrm{i}\gamma_1^*)\Big)\;
\nonumber
\\
&\times \overline{\varphi_3}(\overline{z_3})\;
\left(\varphi_4 U_{\gamma} U_{-\tfrac{
(1-\mathrm{e}^{-\beta_2 \omega})(\gamma-\mathrm{i}\gamma_1^*)
}{(2-\mathrm{e}^{-\beta_1 \omega}-\mathrm{e}^{-\beta_2 \omega})}}
U_{\tfrac{(1-\mathrm{e}^{-\beta_1 \omega} )
(1-\mathrm{e}^{-\beta_2 \omega} )(\mathrm{e}^{-\beta_1 \omega} 
-\mathrm{e}^{-\beta_2 \omega} )}{
2(2-\mathrm{e}^{-\beta_1 \omega}-\mathrm{e}^{-\beta_2 \omega})
(\mathrm{e}^{-\beta_2 \omega} (1-\mathrm{e}^{-\beta_1 \omega} )
+\mathrm{e}^{-\beta_1 \omega} (1-\mathrm{e}^{-\beta_2 \omega} ))
} (\gamma-\mathrm{i}\gamma_1^*)}
\right)
\\
&
\Big(z_3- \tfrac{(1-\mathrm{e}^{-\beta_1 \omega} )
(1-\mathrm{e}^{-\beta_2 \omega} )}{
(2-\mathrm{e}^{-\beta_1 \omega}-\mathrm{e}^{-\beta_2 \omega})} 
( z_1{+}z_3)
- \tfrac{(1-\mathrm{e}^{-\beta_1 \omega} )
(1-\mathrm{e}^{-\beta_2 \omega} )(\mathrm{e}^{-\beta_1 \omega} 
-\mathrm{e}^{-\beta_2 \omega} )}{
2(2-\mathrm{e}^{-\beta_1 \omega}-\mathrm{e}^{-\beta_2 \omega})
(\mathrm{e}^{-\beta_2 \omega} (1-\mathrm{e}^{-\beta_1 \omega} )
+\mathrm{e}^{-\beta_1 \omega} (1-\mathrm{e}^{-\beta_2 \omega} ))
} (\gamma-\mathrm{i}\gamma_1^*)
\Big)
\nonumber
\\
& \times 
\mathrm{e}^{-\frac{\omega |\gamma-\mathrm{i}\gamma_1^*|^2}{
(2-\mathrm{e}^{-\beta_1 \omega}-\mathrm{e}^{-\beta_2 \omega})} 
\big(1-\frac{(1-\mathrm{e}^{-\beta_1 \omega})^2
+(1-\mathrm{e}^{-\beta_2 \omega})^2
}{2(2-\mathrm{e}^{-\beta_1 \omega}-\mathrm{e}^{-\beta_2 \omega})}
+\frac{(1-\mathrm{e}^{-\beta_1 \omega} )^2
(1-\mathrm{e}^{-\beta_2 \omega} )^2(\mathrm{e}^{-\beta_1 \omega} 
-\mathrm{e}^{-\beta_2 \omega} )^2}{
4(2-\mathrm{e}^{-\beta_1 \omega}-\mathrm{e}^{-\beta_2 \omega})
(\mathrm{e}^{-\beta_2 \omega} (1-\mathrm{e}^{-\beta_1 \omega} )
+\mathrm{e}^{-\beta_1 \omega} (1-\mathrm{e}^{-\beta_2 \omega} ))^2
}\big)}\;.
\end{align*}
Since 
$$
1-\frac{(1-\mathrm{e}^{-\beta_1 \omega})^2
+(1-\mathrm{e}^{-\beta_2 \omega})^2
}{2(2-\mathrm{e}^{-\beta_1 \omega}-\mathrm{e}^{-\beta_2 \omega})}
\geq\frac12,
$$
we obtain
\begin{align*}
&\big|\Gamma^{(1)}_{4d,\theta\notin\mathbb{Q},conv}[
\overline{\varphi_1},\varphi_2,
\overline{\varphi_3},\varphi_4]\big|\nonumber\\
& \leq\frac{\lambda^2}{8\theta^2} \sum_{(\gamma_1^*,\gamma)\neq (0,0)}
\int_\eps^\infty \frac{d\beta_1 d\beta_2\, \mathrm{e}^{
-\frac{\mu_0^2}{2} (\beta_1+\beta_2)  }}{
(2-\mathrm{e}^{-\beta_1 \omega}-\mathrm{e}^{-\beta_2 \omega})^2}\,
\mathrm{e}^{-\frac{\omega|\gamma-\mathrm{i}\gamma_1^*|^2}{
2(2-\mathrm{e}^{-\beta_1 \omega}-\mathrm{e}^{-\beta_2 \omega})} 
}\nonumber\\
&\x
\Big|\int d\mu(z_1,\overline{z_1})\,d\mu(z_3,\overline{z_3}) 
\overline{\varphi_1}(\overline{z_1})\overline{\varphi_3}(\overline{z_3})
\nonumber
\\
&\times 
\Big(\varphi_2 U_{-\gamma} U_{f_1(\b_1,\b_2)(\ga-i\ga_1^*)}
U_{f_2(\b_1,\b_2)(\ga-i\ga_1^*)}
\Big)
\big(z_1-f_3(\b_1,\b_2)(z_1+z_3) -f_2(\b_1,\b_2)(\ga-i\ga_1^*)\big) \;
\nonumber
\\
&\times 
\Big(\varphi_4 U_{\gamma} U_{-f_1(\b_2,\b_1)(\ga-i\ga_1^*)}
U_{f_2(\b_1,\b_2)(\ga-i\ga_1^*)}
\Big)
\big(z_3-f_3(\b_1,\b_2)(z_1+z_3) -f_2(\b_1,\b_2)(\ga-i\ga_1^*) \big)\;
\Big|,
\end{align*}
where we have set
\begin{align*}
f_1(\b_1,\b_2)&=\frac{
(1-\mathrm{e}^{-\beta_1 \omega})}{
(2-\mathrm{e}^{-\beta_1 \omega}-\mathrm{e}^{-\beta_2 \omega})}, \\
f_2(\b_1,\b_2)&= \tfrac{(1-\mathrm{e}^{-\beta_1 \omega} )
(1-\mathrm{e}^{-\beta_2 \omega} )(\mathrm{e}^{-\beta_1 \omega} 
-\mathrm{e}^{-\beta_2 \omega} )}{
2(2-\mathrm{e}^{-\beta_1 \omega}-\mathrm{e}^{-\beta_2 \omega})
(\mathrm{e}^{-\beta_2 \omega} (1-\mathrm{e}^{-\beta_1 \omega} )
+\mathrm{e}^{-\beta_1 \omega} (1-\mathrm{e}^{-\beta_2 \omega} ))
} ,\\
f_3(\b_1,\b_2)&= \frac{(1-\mathrm{e}^{-\beta_1 \omega} )
(1-\mathrm{e}^{-\beta_2 \omega} )}{
(2-\mathrm{e}^{-\beta_1 \omega}-\mathrm{e}^{-\beta_2 \omega})}.
\end{align*}
The next step consists in disregarding the term
$f_3(\b_1,\b_2)(z_1+z_3)+f_2(\b_1,\b_2)(\ga-i\ga_1^*)$ in the argument
of $\vf_2,\vf_4$. Indeed, any
$\phi\in\E_B(\C^2)=B\SS(\R^2)$ can be written as 
$\phi(z)=\sum_n \phi_n\,z^n$, where $\{\phi_n\}\in\SS(\N)$. This holds
because any $f\in\SS(\R^2)$ can be expanded as a sum
of the eigen-modes of
the harmonic oscillator, with rapid decreasing coefficients. 
Thus, applying that to
$\varphi_2 U_{-\gamma} U_{f_1(\b_1,\b_2)(\ga-i\ga_1^*)}
U_{f_2(\b_1,\b_2)(\ga-i\ga_1^*)}$, 
the analysis of the first term in the expansion of 
$$\big(z_1-f_3(\b_1,\b_2)(z_1+z_3)
-f_2(\b_1,\b_2)(\ga-i\ga_1^*)\big)^n,$$
is enough since they are all of the same weight, because 
$f_i(\b_1,\b_2)\in
L^\infty(\R^+\x\R^+)$, $i=1,2,3$. Thus, each term can be estimated
along the same lines as the first one, with estimates uniform in
$\b_i,\ga,\ga_1^*$. 
Finally, the Schwartz coefficients
will absorb the overall number of terms of the expansion in $z^n$.
Thus, disregarding the integral from $1$ to $+ \infty$ which gives a
finite contribution $C_1$, we are left with
\begin{align*}
&\big|\Gamma^{(1)}_{4d,\theta\notin\mathbb{Q},conv}[
\overline{\varphi_1},\varphi_2,
\overline{\varphi_3},\varphi_4]\big|
 \leq C_1+
\frac{\lambda^2}{8\theta^2} \sum_{(\gamma_1^*,\gamma)\neq (0,0)}
\int_\eps^1 \frac{d\beta_1 d\beta_2\, \mathrm{e}^{
-\frac{\mu_0^2}{2} (\beta_1+\beta_2)  }}{
(2-\mathrm{e}^{-\beta_1 \omega}-\mathrm{e}^{-\beta_2 \omega})^2}\,
\mathrm{e}^{-\frac{\omega|\gamma-\mathrm{i}\gamma_1^*|^2}{
2(2-\mathrm{e}^{-\beta_1 \omega}-\mathrm{e}^{-\beta_2 \omega})} 
}\\
&\qquad\x\Big|\big\langle\varphi_1,
\varphi_2 U_{-\gamma+(f_1(\b_1,\b_2)+f_2(\b_1,\b_2)(\ga-i\ga_1^*)
}\big\rangle_B
\big\langle\varphi_3,
\varphi_4
U_{\gamma-(f_1(\b_2,\b_1)-f_2(\b_1,\b_2)(\ga-i\ga_1^*)}\big\rangle_B \Big|,
\end{align*}
Since $\langle\phi,\chi U_\eta\rangle_B$ belongs to $\SS(\R^4)$ as a 
function of
$\eta\in\C^2\simeq \R^4$, and since $f_2\to0$, $\b_1,\b_2\to0$ and
$f_1\to1$, $\b_1,\b_2\to0$, $\b_1=\b_2$ (which is the only important
case to treat), we deduce that
$$
\Big|\big\langle\varphi_1,
\varphi_2 U_{-\gamma+(f_1(\b_1,\b_2)+f_2(\b_1,\b_2)(\ga-i\ga_1^*)
}\big\rangle_B
\big\langle\varphi_3,
\varphi_4
U_{\gamma-(f_1(\b_2,\b_1)-f_2(\b_1,\b_2)(\ga-i\ga_1^*)}\big\rangle_B
\Big|
\leq a_{\ga,\ga^*_1},
$$
uniformly in $\b_1,\b_2$ and where
$0\leq\{a_{\ga,\ga_1^*}\}\in\SS(\Z^8)$. Finally, we get
\begin{align*}
&\big|\Gamma^{(1)}_{4d,\theta\notin\mathbb{Q},conv}[
\overline{\varphi_1},\varphi_2,
\overline{\varphi_3},\varphi_4]\big|
 \leq C_1+
 \sum_{(\gamma_1^*,\gamma)\neq (0,0)} a_{\ga,\ga^*_1}
\int_\eps^1 \frac{d\beta_1 d\beta_2\, \mathrm{e}^{
-\frac{\mu_0^2}{2} (\beta_1+\beta_2)  }}{
(2-\mathrm{e}^{-\beta_1 \omega}-\mathrm{e}^{-\beta_2 \omega})^2}\,
\mathrm{e}^{-\frac{\omega|\gamma-\mathrm{i}\gamma_1^*|^2}{
2(2-\mathrm{e}^{-\beta_1 \omega}-\mathrm{e}^{-\beta_2 \omega})} 
}.
\end{align*}
 Thus the estimate \eqref{ucla} of the previous diagram applies as
 well and shows that the limit $\epsilon\to 0$ is finite.

Gathering all result we have proven:

\begin{prop}
When $\th\in\mathbb{Q}$,
the divergent part of the four-point non-planar 1PI Green
function associated with the action \eqref{actiontorus}, 
in its one-loop approximation,  is given by
\begin{align*}
&\Ga^{(1)}_{4,\mathit{non-planar},div}[M_1^\dagger,M_2,M_3^\dagger,M_4]
\nonumber
\\
&=-\frac{3\lambda^2}{8\omega^2\th^2}
\int_0^{\frac 1q}dx\int_0^{\frac1q}dy\,\Tr\big[M_1^\dagger M_2\big](x,y)\,
\Tr\big[M_3^\dagger M_4\big](x,y)\,\ln\frac1\epsilon,
\end{align*}
and when $\th\in\R\setminus\mathbb{Q}$ plus satisfying the weak Diophantine
condition \eqref{weakdioph}, it is given by
\begin{equation*}
\Ga^{(1)}_{4,\mathit{non-planar}div}
[\overline{\vf_1},\vf_2,\overline{\vf_3},\vf_4]=
-\frac{3\lambda^2}{8\omega^2\th^2}\,
\Tr_{\A_\th}\big[(\vf_1,\vf_2)_{\A_\th}\big]\,
\Tr_{\A_\th}\big[(\vf_3,\vf_4)_{\A_\th}\big]\,\ln\frac1\epsilon.
\end{equation*}
\end{prop}

\quad

In the irrational case,
this analysis leads to the introduction of a non-local term in the action
\eqref{actiontorus}, associated to the new divergence and with a
second coupling constant:
\begin{align}
S[\phi,\overline{\phi}]=
g^{\mu\nu}\mathrm{Tr}_{{\cal A}_{\Theta}}\big[
\left( \nabla_{\mu}\phi,\nabla_{\nu}\phi\right)_{{\cal A}_{\Theta}}\big]+
\mu_0^{2}\mathrm{Tr}_{{\cal A}_{\Theta}}
\big[\left( \phi,\phi\right)_{{\cal A}_{\Theta}}\big]+
&\frac{\lambda}{2}\,\mathrm{Tr}_{{\cal A}_{\Theta}}
\big[\left( \phi,\phi\right)_{{\cal A}_{\Theta}}^2\big]
\nonumber\\
+&\frac{\lambda'}{2}\Big[\Tr_{\A_\th}\big[(\phi,\phi)_{\A_\th}\big]
\Big]^2.
\label{action2}
\end{align}
Such a term also appeared in \cite{Becchi:2002kj}.

To show that our theory is now reasonably well defined, we have to
show that the divergences coming from loop diagrams constructed with
the new vertex (and also with mixed vertices) can be absorbed by a
redefinition of $\mu_0$, $\lambda$, $\lambda'$. This is shown to hold in
the one-loop approximation in the next section.

\section{Mixed diagrams}

We symbolize the new vertex by a dotted line:
\begin{align}
\parbox{40mm}{\begin{picture}(35,20)
\put(0,0){\begin{fmfgraph}(35,20)
\fmfleft{l1,l2}
\fmfright{r1,r2}
\fmf{fermion}{r1,i1,l1}
\fmf{fermion}{l2,i2,r2}
\fmf{dashes}{i1,i2}
\end{fmfgraph}}
\put(3,3){\mbox{\small$\overline{z_2}$}}
\put(28,3){\mbox{\small$z_1$}}
\put(30,16){\mbox{\small$\overline{z_4}$}}
\put(4,15.5){\mbox{\small$z_3$}}
\end{picture}}= \frac{\lambda'}{2}
I(z_1,z_2)\,I(z_3,z_4)\;.
\label{vertex-new}
\end{align} 

This interaction $\big(\Tr_{\A_\th}\big[(\phi,\phi)_{\A_\th}\big]\big)^2$
can also
be written using the Hubbard-Stratonovitch transform as
\begin{equation}
e^{-\frac{\lambda'}{2}\big(\Tr_{\A_\th}\big[(\phi,\phi)_{\A_\th}\big]\big)^2}
=\sqrt{\frac{\lambda'}{2\pi}} \int_\R da\,
e^{-\frac{\lambda'}{2}a^2-i\lambda' a\Tr_{\A_\th}
\left[(\phi,\phi)_{\A_\th}\right]},
\end{equation}
where we integrate over a real number $a$. Equivalently, it can also
be obtained by introducing a different coupling for the constant mode
of the field $A$ appearing in the Hubbard-Stratonovitch transform.

\subsection{2-point sector}

The first 2-point graph with the new vertex is thus 
\begin{align}
\parbox{35mm}{\begin{picture}(25,30)
\put(0,2){\begin{fmfgraph}(25,30)
\fmfbottom{l,r}
\fmftop{t}
\fmf{fermion}{r,i1,l}
\fmffreeze
\fmf{dashes,tension=3}{i1,i2}
\fmf{fermion,right}{i2,t}
\fmf{fermion,right}{t,i2}
\end{fmfgraph}}
\put(-1,6){\mbox{\small$\overline{z_2}$}}
\put(24,5){\mbox{\small$z_1$}}
\put(18,13){\mbox{\small$\overline{z_4}$}}
\put(3,13){\mbox{\small$z_3$}}
\end{picture}} 
\Gamma^{(1)}_{2c}[z_1,\overline{z_2}] 
= \frac{\lambda'}{2}\, I(z_1,\overline{z_2})
\int_\eps^\infty \!\!d\beta 
\,\frac{\mathrm{e}^{-\frac{\b \mu_0^2}{2}}}{
(1-\mathrm{e}^{-\beta \omega})^2}\;.
\label{mass-ren-1-new}
\end{align}
Its value is obtained from (\ref{mass-ren-1}) by putting $\gamma\mapsto 0$ 
and $\lambda \mapsto \lambda'$.
This coincides exactly with the previous divergence.  

The second contribution to the one-loop two-point function is the graph
\begin{align}
\parbox{50mm}{\begin{picture}(45,20)
\put(0,3){\begin{fmfgraph}(45,20)
\fmfbottom{l,r}
\fmf{fermion,tension=2}{r,i1}
\fmf{fermion}{i1,i2}
\fmf{fermion,tension=2}{i2,l}
\fmffreeze
\fmf{dashes,right}{i1,i2}
\end{fmfgraph}}
\put(2,5){\mbox{\small$\overline{z_2}$}}
\put(40,5){\mbox{\small$z_1$}}
\put(28,4){\mbox{\small$\overline{z_4}$}}
\put(14,4){\mbox{\small$z_3$}}
\end{picture}} 
\Gamma^{(1)}_{2d}[z_1,\overline{z_2}] 
&= \frac{\lambda'}{2}\,I(z_1,\overline{z_2}) 
\int_\eps^\infty\!\!d\beta \;
\mathrm{e}^{-\frac{\beta \mu_0^2}{2}}\mathrm{e}^{ -\omega
\overline{z_2}(1-\mathrm{e}^{-\beta\omega})z_1}\;.
\label{Gamma12b-1-new}
\end{align}
Again, this is immediately obtained from (\ref{Gamma12b-1}). After
insertion of external fields, the integral is obviously finite.

\subsection{4-point sector: planar graphs}

To each of the previous graphs there correspond three different new
graphs. The mixed analogue of (\ref{G4a-graph}) is 
\begin{align}
2\times \parbox{40mm}{\begin{picture}(35,17)
\put(0,-10){\begin{fmfgraph}(35,25)
\fmfleft{l1,l2}
\fmfright{r1,r2}
\fmf{fermion}{r1,i1,l1}
\fmf{fermion}{l2,i2,r2}
\fmffreeze
\fmf{dashes,tension=3}{i1,i3}
\fmf{photon,tension=3}{i2,i4}
\fmf{phantom_arrow,tension=0}{i2,i4}
\fmf{fermion,right}{i3,i4}
\fmf{fermion,right}{i4,i3}
\end{fmfgraph}}
\put(4,-13){\mbox{\small$\overline{z_2}$}}
\put(28,-13){\mbox{\small$z_1$}}
\put(10,-3){\mbox{\small$z_1'$}}
\put(22,-3){\mbox{\small$\overline{z_2}'$}}
\put(21,7){\mbox{\small$z_3'$}}
\put(10,7){\mbox{\small$\overline{z_4}'$}}
\put(28,17){\mbox{\small$\overline{z_4}$}}
\put(4,17){\mbox{\small$z_3$}}
\put(18.5,11){\mbox{\small$\gamma$}}
\put(7,3){\mbox{\small$\beta_1$}}
\put(25,3){\mbox{\small$\beta_2$}}
\end{picture}}
&\Gamma^{(1)}_{4e}[z_1,\overline{z_2},z_3,\overline{z_4}] 
\nonumber
\\*[-5ex]
&= -\frac{\lambda\lambda'}{4}\sum_{\ga}
I(z_1,\overline{z_2})\,I(z_3,\overline{z_4})
\int_\eps^\infty \frac{d\beta_1d\beta_2\;
\mathrm{e}^{-\frac{\mu_0^2}{2}(\beta_1+\beta_2)}}{
(1-\mathrm{e}^{-(\beta_1+\beta_2)\omega})^2}
\nonumber
\\*[-1ex]
&\times 
\mathrm{e}^{-\omega(\overline{z_4}\gamma-\overline{\gamma}z_3
+\frac{|\ga|^2}{
1-\mathrm{e}^{-(\beta_1+\beta_2)\omega}})}\;.
\label{Gamma-4a-new}
\end{align}
The graph where the new vertex is above has the same value so that we
count this graphs twice. Its value is obtained from (\ref{Gamma-4a})
by setting $\gamma_1=0$ and $\gamma_2=\gamma$. Obviously, only the
mode $\gamma=0$ produces a divergence:
\begin{align}
\Gamma^{(1)}_{4e,div}[z_1,\overline{z_2},z_3,\overline{z_4}] 
&= -\frac{\lambda\lambda'}{4}
I(z_1,\overline{z_2})\,I(z_3,\overline{z_4})
\int_\eps^\infty \frac{d\beta_1d\beta_2\;
\mathrm{e}^{-\frac{\mu_0^2}{2}(\beta_1+\beta_2)}}{
(1-\mathrm{e}^{-(\beta_1+\beta_2)\omega})^2}\;.
\end{align}

There is also the analogue of (\ref{G4a-graph}) where both vertices
are new ones:
\begin{align}
\parbox{40mm}{\begin{picture}(35,17)
\put(0,-10){\begin{fmfgraph}(35,25)
\fmfleft{l1,l2}
\fmfright{r1,r2}
\fmf{fermion}{r1,i1,l1}
\fmf{fermion}{l2,i2,r2}
\fmffreeze
\fmf{dashes,tension=3}{i1,i3}
\fmf{dashes,tension=3}{i2,i4}
\fmf{fermion,right}{i3,i4}
\fmf{fermion,right}{i4,i3}
\end{fmfgraph}}
\put(4,-13){\mbox{\small$\overline{z_2}$}}
\put(28,-13){\mbox{\small$z_1$}}
\put(10,-3){\mbox{\small$z_1'$}}
\put(22,-3){\mbox{\small$\overline{z_2}'$}}
\put(21,7){\mbox{\small$z_3'$}}
\put(10,7){\mbox{\small$\overline{z_4}'$}}
\put(28,17){\mbox{\small$\overline{z_4}$}}
\put(4,17){\mbox{\small$z_3$}}
\put(7,3){\mbox{\small$\beta_1$}}
\put(25,3){\mbox{\small$\beta_2$}}
\end{picture}}
&\Gamma^{(1)}_{4f}[z_1,\overline{z_2},z_3,\overline{z_4}] 
\nonumber
\\*[-2ex]
&= -\frac{{\lambda'}^2}{8} 
I(z_1,z_2)\,I(z_3,z_4)
\int_\eps^\infty \frac{d\beta_1d\beta_2\;
\mathrm{e}^{-\frac{\mu_0^2}{2}(\beta_1+\beta_2)}}{
(1-\mathrm{e}^{-(\beta_1+\beta_2)\omega})^2}\;.
\label{Gamma-4a-new2}
\\
\nonumber
\end{align}
This gives directly the divergent part.

All mixed analogues of the graph (\ref{Gamma4b}) are finite, because there is
no loop summation:
\begin{align}
2\times \parbox{48mm}{\begin{picture}(45,8)
\put(0,-15){\begin{fmfgraph}(45,20)
\fmfleft{l1,l2}
\fmfright{r1,r2}
\fmf{fermion,tension=2}{r1,i1}
\fmf{fermion,tension=1}{i1,i3}
\fmf{fermion,tension=2}{i3,l1}
\fmf{fermion,tension=2}{l2,i4}
\fmf{fermion,tension=1}{i4,i2}
\fmf{fermion,tension=2}{i2,r2}
\fmffreeze
\fmf{dashes}{i3,i4}
\fmf{photon}{i1,i2}
\fmf{phantom_arrow,tension=0}{i2,i1}
\end{fmfgraph}}
\put(4,-18.5){\mbox{\small$\overline{z_2}$}}
\put(38,-18.5){\mbox{\small$z_1$}}
\put(27,-19){\mbox{\small$\overline{z_2'}$}}
\put(15,-18.5){\mbox{\small$z_1'$}}
\put(38,7){\mbox{\small$\overline{z_4}$}}
\put(4,7){\mbox{\small$z_3$}}
\put(15,7){\mbox{\small$\overline{z_4'}$}}
\put(27,7){\mbox{\small$z_3'$}}
\put(18.5,-13){\mbox{\small$\beta_1$}}
\put(18.5,1){\mbox{\small$\beta_2$}}
\put(33,-6){\mbox{\small$\gamma$}}
\end{picture}}  
&\Gamma^{(1)}_{4g}[z_1,\overline{z_2},z_3,\overline{z_4}] 
\nonumber
\\*[-1ex]
&= -\frac{\lambda \lambda'}{4} \sum_{\gamma} 
   \int_\eps^\infty d\beta_1 d\beta_2\;
 \mathrm{e}^{-\frac{\mu_0^2}{2}(\beta_1+\beta_2)}
\mathrm{e}^{ \omega( -|\gamma_1|^2-\overline{\gamma} z_1 
-\overline{z_4}\gamma) }
\nonumber
\\*
&\qquad\times \mathrm{e}^{\omega(
\overline{z_2} \mathrm{e}^{-\beta_1 \omega} \gamma
+ \overline{z_2}\mathrm{e}^{-\beta_1 \omega} z_1 
+ \overline{z_4} \mathrm{e}^{-\beta_2\omega} z_3
+ \overline{\gamma} \mathrm{e}^{-\beta_2\omega} z_3) }.
\end{align}

\begin{align}
\parbox{48mm}{\begin{picture}(45,8)
\put(0,-15){\begin{fmfgraph}(45,20)
\fmfleft{l1,l2}
\fmfright{r1,r2}
\fmf{fermion,tension=2}{r1,i1}
\fmf{fermion,tension=1}{i1,i3}
\fmf{fermion,tension=2}{i3,l1}
\fmf{fermion,tension=2}{l2,i4}
\fmf{fermion,tension=1}{i4,i2}
\fmf{fermion,tension=2}{i2,r2}
\fmffreeze
\fmf{dashes}{i1,i2}
\fmf{dashes}{i3,i4}
\end{fmfgraph}}
\put(4,-18.5){\mbox{\small$\overline{z_2}$}}
\put(38,-18.5){\mbox{\small$z_1$}}
\put(27,-19){\mbox{\small$\overline{z_2'}$}}
\put(15,-18.5){\mbox{\small$z_1'$}}
\put(38,7){\mbox{\small$\overline{z_4}$}}
\put(4,7){\mbox{\small$z_3$}}
\put(15,7){\mbox{\small$\overline{z_4'}$}}
\put(27,7){\mbox{\small$z_3'$}}
\put(18.5,-13){\mbox{\small$\beta_1$}}
\put(18.5,1){\mbox{\small$\beta_2$}}
\end{picture}}  
&\Gamma^{(1)}_{4h}[z_1,\overline{z_2},z_3,\overline{z_4}] 
\nonumber
\\*
&= -\frac{{\lambda'}^2}{8} 
   \int_\eps^\infty d\beta_1 d\beta_2\;
 \mathrm{e}^{-\frac{\mu_0^2}{2}(\beta_1+\beta_2)}
\mathrm{e}^{\omega(
\overline{z_2}\mathrm{e}^{-\beta_1 \omega} z_1 
+ \overline{z_4} \mathrm{e}^{-\beta_2\omega} z_3)}\;.
\\
\nonumber
\end{align}

\subsection{4-point sector: non-planar graphs}

The analogues of the first non-planar graph (\ref{G4c-graph}) have one
possibility where a loop summation remains. Writing directly the
result of the Poisson re-summation, we have
\begin{align}
\parbox{55mm}{\begin{picture}(50,10)
\put(0,-20){\begin{fmfgraph}(50,25)
\fmfleft{l1,l2}
\fmfright{r1,r2}
\fmf{fermion,tension=2}{l2,i1}
\fmf{fermion,tension=1}{i1,i2}
\fmf{fermion,tension=1}{i2,i3}
\fmf{fermion,tension=2}{i3,r2}
\fmf{fermion,tension=1}{r1,i4,l1}
\fmffreeze
\fmf{dashes}{i2,i6,i4}
\fmffreeze
\fmf{phantom}{l2,a1,a2,i6}
\fmffreeze
\fmf{photon,left=0.3}{a2,i1}
\fmf{phantom_arrow,left=0.3}{a2,i1}
\fmf{photon,right=0.6,rubout=5}{a2,i3}
\end{fmfgraph}}
\put(8,-18){\mbox{\small$\overline{z_2}$}}
\put(38,-18){\mbox{\small$z_1$}}
\put(3,7){\mbox{\small$z_3$}}
\put(13,7){\mbox{\small$\overline{z_4'}$}}
\put(20,7){\mbox{\small$z_1'$}}
\put(26,7){\mbox{\small$\overline{z_2'}$}}
\put(34,7){\mbox{\small$z_3'$}}
\put(43,7){\mbox{\small$\overline{z_4}$}}
\put(30,1){\mbox{\small$\beta_1$}}
\put(17,1){\mbox{\small$\beta_2$}}
\put(13,-7){\mbox{\small$\gamma_2$}}
\end{picture}}  
&\Gamma^{(1)}_{4i}[z_1, \overline{z_2},z_3,\overline{z_4}] 
\nonumber
\\*
&= -\frac{\lambda\lambda'}{4\th^2} \sum_{\gamma_2^*} 
\mathrm{e}^{\omega(\overline{z_2}z_1+\overline{z_4}z_3+
\mathrm{i}\overline{\ga^*_2}z_3
+\overline{z_4}i\ga_2^*)}
\nonumber
\\*
&\times 
\int_\eps^\infty d\beta_1d\beta_2
\frac{\mathrm{e}^{-\frac{\mu_0^2}{2}(\beta_1+\beta_2)}}{
(1-\mathrm{e}^{-(\b_1+\b_2)\omega})^2}
\mathrm{e}^{-\frac{\omega}{1-\mathrm{e}^{-(\b_1+\b_2)\omega}}
|\ga_2^*|^2}\;.
\end{align}
Only the mode $\gamma_2^*=0$ contributes to the divergence:
\begin{align}
\Gamma^{(1)}_{4i,div}[z_1,\overline{z_2},z_3,\overline{z_4}] 
&= -\frac{\lambda\lambda'}{4\th^2} I(z_1,\overline{z_2})
\,I(z_3,\overline{z_4})
\int_\eps^\infty d\beta_1d\beta_2
\frac{\mathrm{e}^{-\frac{\mu_0^2}{2}(\beta_1+\beta_2)}}{
(1-\mathrm{e}^{-(\b_1+\b_2)\omega})^2}\;.
\end{align}
The other analogues of (\ref{G4c-graph}) are finite:
\begin{align}
\parbox{55mm}{\begin{picture}(50,10)
\put(0,-20){\begin{fmfgraph}(50,25)
\fmfleft{l1,l2}
\fmfright{r1,r2}
\fmf{fermion,tension=2}{l2,i1}
\fmf{fermion,tension=1}{i1,i2}
\fmf{fermion,tension=1}{i2,i3}
\fmf{fermion,tension=2}{i3,r2}
\fmf{fermion,tension=1}{r1,i4,l1}
\fmffreeze
\fmf{photon}{i2,i6,i4}
\fmf{phantom_arrow}{i6,i4}
\fmffreeze
\fmf{phantom}{l2,a1,a2,i6}
\fmffreeze
\fmf{dashes,left,rubout=5}{i3,i1}
\end{fmfgraph}}
\put(8,-18){\mbox{\small$\overline{z_2}$}}
\put(38,-18){\mbox{\small$z_1$}}
\put(3,7){\mbox{\small$z_3$}}
\put(13,7){\mbox{\small$\overline{z_4'}$}}
\put(20,7){\mbox{\small$z_1'$}}
\put(26,7){\mbox{\small$\overline{z_2'}$}}
\put(34,7){\mbox{\small$z_3'$}}
\put(43,7){\mbox{\small$\overline{z_4}$}}
\put(30,1){\mbox{\small$\beta_1$}}
\put(17,1){\mbox{\small$\beta_2$}}
\put(20,-14){\mbox{\small$\gamma_1$}}
\end{picture}}  
&\Gamma^{(1)}_{4j}[z_1,\overline{z_2},z_3,\overline{z_4}] 
\nonumber
\\*
&= -\frac{\lambda\lambda'}{4} \sum_{\gamma_1} 
\int_\eps^\infty d\beta_1d\beta_2\;
\mathrm{e}^{-\frac{\mu_0^2}{2}(\beta_1+\beta_2) }
\mathrm{e}^{\omega(
-|\gamma_1|^2 +\overline{z_2}\ga_1+\overline{z_2}z_1 -\overline{\ga_1}z_1)}
\nonumber
\\*
& \qquad\times \mathrm{e}^{\omega(-\overline{z_4}
\mathrm{e}^{-\beta_1\omega}\gamma_1
+\overline{z_4}\mathrm{e}^{-(\beta_1+\beta_2)\omega} z_3
+\overline{\gamma_1}\mathrm{e}^{-\beta_2 \omega} z_3)}
\;.
\end{align}
\begin{align}
\parbox{55mm}{\begin{picture}(50,10)
\put(0,-20){\begin{fmfgraph}(50,25)
\fmfleft{l1,l2}
\fmfright{r1,r2}
\fmf{fermion,tension=2}{l2,i1}
\fmf{fermion,tension=1}{i1,i2}
\fmf{fermion,tension=1}{i2,i3}
\fmf{fermion,tension=2}{i3,r2}
\fmf{fermion,tension=1}{r1,i4,l1}
\fmffreeze
\fmf{dashes}{i2,i4}
\fmf{dashes,left,rubout=5}{i3,i1}
\end{fmfgraph}}
\put(8,-18){\mbox{\small$\overline{z_2}$}}
\put(38,-18){\mbox{\small$z_1$}}
\put(3,7){\mbox{\small$z_3$}}
\put(13,7){\mbox{\small$\overline{z_4'}$}}
\put(20,7){\mbox{\small$z_1'$}}
\put(26,7){\mbox{\small$\overline{z_2'}$}}
\put(34,7){\mbox{\small$z_3'$}}
\put(43,7){\mbox{\small$\overline{z_4}$}}
\put(30,1){\mbox{\small$\beta_1$}}
\put(17,1){\mbox{\small$\beta_2$}}
\end{picture}}  
&\Gamma^{(1)}_{4k}[z_1,\overline{z_2},z_3,\overline{z_4}] 
\nonumber
\\*
&= -\frac{{\lambda'}^2}{4} 
\int_\eps^\infty d\beta_1d\beta_2\;
\mathrm{e}^{-\frac{\mu_0^2}{2}(\beta_1+\beta_2) }
\mathrm{e}^{\omega(\overline{z_2}z_1
+\overline{z_4}\mathrm{e}^{-(\beta_1+\beta_2)\omega} z_3)}\;.
\\
\nonumber ~
\\
\nonumber ~
\end{align}

All analogues of the second non-planar graph (\ref{Gamma4d}) are finite:
\begin{align}
2\times \parbox{48mm}{\begin{picture}(45,8)
\put(0,-15){\begin{fmfgraph}(45,20)
\fmfleft{l1,l2}
\fmfright{r1,r2}
\fmf{fermion,tension=2}{r1,i1}
\fmf{fermion,tension=1}{i1,i3}
\fmf{fermion,tension=2}{i3,l1}
\fmf{fermion,tension=2}{l2,i4}
\fmf{fermion,tension=1}{i4,i2}
\fmf{fermion,tension=2}{i2,r2}
\fmffreeze
\fmf{dashes}{i1,i5,i4}
\fmf{photon,rubout=5}{i3,i7}
\fmf{photon}{i2,i7}
\fmf{phantom}{i3,i6,i7,i2}
\fmf{phantom_arrow,tension=0}{i2,i6}
\end{fmfgraph}}
\put(4,-18.5){\mbox{\small$\overline{z_2}$}}
\put(38,-18.5){\mbox{\small$z_1$}}
\put(27,-19){\mbox{\small$\overline{z_2'}$}}
\put(15,-18.5){\mbox{\small$z_1'$}}
\put(38,7){\mbox{\small$\overline{z_4}$}}
\put(4,7){\mbox{\small$z_3$}}
\put(15,7){\mbox{\small$\overline{z_4'}$}}
\put(27,7){\mbox{\small$z_3'$}}
\put(18.5,-13){\mbox{\small$\beta_1$}}
\put(19.5,1){\mbox{\small$\beta_2$}}
\put(28,-2){\mbox{\small$\gamma_2$}}
\end{picture}}
&\Gamma^{(1)}_{4l}[z_1,\overline{z_2},z_3,\overline{z_4}] 
\nonumber
\\[-1ex]
&=-\frac{\lambda \lambda'}{4} \sum_{\gamma_2}
\int_\eps^\infty d\beta_1 d\beta_2\, \mathrm{e}^{
-\frac{\mu_0^2}{2} (\beta_1+\beta_2)  }
\mathrm{e}^{
-\omega |\gamma_2|^2 }
\nonumber
\\
&\times
\mathrm{e}^{\omega(\overline{z_2}\gamma_2
+ (\overline{z_2}-\overline{\gamma_2})\mathrm{e}^{-\beta_1 \omega}z_1
-\overline{z_4}\gamma_2
+ (\overline{z_4}+\overline{\gamma_2})\mathrm{e}^{-\beta_2 \omega}z_3)}\;.
\end{align}

\begin{align}
\parbox{48mm}{\begin{picture}(45,8)
\put(0,-15){\begin{fmfgraph}(45,20)
\fmfleft{l1,l2}
\fmfright{r1,r2}
\fmf{fermion,tension=2}{r1,i1}
\fmf{fermion,tension=1}{i1,i3}
\fmf{fermion,tension=2}{i3,l1}
\fmf{fermion,tension=2}{l2,i4}
\fmf{fermion,tension=1}{i4,i2}
\fmf{fermion,tension=2}{i2,r2}
\fmffreeze
\fmf{dashes}{i1,i5,i4}
\fmf{dashes,rubout=5}{i3,i2}
\end{fmfgraph}}
\put(4,-18.5){\mbox{\small$\overline{z_2}$}}
\put(38,-18.5){\mbox{\small$z_1$}}
\put(27,-19){\mbox{\small$\overline{z_2'}$}}
\put(15,-18.5){\mbox{\small$z_1'$}}
\put(38,7){\mbox{\small$\overline{z_4}$}}
\put(4,7){\mbox{\small$z_3$}}
\put(15,7){\mbox{\small$\overline{z_4'}$}}
\put(27,7){\mbox{\small$z_3'$}}
\put(18.5,-13){\mbox{\small$\beta_1$}}
\put(19.5,1){\mbox{\small$\beta_2$}}
\end{picture}}
&\Gamma^{(1)}_{4m}[z_1,\overline{z_2},z_3,\overline{z_4}] 
\nonumber
\\[-1ex]
&=-\frac{{\lambda'}^2}{8} 
\int_\eps^\infty d\beta_1 d\beta_2\, \mathrm{e}^{
-\frac{\mu_0^2}{2} (\beta_1+\beta_2)  }
\mathrm{e}^{\omega(
\overline{z_2}\mathrm{e}^{-\beta_1 \omega}z_1
+ \overline{z_4}\mathrm{e}^{-\beta_2 \omega}z_3)}\;.
\\
\nonumber
\end{align}

\end{fmffile}

\quad

In conclusion, the new divergences reproduce the action functional
(\ref{action2}) (which includes the new vertex) so that the model is
one-loop renormalizable.

\section{$\b$-functions}

Here we compute the $\beta$-functions of our model in the most
interesting case where $\theta$ is irrational and satisfies the
Diophantine condition \eqref{weakdioph}. We can summarize the divergent
Green's functions to
\begin{align}
\Gamma^{(1)}_2[z_1,\overline{z_2}] &=
\frac{1}{2\omega}\Big( \lambda\Big(1+\frac{1}{\theta^2}\Big) 
+ \lambda'\Big)
I(z_1,\overline{z_2})\Big(\frac{1}{\epsilon\omega}
+\Big(1-\frac{\mu_0^2}{2\omega} \Big)\ln \frac{1}{\epsilon}\Big)\;,
\\
\Gamma^{(1)}_4[z_1,\overline{z_2},z_3,\overline{z_4}] &=
-\frac{\lambda}{4\omega^2} \Big(1+\frac{1}{\theta^2}\Big) 
V(z_1,\overline{z_2},z_3,\overline{z_4}) 
\;\ln \frac{1}{\epsilon} 
\\
& -\frac{1}{8\omega^2}\Big(
{\lambda'}^2 + \frac{3\lambda^2}{\theta^2} 
+ 2\lambda\lambda'\Big(1+\frac{1}{\theta^2}\Big)\Big)
I(z_1,\overline{z_2}) I(z_3,\overline{z_4}) \;\ln \frac{1}{\epsilon} \;.
\end{align}
This leads to the following relation between the renormalized
quantities $\mu_R,\lambda_R,\lambda'_R$ and the bare 
quantities $\mu_0,\lambda,\lambda'$:
\begin{align}
\mu_R& =\mu_0+\frac{1}{2\omega}\Big( \lambda\Big(1+\frac{1}{\theta^2}\Big) 
+ \lambda'\Big)\Big(\frac{1}{\epsilon\omega} 
+\Big(1-\frac{\mu_0^2}{2\omega} \Big)\ln \frac{1}{\epsilon}\Big)\;,
\\
\lambda_R &= \lambda-\frac{\lambda^2}{4\omega^2} 
\Big(1+\frac{1}{\theta^2}\Big) 
\;\ln \frac{1}{\epsilon} \;,
\\
\lambda_R' &= \lambda' -\frac{1}{4\omega^2}\Big(
{\lambda'}^2 + \frac{3\lambda^2}{\theta^2} 
+ 2\lambda\lambda'\Big(1+\frac{1}{\theta^2}\Big)\Big)
\;\ln \frac{1}{\epsilon} \;.
\end{align}
Solving for the bare quantities, we obtain
\begin{align}
\mu_0& =\mu_R-\frac{1}{2\omega}\Big( \lambda_R\Big(1+\frac{1}{\theta^2}\Big) 
+ \lambda_R'\Big)\Big(\frac{1}{\epsilon\omega} 
+\Big(1-\frac{\mu_0^2}{2\omega} \Big)\ln \frac{1}{\epsilon}\Big)\;,
\\
\lambda &= \lambda_R+\frac{\lambda_R^2}{4\omega^2} 
\Big(1+\frac{1}{\theta^2}\Big) 
\;\ln \frac{1}{\epsilon} \;,
\\
\lambda' &= \lambda_R' +\frac{1}{4\omega^2}\Big(
{\lambda_R'}^2 + \frac{3\lambda_R^2}{\theta^2} 
+ 2\lambda_R\lambda_R'\Big(1+\frac{1}{\theta^2}\Big)\Big)
\;\ln \frac{1}{\epsilon} \;.
\end{align}
The $\beta$-functions are therefore given by
\begin{align}
\beta&= \frac{\lambda_R^2}{4\omega^2} 
\Big(1+\frac{1}{\theta^2}\Big)\;, &
\beta'&= \frac{1}{4\omega^2}\Big(
{\lambda_R'}^2 + \frac{3\lambda_R^2}{\theta^2} 
+ 2\lambda_R\lambda_R'\Big(1+\frac{1}{\theta^2}\Big)\Big)\;.
\end{align}
We thus conclude that the model is neither asymptotically free nor has
a finite fixed point as it was the case for the renormalizable
$\phi^4_4$-model on the Moyal plane \cite{Grosse:2004ik}. The result
also shows that there does not exist a ($\theta$-independent) relation
between the coupling constants $\lambda',\lambda'$ which is preserved
over all scales.

\section*{Conclusion}

In the general formalism of noncommutative geometry, we have proposed
a definition of a field theory on a projective module which is the
noncommutative analogue of a scalar field theory with non-trivial
topology. In the case of a noncommutative torus, the simplest
non-trivial projective modules can be constructed using
representations of the Heisenberg commutation relations. We have given
a detailed account of the corresponding field theory, whose properties
are reminiscent of those of a rectangular matrix model, whereas 
algebra-valued fields correspond to square matrices.

In particular, we have shown that the model suffers from an
UV/IR-mixing. Contrarily to what happens for field theories on the
Moyal plane, here the arithmetical nature of the deformation
parameters plays a central role.  For $\theta$ satisfying a
Diophantine condition, the theory is one-loop renormalizable at the
price of introducing the extra counterterm 
\begin{equation}
\left[\mathrm{Tr}_{\cal A_{\theta}}
\left(\phi,\phi\right)_{\cal A_{\theta}}\right]^{2}
\end{equation}
in the action, whereas the basic action we started with is
\begin{equation}
\mathrm{Tr}_{\cal A_{\theta}}
\left[\left(\phi,\phi\right)_{\cal A_{\theta}}^2\right].
\end{equation}
In a commutative field theory, such a counterterm would break the
locality of the action but in a matrix model it corresponds to
$[\mathrm{Tr}(M^{\dagger}M)]^{2}$ and is perfectly acceptable. 
This is the only way we found to cure the UV/IR-mixing problem. Note
that the adjunction of such a counterterm fits perfectly with the
spirit of \cite{Grosse:2004yu}. The only difference is that they need
to modify the propagator, whereas we need to change the interaction.

The occurrence of the Diophantine condition is not so surprising
here. In fact, after suitable Poisson re-summations, divergences in the
$\beta$ integrals appear whenever the lattice $\Gamma$ and its dual
$\widehat{\Gamma}$ have elements that are close enough. In the
rational case, the two lattices have non-trivial intersections that lead
to sharp divergences corresponding to the usual counterterms but
other\-wise the distance is bounded by a fixed number. In the
irrational case, $\Gamma\cap\widehat{\Gamma}=\left\{0\right\}$ but the
two lattices  have elements that are as close as possible. The Diophantine
condition allows to set a lower bound for the distance between
$\Gamma$ and $\widehat{\Gamma}$ when restricted to a ball of radius
$R$ as a function of $R$.

\section*{Acknowledgments}
RW thanks the Universit\'e de Provence. 
J.-H. J. gratefully acknowledges his fellowship of the
Friedrich-Ebert-Stiftung and wants to thank Prof. G. Grensing for
helpful discussions.

\appendix
\section{Comments on the duality covariant 
model}
\label{dualcov}
In the duality covariant $\phi_4^4$-theory on the Moyal plane
\eqref{eq:classaction}, the planar one-loop two-point graph  
reads:

\begin{align}
&\Ga^{(1)}_{1,P}(x_1,x_2)=\int dx_3\,dx_4\,V_\star(x_1,x_2,x_3,x_4)
\,H^{-1}_\epsilon(x_3,x_4)
\nonumber\\
&\quad=\frac{\lambda\Omega}{4!4\pi^6\th^5}\int dx_3\,dx_4\,
\delta(x_1-x_2+x_3-x_4)\,e^{-2i\th^{-1}(x_1,x_2)}\,e^{-2i\th^{-1}(x_3,x_4)}
\nonumber\\
&\hspace{6cm}\x\int_\epsilon^\infty d\b\,\frac{e^{-\th  \mu_0^2\b/4\Omega}}
{\sinh^2(\b)}
\,e^{-\frac{\Omega}{2\th}\big(\coth(\b/2)|x_3-x_4|^2
+\tanh(\b/2)|x_3+x_4|^2\big)}
\nonumber\\
&\quad=\frac{\lambda\Omega}{4!4\pi^6\th^5}\,e^{-2i\th^{-1}(x_1,x_2)}\,
\int_\epsilon^\infty d\b\,\frac{e^{-\th  \mu_0^2\b/4\Omega}}{\sinh^2(\b)}
\int dx_3\,e^{-2i\th^{-1}(x_3,x_1-x_2)}\nonumber\\
&\hspace{9.5cm}\x e^{-\frac{\Omega}{2\th}\big(\coth(\b/2)|x_1-x_2|^2
+4\tanh(\b/2)|x_3|^2\big)}
\nonumber\\
&\quad=\frac{\lambda}{4!4^2\pi^4\th^3\Omega}\,e^{-2i\th^{-1}(x_1,x_2)}\,
\int_\epsilon^\infty d\b\,\frac{e^{-\th
    \mu_0^2\b/4\Omega}}{\sinh^2(\b)
\tanh^2(\b/2)}
\,e^{-\frac{1+\Omega^2}{2\th\Omega}\coth(\b/2)|x_1-x_2|^2}.
\end{align}
To obtain the third equality we used the translation $x_3\to
x_3+\thalf(x_1-x_2)$, which leaves unchanged the phase factor
due to the skew-symmetry of $S$.\\
This amplitude is divergent in the coinciding-points limit, i.e.\ it
is UV-divergent.  We now extract local divergences from the previous
expression.  They will correspond to the mass, wave-function and
oscillator frequency renormalization.  In the same manner that
momentum space renormalization corresponds to subtract Feynman
amplitude with zero external momenta, configuration space
renormalization is done by attaching all the external legs at the same
point. To be able to apply this guiding principle, we have to go to
the quantum effective action level, that is to smear the regularized
Green functions with external or background fields, and to expand them
on a neighborhood of a given point via a Taylor expansion with
integral remainder.

\begin{align}
\label{eq:2ppeffaction}
\Ga^{(1)}_{1,P}[\vf_1,\vf_2]&:=\int dx\,dy\,\Ga^{(1)}_{1,P}(x,y)\,
\vf_1(x)\,\vf_2(y)
\nonumber\\
&=\frac{\lambda}{4!4^2\pi^4\th^3\Omega}\,\int dx\,dy\,\vf_1(x)\,
\vf_2(x+y)\,e^{-2i\th^{-1}(x,y)}\nonumber\\
&\hspace{4cm}\x\int_\eps^\infty d\b\,
\frac{e^{-\th \mu_0^2\b/4\Omega}}{\sinh^2(\b)\tanh^2(\b/2)}\,
e^{-\frac{1+\Omega^2}{2\th\Omega}\coth(\b/2)|y|^2}\nonumber\\
&=\frac{\lambda}{4!4^2\pi^4\th^3\Omega}\,\int dx\,dy\,\vf_1(x)\,
\Big(\vf_2(x)+y^\mu\pa_\mu\vf_2(x)+\thalf y^\mu y^\nu\pa_\mu\pa_\nu\vf_2(x)
\nonumber \\
&\hspace{4cm}+\thalf y^\mu y^\nu y^\rho\, \int_0^1d\xi\,(1-\xi)^2
\big(\pa_\mu\pa_\nu\pa_\rho\vf\big)(x+\xi y)\Big)
\,e^{-2i\th^{-1}(x,y)}\nonumber\\
& \hspace{4cm}
\x\int_\eps^\infty d\b\,
\frac{e^{-\th \mu_0^2\b/4\Omega}}{\sinh^2(\b)\tanh^2(\b/2)}
\,e^{-\frac{1+\Omega^2}{2\th\Omega}\coth(\b/2)|y|^2},
\end{align}

Let us label by $X_i$, $i=1,\cdots,4$ the four different terms
coming from the Taylor expansion.\\
The first summand reads:
\begin{align}
\label{eq:X1}
X_1&=\frac{\lambda}{4!4^2\pi^4\th^3\Omega}\,\int dx\,dy\,\vf_1(x)\,
\vf_2(x)\,e^{-2i\th^{-1}(x,y)}
\int_\eps^\infty d\b\,\frac{e^{-\th \mu_0^2\b/4\Omega}}
{\sinh^2(\b)\tanh^2(\b/2)}
\,e^{-\frac{1+\Omega^2}{2\th\Omega}
\coth(\b/2)|y|^2}\nonumber\\
&=\frac{\lambda\Omega}{96\pi^2\th(1+\Omega^2)^2}\int dx\,\vf_1(x)\,\vf_2(x)
\int_\eps^\infty d\b\,\frac{e^{-\th \mu_0^2\b/4\Omega}}{\sinh^2(\b)}
\,e^{-\frac{2\Omega}{\th(1+\Omega^2)}\tanh(\b/2)|x|^2}\nonumber\\
&=\frac{\lambda\Omega}{96\pi^2\th(1+\Omega^2)^2}\int dx\,\vf_1(x)\,\vf_2(x)
\int_\eps^\infty d\b\,\frac{e^{-\th \mu_0^2\b/4\Omega}}{\sinh^2(\b)}\nonumber\\
&\quad -\frac{\lambda\Omega^2}{96\pi^2\th^2(1+\Omega^2)^3}\int dx
\,\vf_1(x)\,|x|^2\,\vf_2(x)
\int_\eps^\infty d\b\,\frac{e^{-\th \mu_0^2\b/4\Omega}}{\sinh(\b)\cosh^2(\b/2)}
+O(\eps^0).
\end{align}

For the second summand, we obtain using integration by parts:

\begin{align}
\label{eq:X2}
X_2&=\frac{\lambda}{4!4^2\pi^4\th^3\Omega}\,\int dx\,dy\,\vf_1(x)\,
y^\mu\pa_\mu\vf_2(x)\,e^{-2i\th^{-1}(x,y)}
\nonumber\\
& \hspace{6cm}
\x\int_\eps^\infty d\b\,\frac{e^{-\th \mu_0^2\b/4\Omega}}{\sinh^2(\b)
\tanh^2(\b/2)}
\,e^{-\frac{1+\Omega^2}{2\th\Omega}\coth(\b/2)|y|^2}\nonumber\\
&=\frac{\lambda}{4!4^2\pi^4\th^3\Omega}\,\int dx\,dy\,\vf_1(x)\,
\frac{i}{2}\th^{\mu\nu} \pa_\mu\pa_\mu\vf_2(x)\,e^{-2i\th^{-1}(x,y)}\nonumber\\
& \hspace{6cm}
\x\int_\eps^\infty d\b\,\frac{e^{-\th \mu_0^2\b/4\Omega}}{\sinh^2(\b)
\tanh^2(\b/2)}
\,e^{-\frac{1+\Omega^2}{2\th\Omega}\coth(\b/2)|y|^2}\nonumber\\
&=0.
\end{align}
Thus, $X_2$ identically vanishes due to
the skew-symmetry of the deformation matrix. For the third one,
we get:

\begin{align}
\label{eq:X3}
X_3&=\frac{\lambda}{4!4^2\pi^4\th^3\Omega}\,\int dx\,dy\,\vf_1(x)\,
\thalf y^\mu y^\nu\pa_\mu\pa_\nu\vf_2(x)
\,e^{-2i\th^{-1}(x,y)}\nonumber\\
& \hspace{6cm}
\x\int_\eps^\infty d\b\,
\frac{e^{-\th \mu_0^2\b/4\Omega}}{\sinh^2(\b)\tanh^2(\b/2)}
\,e^{-\frac{1+\Omega^2}{2\th\Omega}\coth(\b/2)|y|^2}\nonumber\\
&=-\frac{\lambda\Omega}{4!4\pi^2\th(1+\Omega^2)^2}
\frac{1}{8}\th^{\mu_1\nu_1}\th^{\mu_2\nu_2}\int dx\,
\vf_1(x)\,\pa_{\mu_1}\pa_{\mu_2}\vf_2(x)
\int_\eps^\infty d\b\,\frac{e^{-\th \mu_0^2\b/4\Omega}}{\sinh^2(\b)}\nonumber\\
&\quad\x
\Big(-\frac{4\Omega}{\th(1+\Omega^2)}\,\tanh(\b/2)\,\delta_{\nu_1\nu_2}
+\frac{16\Omega^2}{\th^2(1+\Omega^2)^2}\tanh^2(\b/2)x_{\nu_1}x_{\nu_2}\Big)
e^{-\frac{2\Omega}{\th(1+\Omega^2)}\tanh(\b/2)|x|^2}
\nonumber\\
&=-\frac{\lambda\Omega^2}{384\pi^2(1+\Omega^2)^3}\int dx\,\vf_1(x)\,
\tri\vf_2(x)
\,\int_\eps^\infty d\b\,
\frac{e^{-\th \mu_0^2\b/4\Omega}}{\sinh(\b)\cosh^2(\b/2)}
+O(\eps^0).
\end{align}

Let us show that $X_4$, the integral remainder of the Taylor
expansion, gives a finite contribution. Indeed, we have

\begin{align}
\label{eq:X4}
X_4&=\frac{\lambda}{4!4^2\pi^4\th^3\Omega}\,\int dx\,dy\,\vf_1(x)\,
\thalf y^\mu y^\nu y^\rho\, \int_0^1d\xi\,(1-\xi)^2
(\pa_\mu\pa_\nu\pa_\rho\vf_2(x+\xi y)\,e^{-2i\th^{-1}(x,y)}\nonumber\\
& \hspace{5cm}
\x\int_\eps^\infty d\b\,
\frac{e^{-\th \mu_0^2\b/4\Omega}}{\sinh^2(\b)\tanh^2(\b/2)}
\,e^{-\frac{1+\Omega^2}{2\th\Omega}\coth(\b/2)|y|^2},
\end{align}
thus,

\begin{align*}
|X_4|&\leq C\,\|\pa_\mu\pa_\nu\pa_\rho\vf_2\|_\infty\,\|\vf_1\|_1\,
\int_0^1d\xi\,(1-\xi)^2\,\int dy\,| y^\mu y^\nu y^\rho|
\\
&\hspace{5.5cm}\x
\int_\eps^\infty d\b\,
\frac{e^{-\th \mu_0^2\b/4\Omega}}{\sinh^2(\b)\tanh^2(\b/2)}
\,e^{-\frac{1+\Omega^2}{2\th\Omega}\coth(\b/2)|y|^2}\\
&\quad=C'\,\|\pa_\mu\pa_\nu\pa_\rho\vf\|_\infty\,\|\vf\|_1\,
\int dy\,| y^\mu y^\nu y^\rho|e^{-\frac{1+\Omega^2}{2\th\Omega}|y|^2}
\int_\eps^\infty d\b\,e^{-\th \mu_0^2\b/4\Omega}\,
\frac{\tanh^{3/2}(\b/2)}{\sinh^2(\b)},
\end{align*}
which is finite in the limit $\eps\to 0$ since $\vf$ is a Schwartz
function and since\\ $\tanh^{3/2}(\b/2)\sinh^{-2}(\b)$ is integrable in
$\b=0$. \\

Putting all together, we finally obtain:

\begin{align*}
\Ga^{(1)}_{1,P}[\vf_1,\vf_2]&=
\frac{\lambda\Omega}{96\pi^2\th(1+\Omega^2)^2}\int
dx\,\vf_1(x)\,\vf_2(x)\,\int_\eps^\infty d\b\,
\frac{e^{-\th \mu_0^2\b/4\Omega}}{\sinh^2(\b)}\\
&\quad-\frac{\lambda\Omega^2}{384\pi^2(1+\Omega^2)^3}\int
dx\,\vf_1(x)\,\big(\tri+\frac{X^2}{\Omega^2}\big)\vf_2(x)\,
\int_\eps^\infty d\b\,
\frac{e^{-\th \mu_0^2\b/4\Omega}}{\sinh(\b)\cosh^2(\b/2)}\\
&\quad+O(\eps^0).
\end{align*}
The last two lines are the relevant contribution to the mass
renormalization and marginal contribution to the wave-function and to
the oscillator frequency renormalization.

\end{document}